%% file: main.tex
\documentclass[12pt,american,round,hidelinks]{article}
\usepackage[T1]{fontenc}
\usepackage[utf8]{inputenc}
\usepackage{empheq}
\usepackage{makecell}
\usepackage{float}
\usepackage{rotfloat}
\usepackage{subcaption}
\usepackage{babel}
\usepackage[ruled]{algorithm2e}
\usepackage{tabularx}
\newcolumntype{Y}{>{\centering\arraybackslash}X}
\usepackage[dvipsnames]{xcolor}
\usepackage{array}
\usepackage{microtype}
\usepackage{fancyhdr}
\usepackage[flushleft]{threeparttable}
\usepackage{rotating}
\usepackage{longtable}
\usepackage{booktabs,multirow,multicol} 
\usepackage{bm}
\usepackage{mathptmx}
\usepackage{amsmath}
\usepackage{amssymb}
\usepackage{graphicx}
\usepackage{setspace}
\usepackage{pdflscape}
\usepackage{empheq}
\usepackage[export]{adjustbox}
\usepackage{enumitem}
\usepackage{amsthm}
\usepackage{epigraph}
\usepackage{rotating}
\DeclareMathAlphabet{\mathcal}{OMS}{cmsy}{m}{n}

\usepackage[a4paper]{geometry}
\usepackage{color}

\usepackage[authoryear]{natbib}
\usepackage[unicode=true,
 bookmarks=true,bookmarksnumbered=false,bookmarksopen=false,
 breaklinks=false,pdfborder={0 0 1},backref=false,colorlinks=false]
 {hyperref}

\newcommand{\red}{\textcolor{red}}

\newtheorem{theorem}{Theorem}
\newtheorem*{theorem*}{Theorem}
\newtheorem{proposition}{Proposition}
\newtheorem{assumption}{Assumption}
\newtheorem{manualassumptioninner}{Assumption}
\newenvironment{manualassumption}[1]{%
  \renewcommand\themanualassumptioninner{#1}%
  \begin{manualassumptioninner}%
}{%
  \end{manualassumptioninner}%
}
\newtheorem{definition}{Definition}
\newtheorem{lemma}{Lemma}

\newcommand{\mask}[1]{}
\newcolumntype{L}[1]{>{\raggedright\let\newline\\arraybackslash\hspace{0pt}}m{#1}}
\newcolumntype{C}[1]{>{\centering\let\newline\\arraybackslash\hspace{0pt}}m{#1}}
\newcolumntype{R}[1]{>{\raggedleft\let\newline\\arraybackslash\hspace{0pt}}m{#1}}

\usepackage[flushleft]{threeparttable}
\usepackage{booktabs,multirow,multicol} 

\usepackage{tikz}
\input{TikzPics/tikz_setup.tex}

\hypersetup{colorlinks=true, linkcolor= BrickRed, citecolor = BrickRed, filecolor = BrickRed, urlcolor = BrickRed, hypertexnames = true}

\usepackage{caption}
\usepackage{bbm}

\def \firstdraftdate {November 15, 2024}
\def \thisdraftdate {May 1, 2026}

\begin{document}

\begin{titlepage}
\title{\mbox{A Ranking Representation of Optimal Sequential Search}}
\author{Tinghan Zhang\thanks{Tilburg University, Department of Econometrics and Operations Research. \href{mailto:t.zhang_2@tilburguniversity.edu}{T.Zhang\_2@tilburguniversity.edu}. This paper was previously circulated as “Estimating Sequential Search Models Based on a Partial Ranking Representation”. I am indebted to Tobias Klein and Christoph Walsh for their careful supervision and encouragement in accomplishing this paper. I thank Jaap Abbring, Bart Bronnenberg, Rafael Greminger, Elisabeth Honka, Johannes Kasinger, Jos\'e L. Moraga, Ao Wang, Yifan Yu, Kun Zhang, and two anonymous reviewers for WoPA 2025, as well as audiences from Tilburg Structural Econometrics Group and EWMES 2025, for their helpful comments. All errors are my own. }}
\date{First version: \firstdraftdate \\ Current version: \thisdraftdate \\ \vspace{0.1em}}
\maketitle
\begin{center}
\textbf{Please check the \href{https://www.dropbox.com/scl/fi/1fusn7428ic8kp92cle4b/Partial_Ranking.pdf?rlkey=k9yj6yoztjrdgql0i3q02vo1r&st=g6rc41ul&dl=0}{latest version}.}
\end{center}
\bigskip
\begin{abstract} \singlespacing \vspace{-1em} 
\noindent Sequential search models provide a powerful framework for studying consumer search using rich data that records the sequence of consumer actions taken during the search process. In existing empirical applications, their implementation often builds on optimal policies, in which later decisions depend on outcomes from earlier actions that are often fully observed by researchers. Therefore, implementation is largely restricted by computation burden and limited model flexibility. This paper establishes a theoretical equivalence showing that, under common and mild assumptions of Independence and Invariance, a sequential search process is optimal if and only if a corresponding ranking over all feasible actions throughout the process holds, thereby introducing a ranking representation of optimal sequential search. This representation enables a novel, simple, and unified empirical strategy for implementing sequential search models. For the classic \cite{weitzman1979optimal} model, the proposed approach reduces simulation requirements while improving accuracy, computational efficiency, and ease of implementation. We further show that the same strategy extends to a broad class of sequential search settings, including partially observed action sequences and multi-stage information acquisition, such as discovery. Overall, the results enhance both the tractability and the empirical applicability of sequential search models.\\
\bigskip
\vspace{0in}
\vspace{0in}\\
\end{abstract}

\setcounter{page}{0}
\thispagestyle{empty}
\end{titlepage}
\pagebreak 

\newpage

\onehalfspacing

\section{Introduction} \label{sec:introduction}

In markets with a proliferation of product variety, consumers often lack sufficient information to determine which alternative best aligns with their preferences. To reduce the impact of this uncertainty, they can engage in search to acquire product information and more accurately assess the utility of searched products, enabling better-informed purchases. Consumers’ search process, hence, reflects trade-offs between the expected gains from extra product information and the associated search costs. Structurally analyzing the sequence of actions recording the search process can reveal consumer preferences, characterize market demand, and also help assess how market strategies, such as platform design \citep[e.g.,][]{chen2017sequential,gu2022consumer,donnelly2024welfare}, advertising \citep[e.g.,][]{chan2015consumer}, product ranking \citep[e.g.,][]{ursu2018power, compiani2024online}, and recommendation systems \citep[e.g.,][]{kim2011mapping, wan2024product}, interact with the search process to influence firm profitability and consumer welfare.

Sequential search models are among the most commonly applied frameworks in empirical studies. The setup of these models is that, when facing product-level uncertainty, consumers acquire product information step by step through a series of costly actions that are subject to feasibility constraints;\footnote{For example, entering a product page reveals product details and enables checkout, while entering the checkout page reveals shipping costs and enables purchase.} at each step, they select the next action based on all information obtained from previous actions, until selecting a purchase to a product whose information is fully revealed.\footnote{Two well-known models that stand in contrast to the sequential search models are the simultaneous search model \citep[e.g.][]{stigler1961economics} and the learning model \citep[e.g.][]{erdem1996decision}. The former assumes that consumers determine the products to be searched \textit{a priori} and purchase one only after search is completed. The latter assumes that consumers repeatedly learn about products and update their beliefs, and may make a purchase at any time.} In a milestone paper, \cite{weitzman1979optimal} studies sequential search in a process involving two action types, inspections (obtaining full product information) and purchases, and proves that its optimal solution is characterized by a set of policies, called optimal rules. Accordingly, conducting a sequential search process optimally is equivalent to following these rules in making all search and purchase decisions throughout the process. This representation has been widely adopted in subsequent research to describe the optimality of other sequential search models. 

However, the policy-based representation poses substantial difficulties for empirical applications, primarily because it introduces a sequential dependence issue: under optimal rules, consumers’ decisions at each step depend on information from all previous search actions, often including private evaluations that researchers cannot observe or identify.\footnote{The original paper considers search only over identifiable product attributes, such as price. While analytically convenient, such a setting struggles to explain divergence in subsequent decisions arising from identical prior search processes. Empirical work, therefore, often allows for unobserved idiosyncratic post-inspection evaluations.} Thus, without further assumptions, later search and purchase decisions cannot be analyzed unless all previous search outcomes are simulated. This imposes a large dimensionality requirement on simulation and a substantial implementation burden, and makes the model for the process (the Weitzman model hereafter) dependent on a restricted specification and complete information about the search process, thereby limiting its practical simplicity and applicability.

To address these empirical challenges, we propose a novel representation of optimal sequential search that departs from the optimal rules. Our main conclusion is that, under two foundational assumptions that are already made in the classic Weitzman model, Independence and Invariance, an optimal sequential search process is observationally equivalent to a partial ranking of all actions feasible throughout the search process. That is, for an action sequence that can be interpreted as the sequence of selected actions in an optimal sequential search process, we can equally interpret it as a partial ranking of all actions the consumer can select during that process. The two interpretations have identical probabilities and are empirically indistinguishable. Hence, an optimal sequential search process admits a ranking representation.

The intuition for this equivalence is that, although the private information underlying each search decision is not directly observed, subsequent decisions in the search process can be used to infer which information is relevant at earlier stages. This generates a sequence of pairwise comparisons that, when combined, induces a ranking of feasible actions. Consider a Weitzman-style sequential search process in which a consumer inspects $A$, $B$, and $C$, and then purchases $A$. The decision to inspect $C$ depends on whether its expected gain exceeds the highest utility among $A$ and $B$. Since utilities are unobserved, an analyst must consider both cases in which $A$ is preferred to $B$ and vice versa. However, the final purchase of $A$ reveals that $A$’s utility exceeds that of $B$, so the decision to inspect $C$ should depend only on $A$. This yields a local ranking in which inspecting $C$ is preferred to purchasing $A$, which in turn is preferred to purchasing $B$.

The ranking representation alleviates the complexities induced by sequential dependence and offers two main advantages for empirical application. First, it leads to a simple empirical strategy: implementing a sequential search model via its ranking representation. Compared with the policy-based, the ranking representation provides an equally complete but more parsimonious set of inequality conditions to capture optimal sequential search. Moreover, as it shares the probability structure of ranking models, researchers can leverage the latter’s established econometric tools to streamline the application of sequential search models. Using the Weitzman model as a benchmark, we show that its likelihood function can be expressed in a value-difference form. Building on this expression, we provide direct identification arguments and develop a GHK-style simulator, termed the rank-based GHK simulator, for simulated maximum likelihood estimation.\footnote{The GHK-style simulator \citep{geweke1989bayesian, hajivassiliou1998method, keane1994computationally} was originally proposed as an importance sampling method for simulating choice probabilities in multivariate probit models, and it subsequently became a primary tool for estimating ranking models.} Compared with the existing policy-based simulator, the proposed approach allows the direct computation of the conditional distribution of search outcomes for products not purchased, thereby reducing the required simulation dimension without loss of efficiency. Monte Carlo experiments show that our method preserves the accuracy advantages of policy-based GHK simulators while substantially lowering their high implementation complexity.

Second, the ranking representation broadens the empirical applicability of sequential search models. It is not limited to the classic Weitzman-style process, but applies to any sequential search environment that satisfies Independence and Invariance and in which selection-revealed pairwise comparisons among feasible actions form a unique directed acyclic graph. This generality ensures that the representation remains widely applicable and can accommodate more realistic search processes in modern digital markets. We consider two illustrative extensions. The first concerns partially observed search data, where the inspection order, the set of searched products, or the final purchase may not be fully recorded. The second concerns search processes with multi-stage information acquisition, such as sequential search with discovery \citep{greminger2022optimal}. In these settings, conventional approaches often rely on ad hoc treatments, such as imputing missing data, discarding partially observed observations, or re-deriving optimal rules. By contrast, our empirical strategy based on the ranking representation avoids these issues, allowing researchers to exploit the data in a complete, parsimonious, and unified way. Monte Carlo experiments further show that, using modified rank-based GHK methods, both extensions can be implemented efficiently without the practical difficulties associated with traditional approaches.

Our findings hold importance for both theory and empirical practice. On the theoretical side, although sequential search has been widely considered closely related to discrete choice problems, a formal linkage has been limited to its purchase outcome.\footnote{Prior studies \cite[e.g.][]{choi2018consumer} show that the final purchase outcome in an optimal sequential search process can be viewed as the result of a simple static discrete choice problem. We demonstrate that such results follow directly from the equivalence between sequential search and ranking.} We extend the connection by establishing a complete equivalence between sequential search and ranking models that applies to a broad class of model settings. This link provides a rigorous theoretical foundation for treating consumers’ search processes as directly informative choice data in demand analysis. On the empirical side, the ranking representation breaks the long-standing constraints on empirical strategies in this literature, allowing researchers to fully exploit the rich information contained in search processes in a simple and transparent way. Although with the policy-based representation, the baseline Weitzman model remains solvable with numerical methods, it often proves difficult in extended settings that reflect complex market environments and diverse data structures. The ranking representation provides a robust, flexible, and operational foundation. Drawing on the extensive literature on the tractability of discrete-choice and ranking models, we can apply sequential search models straightforwardly to the analysis of complex consumer search.

The remainder of the paper is organized as follows. Sections \ref{sec:OSR_representation} through \ref{sec:empirics} establish the ranking representation within the baseline Weitzman-style sequential search process. Specifically, Section \ref{sec:OSR_representation} formalizes the process and illustrates the empirical bottleneck inherent in the current policy-based representations. Section \ref{sec:ranking_representation} introduces the ranking representation. Section \ref{sec:empirics} details the rank-based empirical strategy, covering the probability function, the identification arguments, and the rank-based GHK simulator. Section \ref{sec:extension} generalizes the ranking representation to a broader class of model variations, including those with partial observability or with multi-stage information acquisitions, and provides formal conditions for applicability. Section \ref{sec:discussion} discusses the case when the Independence or Invariance assumption fails. Section \ref{sec:conclusion} concludes. 

\subsection{Related Literature}

This paper contributes to three strands of literature. The first concerns the interpretation of consumers' decision-making under the sequential search setting. A large body of work follows \cite{weitzman1979optimal}, treating observed actions as outcomes of an optimal sequential search process governed by optimal rules. Some recent studies interpret the sequence as the outcome of an optimal multi-stage discrete-choice problem, as proposed by \cite{keller2003branching}. In addition to these two policy-based and choice-based representations, this paper proposes a rank-based interpretation in which the observed sequence of actions reveals the consumer’s ranking of feasible actions, offering a new theoretical perspective on consumer search. 

Second, this paper develops a simple empirical strategy for sequential search models that fully exploits search data while simplifying identification and estimation. It extends the identification arguments in \cite{morozov2021estimation} and \cite{ursu2024sequential} from subsets of search decisions to the full set of decisions, with results following directly from standard discrete choice models. In estimation, It provides a novel and simple construction for the likelihood function, which ensures that high-dimensional integrals can be naturally decomposed into iteratively computable low-dimensional integrals, thereby reducing the simulation dimensionality and simplify the computation issue faced by all simulation-based maximum likelihood estimation methods, including the Crude Frequency Simulator \citep{chen2017sequential, ghose2019modeling}, the Kernel-smoothed Frequency Simulator \citep{honka2017simultaneous, ursu2018power, ursu2020search, elberg2019dynamic, yavorsky2021consumer, ursu2023search, zhang2024product, ursu2025online}, and the GHK-style simulator \citep{jiang2021consumer, chung2025simulated}.\footnote{Other estimation approaches have been developed for specific contexts. For example, \cite{morozov2021estimation} applies importance sampling to handle high-dimensional preference heterogeneity; \cite{morozov2023measuring} and \cite{onzo2025bayesian} employ Bayesian MCMC methods to capture unobserved post-search uncertainty. } Drawing on econometric tools for ranking models, we employ a GHK-style simulator to compute the newly constructed likelihood function and obtain a substantial performance improvement. The empirical approach most closely related to ours is the ``double logit'' method used in \cite{compiani2024online}. Compared with theirs, our method advances by not relying on logit assumptions, not requiring fully observed sequence of actions, and having broader applicability in extended settings.

Lastly, this paper develops a unified empirical framework applicable to a broad class of sequential search models, including the classic \cite{weitzman1979optimal} model and its various extensions. These extensions include settings with partial search data, such as when only the final purchase is used \cite[e.g.,][]{moraga2023consumer} or when only the set of searched products is observed \cite[e.g.,][]{jolivet2019consumer}\footnote{\cite{moraga2023consumer} exploits individual-level search information as moment conditions to disentangle the utility component from the search cost related disutility effects.}, as well as settings with richer action types and multiple stages \cite[e.g.,][]{gibbard2023search, greminger2024heterogeneous, zhang2024product}. We show that in each case, the underlying search process admits a ranking representation. This unifying property allows a common class of empirical strategies to be applied naturally across all these models.\footnote{Appendix \ref{app:literature} provides a detailed comparison across models in recent empirical studies in terms of assumptions, data observability, and estimation methods.}

\section{The Weitzman Model and the Optimal Search Rules} \label{sec:OSR_representation}

We begin with the sequential search process studied in \cite{weitzman1979optimal}, and the model developed to analyze it, which already entails the empirical challenges addressed in this paper and is shared by a broad class of related settings. We first introduce the process and show how it delivers an observed sequence of consumer actions. Then, we summarize the optimal rules that characterize its optimal solution and discuss the difficulties in empirical applications.\footnote{For a comprehensive empirical review of the Weitzman model, see \cite{ursu2024sequential}.}

\subsection{The Weitzman-style Sequential Search Process}
 
Consider a representative consumer $i$ who plans to buy at most one from a set of available products in the market denoted by $\mathcal{M}_i$. While the consumer is aware of the existence of all items $\mathcal{M}_i$, she has only partial information on each product, which makes her evaluations on products uncertain and prevents her from determining the exact utility of any product. 

To fully resolve uncertainty about product information, a consumer can \textit{inspect} a product to disclose its details and determine its utility. A product can only be purchased after its utility has been fully revealed through inspection. After each step, the consumer can decide to stop searching. The set of all inspected products at the time of stopping is denoted by $\mathcal{S}_i$, and $J_i$ denotes its size. Upon stopping, the consumer then selects one from $\mathcal{S}_i$ for \textit{purchase}. The set of uninspected products is denoted by $\bar{\mathcal{S}}_i = \mathcal{M}_i \backslash \mathcal{S}_i$. 

Inspections are costly. Consumers must weigh the cost of inspections against their potential benefits to make optimal decisions, including which products to inspect or purchase and when to stop searching. As shown in Figure \ref{fig:SSMDecisions}, a Weitzman-style sequential search process is captured by a multi-stage decision-making process, with each decision (green boxes) made based on all information obtained at that moment. The candidate actions (white boxes) capture, at each stage, the set of feasible alternatives and the consumer's sequential selections. 

\begin{figure}[h]
    \centering
    \input{TikzPics/figure_ssm.tex}
    \caption{Decision-Making in a Weitzman-style Sequential Search Process}
    \label{fig:SSMDecisions}
\end{figure}

The order observed in consumer $i$'s search process is denoted by $\mathcal{R}_i$. Following this order, the products in $\mathcal{S}_i$ are indexed sequentially, with the $j$-th inspected product labeled as Product $j$. Let $J_i$ denote the index of the last inspected product, so that products with $j \leq J_i$ are exactly those inspected and thus belong to $\mathcal{S}_i$. Accordingly, the inspected set can be written as $\{1,2,\dots,J_i\}_{\mathcal{R}_i}$. Let $\mathcal{H}_i$ denote the singleton containing the purchased product, and let $h_i$ denote its index in $\mathcal{R}_i$. Since only inspected products can be purchased, $\mathcal{H}_i \subset \mathcal{S}_i$ and $h_i \leq J_i$ always hold.

The tuple $\{\mathcal{H}, \mathcal{S}, \mathcal{R}, \mathcal{M}\}_i$ summarizes the complete set of actions taken by consumer $i$ in a Weitzman-style search process and constitutes an \textbf{\textit{action sequence}}. For simplicity, we omit $\mathcal{R}_i$ when products are indexed numerically, and suppress the subscript $i$ when referring to sets or indices in isolation.

\subsection{Optimal Search Rules} \label{subsec:osr}
We now formalize how \cite{weitzman1979optimal} maps observed action sequences into quantifiable relationships in the setting above. Let $u_{ij}$ denote the \textbf{\textit{purchase value}}, defined as the utility that consumer $i$ obtains from purchasing product $j$. These values are initially unknown to the consumer, who instead holds knowledge about their distributions. By incurring a search cost $c_{ij} > 0$, the consumer can inspect product $j$ and learn $u_{ij}$. After inspecting the $(j-1)$th product, given that stopping search yields a utility of $\bar{u}$, the consumer solves the Bellman equation below:
{\normalsize
\begin{align}\label{eq:dp}
    W(\bar{u}, \bar{\mathcal{S}}_{ij}) = \max\left\{\bar{u}, \ \max_{j \in \bar{\mathcal{S}}_{ij}} \left\{ -c_{ij} + F^u_{ij}(\bar{u}) W(\bar{u}, \bar{\mathcal{S}}_{ij}\backslash\{j\}) + \int_{\bar{u}}^\infty W(u, \bar{\mathcal{S}}_{ij} \backslash \{j\}) f^u_{ij}(u)du\right\}\right\}
\end{align}
}%
Here, $\bar{S}_{ij}$ is the set of uninspected products at the beginning of step $j$. $F^u_{ij}(\cdot)$ and $f^u_{ij}(\cdot)$ are the cumulative distribution and probability density functions of $u_{ij}$. At every step $j$, the consumer chooses between stopping search and accepting $\bar{u}$, or inspecting the uninspected product with the highest expected inspection gain.

To make the problem tractable, \cite{weitzman1979optimal} imposes two key assumptions:
\begin{assumption}
    Purchase values are independent across products, and inspecting a product reveals only its purchase value.
\end{assumption}
\begin{assumption}
    Unrealized inspections and purchases remain feasible at all stages with identical search costs and purchase values.\footnote{This assumption is not stated explicitly but is implicitly adopted in \cite{weitzman1979optimal} and in many subsequent empirical studies using the Weitzman model.}
\end{assumption}

Under these assumptions, \cite{weitzman1979optimal} shows that the optimal solution of Equation \eqref{eq:dp} reduces to a static decision problem: given a fallback value $\bar{u}$ available for final selection, should the consumer inspect an additional product? The decision depends on the trade-off between the expected incremental benefit of inspection and the associated search cost. For any product $j$, consumer $i$ is indifferent to inspection when $\bar{u}$ satisfies:
\begin{align} \label{eq:rv}
    c_{ij} = \int_{\bar{u}} (u - \bar{u}) \ d F^u_{ij}(u) 
\end{align}

Equation \eqref{eq:rv} defines $\bar{u}$ implicitly. Since the right-hand side is monotonically decreasing in $\bar{u}$, ranging from infinity to zero, a unique solution exists, denoted by $z_{ij}$ and defined as the \textbf{\textit{reservation value}} of product $j$ for consumer $i$. This reservation value depends only on the search cost $c_{ij}$ and the pre-inspection distribution of $u_{ij}$.

With purchase and reservation values defined and Assumptions 1 and 2 made, \cite{weitzman1979optimal} proposes a set of stepwise policies, known as ``Pandora’s Rules,'' to describe the optimal solution to Equation \eqref{eq:dp}. We restate them below using the terminology of this paper:

SELECTION RULE: \textit{If a product is to be inspected, it should be the uninspected product with the highest reservation value. }

STOPPING RULE: \textit{Terminate search whenever the maximum purchase value of inspected products exceeds the reservation value of every uninspected product. }

Existing studies often reformulate ``Pandora’s Rules'' into four optimal rules, each expressed as a set of inequalities involving reservation and purchase values. Together, these rules capture how decisions are made sequentially in an optimal Weitzman-style sequential search process.\footnote{For simplicity and consistency, we assume that there are no ties among actions or values throughout the paper.} 
\begin{enumerate}
    \item Optimal Ranking: The consumer inspects products in decreasing order of reservation values, so each previously inspected product has a higher reservation value than any inspected later.
    \begin{equation}
        t^1_{ij} \equiv z_{ij} - \max_{k \in \mathcal{M}_i\backslash\{1,2,\cdots,j\}} z_{ik} > 0, \quad \forall j \leq J. \label{eq:pandora_start} 
    \end{equation}
    With Optimal Ranking, we index products not in $\mathcal{S}_i$ as $J+1, J+2, \ldots$ in descending order of their reservation values. Hence, any uninspected product is assigned an index $k > J$.
    \item Optimal Continuing: The consumer continues inspecting if the highest reservation value among uninspected products is larger than the highest purchase value among inspected ones.
    \begin{equation}
        t^2_{ij} \equiv \max_{\ell \geq j} z_{i\ell} - \max_{\ell = 1}^{j-1} u_{i\ell} > 0, \quad \forall j \leq J.
    \end{equation}
    \item Optimal Stopping: The consumer stops inspecting when the maximum purchase value among inspected products is smaller than the highest reservation value among uninspected ones.
    \begin{equation}
        t^3_{i} \equiv \max_{j \leq J} u_{ij} - \max_{k > J} z_{ik} > 0. 
    \end{equation}
    \item Optimal Purchasing: The consumer purchases the product with the highest purchase value among all inspected ones.
    \begin{equation}
        t^4_i \equiv u_{ih} - \max_{j \leq J, j \not= h} u_{ij} > 0. \label{eq:pandora_end}
    \end{equation}
\end{enumerate}

In Figure \ref{fig:OSRrepresentation}, we illustrate how the optimal rules operate in a Weitzman-style sequential search process using an action sequence example: $\{\mathcal{H}_i=\{3\}, \mathcal{S}_i = \{1, 2, 3, 4\}, \mathcal{R}_i = \{1 \succ 2 \succ 3 \succ 4\}, \mathcal{M}_i = \{1,2,3,4,5\} \}$. In Step 1, consumer $i$ inspects product 1, which, under Optimal Ranking, implies that $z_{i1}$ exceeds the reservation values of all other products (red line). In Step 2, she inspects product 2, which requires $z_{i2}$ to exceed the reservation values of the remaining uninspected products by Optimal Ranking (red line) and, by Optimal Continuing, to exceed the highest inspected purchase value (blue line), which is $u_{i1}$ in this step. Both rules also apply in Steps 3 and 4, except that the maximum purchase values among inspected products are not observable in these steps. In Step 5, a purchase occurs, indicating that $u_{i3}$ exceeds both the purchase values of all previously inspected products, as required by Optimal Purchasing, and the highest reservation value among uninspected products, $z_{i5}$, as required by Optimal Stopping.

\begin{figure}[h]
    \centering
    \makebox[\textwidth][c]{
    \input {TikzPics/figure_osr.tex}
    }
    \caption{The Optimal Rules in a Weitzman-style Sequential Search Process}
    \label{fig:OSRrepresentation}
\end{figure}

If a Weitzman-style sequential search process is optimal, then all decisions throughout the search process must satisfy Equations \eqref{eq:pandora_start} to \eqref{eq:pandora_end}. Therefore, the probability that a search process described by an action sequence is optimal is the joint probability that all of these inequalities hold throughout the process. For the example shown in Figure \ref{fig:OSRrepresentation}, the probability function takes the following form: 
\begin{align}
    & \mathrm{Pr}(\{\mathcal{H},\mathcal{S}, \mathcal{R}, \mathcal{M}\}_i) \nonumber \\ 
    = & \ \mathrm{Pr}(t^1_{ij} > 0, \forall j < J \ \cap \ t^2_{ij} > 0, \forall j < J \ \cap \ t^3_{i} > 0 \ \cap \ t^4_i > 0) \label{eq:osr_jp} \\
    = & \ \underbrace{\int_{z_{i1}} \mathbb{I}(t_{i1}^1 > 0)}_{\mbox{\scriptsize Optimal Ranking (Step 1)} } \cdot \Big( \underbrace{\int_{u_{i1}, z_{i2}} \mathbb{I}(t_{i2}^1 > 0, t_{i2}^2 > 0 \mid z_{i1})}_{\mbox{\scriptsize Optimal Ranking and Continuing (Step 2)} } \cdot \Big(  \underbrace{\int_{\max\{u_{i1}, u_{i2}\}, z_{i3}} \mathbb{I}(t_{i3}^1 > 0, t_{i3}^2 > 0 \mid u_{i1}, z_{i2})}_{\mbox{\scriptsize Optimal Ranking and Continuing (Step 3)} } \nonumber \\
    & \cdot\Big(  \underbrace{\int_{\max\{u_{i1}, u_{i2}, u_{i3}\}, z_{i4}} \mathbb{I}(t_{i4}^1 > 0, t_{i4}^2 > 0 \mid u_{i1}, u_{i2}, z_{i3})}_{\mbox{\scriptsize Optimal Ranking and Continuing (Step 4)} } \cdot \Big(  \underbrace{\int_{u_{i4}, z_{i5} } \mathbb{I}(t_{i}^3 > 0, t_{i}^4 > 0 \mid u_{i3}, z_{i4})}_{\mbox{\scriptsize Optimal Stopping and Purchasing (the last step)}}  \nonumber \\ 
    & \quad d \mathcal{F}_5^c (u_{i4}, z_{i5}) \Big) \ d \mathcal{F}_{4}^c (u_{i3}, z_{i4}) \Big) \ d \mathcal{F}_{3}^c (u_{i2}, z_{i3}) \Big) \ d \mathcal{F}_{2}^c (u_{i1}, z_{i2}) \Big) \ d \mathcal{F}_1(d z_{i,1}) \label{eq:osr_jp_int}
\end{align}
Here, $\mathbb{I}(\cdot)$ denotes the indicator function, $\mathcal{F}_{j}^c(u_{i,j-1}, z_{ij})$ denotes the joint cumulative distribution function of $u_{i,j-1}$ and $z_{ij}$ conditional on all previous steps, and $\mathcal{F}_{1}(z_{i1})$ denotes the cumulative distribution function of $z_{i1}$. The joint integrals of $u_{i,j-1}$ and $z_{ij}$ provide a general representation: when $z_{ij}$ is deterministic under a given specification, the associated measure with respect to $z_{ij}$ degenerates to a Dirac measure.

Equation \eqref{eq:osr_jp} involves a high-dimensional integral over the reservation values of all products and the purchase values of all inspected products. Equation \eqref{eq:osr_jp_int} presents a stepwise decomposition that highlights the main empirical challenge under optimal rules. At any step $j$ and thereafter, the consumer’s optimal decision depends on $\max_{\ell = 1}^{j-1}{u_{i\ell}}$, the highest purchase value among previously inspected products. Because this threshold is often unobserved (e.g., at Step 3 in Figure \ref{fig:OSRrepresentation}), whether the Optimal Continuing condition holds depends on the entire history of prior search outcomes, which must be treated as stochastic. We refer to this structural link between past unobserved outcomes and future search decisions as the model's \textbf{\textit{sequential dependence}} inherent in the optimal rules.

Sequential dependence introduces cross-stage correlation to Equation \eqref{eq:osr_jp_int}. As a result, computing the full likelihood requires computing the distributions of $\max_{\ell = 1}^{j-1}\{u_{i\ell}\}$ at all steps $j$, or at least up to the inspection of the purchased product. This leads to two main difficulties. 

First, computation is demanding. The stepwise decomposition provided in Equation \eqref{eq:osr_jp_int} does not effectively reduce the computational burden. In the absence of a closed-form solution, the purchase values of all $J$ inspected products must be simulated. If reservation values are also stochastic, their realizations for these $J$ products must also be simulated. Moreover, because the likelihood contribution of a given action sequence can be very small, achieving adequate simulation accuracy typically requires a large number of draws in a high-dimensional space.

Second, empirical applicability is fragile. The probability function is highly sensitive to both the Weitzman specification and the completeness of the data. Missingness to the search process or the introduction of additional actions alters the distribution of $\max_{\ell = 1}^{j-1}\{u_{i\ell}\}$ in subsequent steps. Even minor specification changes, such as adding an outside option, can change these distributions throughout the process. As a result, empirical extensions often require different ways to construct the likelihood function under new specifications and, in some cases, modifying the optimal rules, which can be difficult or impossible.\footnote{Outside consumer search, some homogeneous sequential search models have been studied in contexts such as financial product search and labor search \citep[e.g. ][]{mccall1970economics}. Subjects in these models evaluate only the currently sampled option, assuming that previously rejected options cannot be recalled and therefore do not involve sequential dependence. Therefore, they fall outside the scope of our discussion.}

\section{The Ranking Representation of the Weitzman Model} \label{sec:ranking_representation}

To overcome the empirical challenges posed by sequential dependence, this paper proposes a novel representation of the optimal solution to sequential search, which leads to a new empirical strategy for a broad class of sequential search models. 

\subsection{Illustrative Example}

Consider the example in Figure \ref{fig:OSRrepresentation}. Taking the process as optimal, then in Step 5, Optimal Purchasing implies that $u_{i3} > \max \{ u_{i1}, u_{i2}, u_{i4} \}$, and Optimal Stopping implies that $u_{i3} > z_{i5}$.

Next, consider Step 4. Optimal Ranking implies that $z_{i4} > z_{i5}$, and Optimal Continuing implies that $z_{i4} > \max \{ u_{i1}, u_{i2}, u_{i3} \}$. Given that Optimal Purchasing and Optimal Stopping are satisfied in Step 5, we already know that $u_{i1}, u_{i2}, z_{i5}$ are all no greater than $u_{i3}$. Therefore, among all conditions in Step 4, it suffices that $z_{i4} > u_{i3}$, as the remaining conditions then follow trivially.

Now suppose the optimal rules in Steps 4 and 5 hold. Consider Step 3. The optimal rules imply that $z_{i3} > \max \{ z_{i4}, z_{i5} \}$ and $z_{i3} > \max \{ u_{i1}, u_{i2} \}$. Since Steps 4 and 5 establish that $z_{i4}$ is the largest among $z_{i4}, z_{i5}, u_{i1}, u_{i2}$, the only additional condition required at Step 3 is $z_{i3} > z_{i4}$. 

Proceeding backward to Step 1 yields the relation shown in Figure \ref{fig:optimality}:

\begin{figure}[h]
    \centering
    \makebox[\textwidth][c]{
    \input{TikzPics/figure_example.tex}
    }
    \caption{The (Partial) Ranking Corresponding to the Search Process in Figure \ref{fig:OSRrepresentation}} 
    \label{fig:optimality}
\end{figure}

This defines a partial ranking over all feasible actions across the sequence, capturing the full empirical content implied by the search process in Figure \ref{fig:OSRrepresentation}. Formally,
\begin{align}
\mathrm{Pr}(\mbox{The process in  } & \mbox{Figure \ref{fig:OSRrepresentation} is optimal}) \nonumber \\ 
&  = \mathrm{Pr}(z_{i1} > z_{i2} > z_{i3} > z_{i4} > u_{i3}, \mbox{ and } u_{i3} > \max\{ z_{i5}, u_{i1}, u_{i2}, u_{i4} \}). \label{eq:illustration}
\end{align}

We verify Equation \eqref{eq:illustration} using a Monte Carlo simulation. In a 5-product market as depicted in Figure \ref{fig:OSRrepresentation}, we assume that the reservation and purchase values of each product follow normal distributions with means $\mu = [5, 4, 3, 2, 1]^\top$ and a standard deviation of $\sigma=2$. We simulate 50 million consumers search following the optimal rules, compute the brute-force frequencies of all 1240 possible action sequences, and compare the top 400 sequences, which account for 99.52\% of all observations, with the theoretical probabilities of their corresponding rankings\footnote{The probabilities are computed with the GHK-style simulator introduced in Section \ref{subsec:estimation}.}. The results show close agreement between brute-force frequencies and computed probabilities, with a cosine similarity of 0.999999 (see Appendix \ref{app:bruteforce} for more comparison details). For the process in Figure \ref{fig:OSRrepresentation}, the brute-force frequency is $1.2148 \times 10^{-4}$, and the computed probability of the ranking in Figure \ref{fig:optimality} is $1.1726 \times 10^{-4}$, with a marginal relative error of 3.48\%.

Compared to the optimal rules, we do not induce any changes in the data but simply exploit ex post information that is already contained in the data. The optimal rules are established from the consumer’s perspective, in which each decision depends only on information revealed in previous steps. However, in an action sequence, researchers observe subsequent decisions in the same search process, which provide additional information on the ranking of actions. As long as these ranking relations hold in the stages preceding the subsequent decisions, researchers can use ex post ranking information to identify constraints in earlier decisions that are already implied by other conditions and exclude them when constructing the probability function.

A direct echo of this intuition is that, among the optimal rules, Optimal Continuing before step $J$ is effectively redundant. Given that a Weitzman-style sequential search process is known to continue until step $J$ and that Optimal Stopping and Optimal Ranking hold, for any step $j < J$, the inequality $\max_{s=1}^{j}\{u_{is}\} < \max_{s=1}^{J-1}\{u_{is}\} < z_{iJ} < z_{ij}$ is satisfied by construction. 

\subsection{Theoretical Formalization} \label{subsec:formalization}

We next formalize the theory illustrated by the example and state the conditions under which it holds. Moving away from the optimal rules, we view each step of a Weitzman-style sequential search process as a discrete choice problem among current feasible actions. At each step, the consumer chooses either to inspect an uninspected product $k$ ($I_k$) or to purchase an inspected product $j$ ($P_j$). Each observed action can be represented as a discrete choice outcome $(a^*, \mathcal{A})$, where $a^*$ denotes the selected action and $\mathcal{A}$ the current feasible action set.

We restate Assumptions 1 and 2 in terms of actions, consistent with the rest of the paper.
\begin{manualassumption}{1a}
    \textbf{(INDEPENDENCE)} Selecting an action reveals information only about the payoffs of actions that cannot be selected unless that action is selected, and does not alter the information state of payoffs of any other unselected action.\footnote{This assumption does not require the actions to be substantively unrelated, but any such relationship must not be incorporated into the consumer's belief. For example, a consumer's belief about the value of purchasing a product should remain unchanged, irrespective of whether another inspection action is selected.}
\end{manualassumption}
\begin{manualassumption}{2a} 
    \textbf{(INVARIANCE)} A feasible action remains feasible at all subsequent stages until it is selected, and its payoff is invariant throughout the entire process.
\end{manualassumption}

With the two assumptions held, we present the following theorem:
\begin{theorem}\label{theorem:main}
    Consider an action sequence $\{\mathcal{H}, \mathcal{S}, \mathcal{R}, \mathcal{M}\}_i$ describing consumer $i$'s search process, which satisfies the assumptions of Independence and Invariance. Suppose consumer $i$ ranks all actions that are feasible at any stage throughout the process, including inspecting any product $j \in \mathcal{M}_i$ ($I_j$) and purchasing any inspected product $j \in \mathcal{S}_i$ ($P_j$). Then the search process is optimal if and only if, in this ranking: 
    \begin{enumerate}
        \item $I_{1} \succ I_{2} \succ \cdots \succ I_{J}$, and $I_{J} \succ P_{h}$ if $h < J$.
        \item All unselected actions are ranked below the lower-ranked of $I_{J}$ and $P_{h}$.
    \end{enumerate}
\end{theorem}

We refer to Theorem \ref{theorem:main} as \textbf{\textit{the ranking representation}} of optimal Weitzman-style sequential search. It implies that, under the Independence and Invariance assumptions, for any observed action sequence, the two events below are informationally equivalent and occur with the same probability. First, the consumer conducts an optimal Weitzman-style sequential search process, and her sequence of actions coincides with the observed sequence. Second, the consumer ranks all feasible actions according to their values, with the top $J+1$ positions corresponding to the observed sequence in the same order, except for the last-selected action if it is the purchase of the last inspected product. 

As an illustration, consider the process shown in Figure \ref{fig:OSRrepresentation}, in which the feasible actions include inspecting products 1 through 5 ($I_1$ to $I_5$), as well as actions that become feasible in subsequent steps, namely purchasing products 1 through 4 ($P_1$ to $P_4$). We assume that the nine actions constitute a unified set and the consumer ranks them according to their associated values. Then the ranking in Figure \ref{fig:optimality} holds if and only if the process in Figure \ref{fig:OSRrepresentation} is optimal. 

To prove Theorem \ref{theorem:main}, having characterized each step of the Weitzman-style sequential search process as an action-level discrete choice problem, we further represent the entire process as a multi-stage discrete choice process $\{(a^*_j, \mathcal{A}_j)\}_{j=1}^J$, where $J$ denotes the length of the action sequence. Formally, this process is referred to as a branching project \citep{keller2003branching}. In a branching project, the decision-maker chooses one action from a set of feasible actions at each stage and obtains an immediate payoff. The payoffs of feasible actions at the initial stage are fully known to the consumer, whereas those of infeasible actions are only partially revealed. Once an action is selected, its payoff is realized immediately, and the action is removed from the feasible set. The selected action may render some previously infeasible actions feasible in subsequent stages and fully reveal their payoffs, thereby inducing parent-child relationships among actions. At the same time, it may partially reveal the payoff information of actions that remain infeasible. Under the Independence assumption, selecting an action only changes the information on payoff of actions that can become feasible only if that action is selected. Under the Invariance assumption, true payoffs are determined ex-ante, and the feasibility of actions is not affected by any exogenous factors other than the decision-maker’s own selections.

Figure \ref{fig:single_branching} shows a branching project representation corresponding to the search process as in Figure \ref{fig:OSRrepresentation}, which can be taken as a depth-two branching project satisfying both the Independence and Invariance assumptions, as all inspection actions are initially feasible with known and given search costs, while each purchase action becomes feasible and its underlying payoff is revealed only after the corresponding inspection being conducted. 

\begin{figure}[htbp]
    \centering
    \begin{minipage}{\linewidth} 
        \input{TikzPics/figure_single_branching.tex}
        
        \caption{Representing the Search Process in Figure \ref{fig:OSRrepresentation} as a Branching Project}
        \label{fig:single_branching}
        
        \vspace{5pt}
        {\footnotesize \emph{Notes:} The figure illustrates the branching project corresponding to the search process in Figure \ref{fig:OSRrepresentation}. At each step, the consumer chooses one action from the currently feasible set. The selected actions $\{I_1, I_2, I_3, I_4, P_3\}$ are highlighted in blue and other feasible actions are shown in green in the corresponding steps, while the actions selected in previous steps are marked with dashed circles.\par}
    \end{minipage}
\end{figure}

Next, building on Luce’s Choice Axioms and Ranking Postulates \citep{luce1959individual}, we present the following equivalence lemma.\footnote{Note that the Axioms and Ranking Postulates are applied algebraically at the level of individual utility realizations, where rankings are observed but not probabilistic. Hence, the lemma remains distribution-agnostic.}

\begin{lemma} \label{lem:equivalence}
    Consider any two consecutive stages in a branching project that satisfy the Independence and Invariance assumptions. Let the first-stage choice be $(a_1, \mathcal{A}_1)$ and the second-stage choice be $(a_2, \mathcal{A}_2 \setminus {a_1})$, where $\mathcal{A}_1 \subset \mathcal{A}_2$. Let $\rho_j = \{a_{j+1} \succ \rho_{j+1}\}$ for $j \geq 1$, where $\rho_1 = \{a_2 \succ a_3 \succ \dots\}$ is the complete ranking over $\mathcal{A}_2 \setminus {a_1}$. Then the following identity holds:
    \begin{align}
        P_{\mathcal{A}_1}(a_1 \mid \rho_1) \cdot & R_{\mathcal{A}_2 / \{a_1\}}(\rho_1) = R_{\mathcal{A}_2}(a_1 \succ \rho_1) + \sum_{\ell=2}^{N-1} R_{\mathcal{A}_2} (a_2 \succ \cdots \succ a_{\ell} \succ a_1 \succ \rho_{\ell}), \label{eq:lemma}
    \end{align}
    where $a_N$ denotes the first action in $\rho_1$ that belongs to $\mathcal{A}_1$, $P_\mathcal{A}(a)$ is the probability that action $a$ is selected from set $\mathcal{A}$, and $R_\mathcal{A}(\rho)$ is the top-down probability of the full ranking $\rho$ over $\mathcal{A}$.\footnote{The top-down ranking probability refers to the product of conditional probabilities that each item in the ranking is selected sequentially from the set of unranked alternatives, conditional on the exclusion of all previously ranked items from the original choice set.}
\end{lemma}

We illustrate how the lemma applies in a Weitzman-style sequential search process with a toy example. Consider a consumer searching between two products, 1 and 2. In the first step, her choice set $\mathcal{A}_1$ consists of two alternative actions, ``inspect product 1'' ($I_1$) and ``inspect product 2'' ($I_2$). Suppose she selects $I_1$, that is, $a_1$ is $I_1$. In the second step, the choice set $\mathcal{A}_2 \setminus \{a_1\}$ contains $I_2$ and ``purchase product 1'' ($P_1$), so that $\mathcal{A}_2 = \{I_1, I_2, P_1 \}$.  

Now consider the only two possible cases in the second step. If $\rho_1 = \{I_2 \succ P_1\}$, since $I_2$ already belongs to $\mathcal{A}_1$, adding $P_1$ to $\mathcal{A}_1$ does not change the probability that $I_1$ is selected in the first stage. The reason is that $P_1$ is strictly dominated by a feasible alternative, so including it in the choice set neither leads to its selection nor affects the selection probabilities of other alternatives.\footnote{Led by the Independence assumption, this irrelevance is consistent with Choice Axiom 1(ii), \citet{luce1959individual}. } Therefore, this two-step process is probabilistically equivalent to the consumer selecting $I_1$ and then $I_2$ from $\mathcal{A}_2$, which corresponds to a ranking of $\{I_1 \succ I_2 \succ P_1\}$. The probability of this ranking is given by the first term on the right-hand side of Equation \eqref{eq:lemma}, while the second term does not apply. In the other case, if $\rho_1 = \{P_1 \succ I_2\}$, the consumer selects $P_1$ in the second step, which is not in $\mathcal{A}_1$. This yields a partial ranking over $\mathcal{A}_2$ where $I_1$ and $P_1$ both rank above $I_2$, but their relative order is unspecified. Adding $P_1$ to $\mathcal{A}_1$ thus corresponds to two possible rankings, namely $\{I_1 \succ P_1 \succ I_2\}$ and $\{P_1 \succ I_1 \succ I_2\}$, which correspond to the two terms in the right-hand side of Equation \eqref{eq:lemma}.

Hence, Lemma \ref{lem:equivalence} proves that the joint probability of two adjacent stages in a branching project is equal to the probability of a ranking over a unified choice set. By repeatedly applying Lemma \ref{lem:equivalence}, this equivalence extends backward from the final two stages, the last inspection and final purchase, to the entire branching project corresponding to a sequential search process. This establishes Theorem \ref{theorem:main}. Other proof details are provided in Appendix \ref{app:proof_optimality}.

We emphasize the central role of the Independence and Invariance assumptions in establishing Theorem \ref{theorem:main}. Independence requires that a consumer’s information about the value of any action remain unaffected by prior selections, except in cases where the action would be infeasible without such a selection. Invariance requires that neither the structure of the branching project nor the value of any action vary with external conditions. Together, these assumptions imply that the binary dominance relations revealed at any local step hold globally across the entire search process. For example, reconsider the two-product example where $I_1$ and $I_2$ are sequentially selected. Under both assumptions, the ranking $\{I_{1} \succ I_{2} \succ P_{1}\}$ holds across stages. If either assumption fails, only stage-specific relations ${I_{1} \succ I_{2}^{pre}}$ and ${I_{2}^{post} \succ P_{1}}$ can be inferred, where $I_{2}^{pre}$ and $I_{2}^{post}$ denote inspecting product 2 before and after inspecting product 1. Whether these two actions are equivalent, and thus whether a consistent ranking exists, may depend on the selection of $I_1$ or on external conditions. 


Finally, we establish value measures for actions for quantitative applications. Under the same Independence and Invariance assumptions, \cite{keller2003branching} prove that with bounded payoffs, the optimality of a branching project is fully characterized by a Gittins index policy. In the Weitzman model, this index corresponds to the purchase value for purchase actions and to the reservation value for inspection actions. Therefore, an optimal branching project is equivalent to selecting actions sequentially based on these values. Since a branching project process is equivalent to a corresponding ranking of feasible actions, the resulting action ranking is thus equivalent to the ranking of action values under optimality. Hence, similar to Pandora’s Rule, Theorem \ref{theorem:main} can also be expressed as a system of inequalities over the values of all feasible actions, which is equivalent to the optimal rules summarized in the following proposition.

\begin{proposition} \label{Prop:weitzman}
    For a sequence observation $\{\mathcal{H},\mathcal{S}, \mathcal{R}, \mathcal{M}\}_i$, Equations \eqref{eq:pandora_start} - \eqref{eq:pandora_end} hold if and only if the following conditions are fulfilled: 
    \begin{enumerate}
        \item \textit{Distribution Condition}: $u_{ih} < z_{iJ}$ if $h < J$; 
        \item \textit{Ranking Condition}: $z_{i1} > z_{i2} > ... > z_{iJ}$; 
        \item \textit{Purchase Choice Condition}: $u_{ij} < \min\{u_{ih}, z_{iJ}\}, \ \forall j \leq J, j \not= h$; 
        \item \textit{Inspection Choice Condition}: $z_{ik} < \min\{u_{ih}, z_{iJ}\} , \ \forall k > J$. 
    \end{enumerate}
    \end{proposition}
\begin{proof}
    See Appendix \ref{app:proof_weitzman}. 
\end{proof}

Proposition \ref{Prop:weitzman} is no longer constrained by sequential dependence, as all conditions are conditioned on the lowest selected action value, $u_{ih}$ or $z_{iJ}$. Given this value, the three restrictions excluding the Distribution Condition correspond respectively to the reservation values of inspected products, the purchase values of inspected products, and the reservation values of uninspected products. These components are conditionally independent, which allows the probability function to be factorized and evaluated separately.

The ranking representation improves the tractability of the Weitzman model. Theorem \ref{theorem:main}, together with its subsequent extensions, shows that in empirical applications, researchers can treat the ranking representation of an optimal sequential search process as an alternative empirical foundation. Econometrically, ranking models and standard discrete choice models exhibit a close structural correspondence: their probability functions can both be expressed in terms of utility differences across alternatives, implying that identification arguments and estimation methods largely carry over. Through this connection, these empirical strategies extend naturally to the Weitzman model. By leveraging tools from the discrete choice literature, the implementation complexity of a Weitzman model can be substantially reduced without sacrificing informational richness provided by the observed search process.\footnote{Figure \ref{fig:optimality} illustrates that a sequential search process provides a richer set of inequality conditions than standard discrete choice data. While a purchase observation identifies that the selected alternative has higher utility than the remaining alternatives, a search sequence can be translated to a partial ranking among selected actions and relative to unselected actions. This increased number of inequality conditions per observation allows for sharper identification of model primitives, such as the covariance structures, without relying on functional form assumptions.}

Meanwhile, note that the ranking representation is established by Lemma \ref{lem:equivalence}, which rests solely on the basic assumptions of Independence and Invariance, and not on any distributional assumptions or functional forms. This feature implies that the ranking representation is not tied to the baseline Weitzman model and can be applied in more general settings. 

\section{Empirical Strategy of the Weitzman Model} \label{sec:empirics}

This section develops a novel, simple empirical strategy for the sequential search model by implementing it via its ranking representation, using the Weitzman setting as an example. In this section, we confine our discussion to an additive specification of purchase values, which combines the expected value determined by observed attributes with other components that may be stochastic and driven by unobservables. 

\subsection{Probability Function} \label{subsec:joint_prob}
Different from Equation \eqref{eq:osr_jp}, the probability function of the Weitzman model based on Proposition \ref{Prop:weitzman} is given as follows: 
\begin{equation}
    \mathrm{Pr}(\{\mathcal{H},\mathcal{S}, \mathcal{R}, \mathcal{M}\}_i) = \mathrm{Pr}( z_{iJ} > u_{ih} \mbox{ if } h < J \ \cap \ z_{i1} > \cdots > z_{iJ} \ \cap \ \max_{j \leq J} u_{ij} < y_i \ \cap \ \max_{k > J} z_{ik} < y_i ) \label{eq:pr_jp}
\end{equation}
where $y_i \equiv \min\{u_{ih}, z_{iJ}\}$. 

We consider a general baseline specification as follows: 
\begin{align}
    u_{ij} \ & = \delta_{i}^u(X_{ij}^u) + \xi^u_{ij} + \varepsilon_{ij} \label{eq:base_model_start}    \\ 
    z_{ij} \ & = \delta_{i}^z(X_{ij}^z) + \xi^u_{ij} + \xi^z_{ij}       \label{eq:base_model_end}
\end{align}    

Here, $\delta_{i}^u(X_{ij}^u)$ and $\delta_{i}^z(X_{ij}^z)$ denote the data-specified components of the purchase value and the reservation value, respectively. $X_{ij}^u$ and $X_{ij}^z$ represent the factors affecting consumer $i$’s evaluation of purchasing and inspecting product $j$. The two vectors may be identical; however, if certain factors influence the reservation value but not the purchase value, such as advertising exposure, they may enter only $X_{ij}^z$, allowing the two vectors to differ.\footnote{A variable that enters $X_{ij}^u$ but not $X_{ij}^z$ is generally inconsistent with the basic assumption of a rational consumer.}

We introduce three components that can be stochastic to account for unobserved factors in the model. The first is a pre-inspection taste shock, denoted by $\xi^u_{ij}$, which affects both inspection and purchase decisions for product $j$. It captures the part of the purchase value known to the consumer before inspection but unobserved by the researcher. The second is the post-inspection taste shock, denoted by $\varepsilon_{ij}$, which represents the portion of the purchase value revealed only upon inspection. We model $\varepsilon_{ij}$ as a one-dimensional stochasticity added to the deterministic component. In alternative formulations, it may also include product attributes that become known only through search \citep{honka2017simultaneous, kaye2024personalization, compiani2024online, greminger2024heterogeneous}. Without loss of generality, we assume that $\varepsilon_{ij}$ is independently and identically distributed across consumers and products, with mean zero.

The third component, $\xi^z_{ij}$, is referred to as the inspection propensity. It enters only the reservation value and influences the consumer's inspection decisions.\footnote{We use the term ``propensity'' to describe the difference between the reservation value and the conditional expectation of the purchase value, following \cite{morozov2023measuring} and \cite{onzo2025bayesian}.} Search cost $c_{ij}$ is typically considered the primary determinant of inspection propensity. When $c_{ij}$ is independent and known to consumer, its impact to the inspection propensity depends solely on the distribution of $\varepsilon_{ij}$ and the magnitude of $c_{ij}$, expressed as $m_\varepsilon(c_{ij})$, a strictly decreasing function derived from Equation \eqref{eq:rv}.\footnote{The additivity of inspection propensity and the monotonicity of $m_\varepsilon(\cdot)$ are proved in Appendix \ref{app:linear_and_monotonicity}.} Beyond search costs, $\xi^z_{ij}$ may also capture other sources of stochasticity that are not directly tied to cost but influence consumers' inclination to inspect.\footnote{Existing empirical studies often do not distinguish between these two components, particularly for observable factors that influence search actions, such as list-page positions \citep{ursu2018power}, store distances \citep{yavorsky2021consumer}, and search refinement tools \citep{chen2017sequential}. Some motives, however, are less appropriately attributed to search costs, such as searches undertaken for entertainment or leisure \citep{moe2003buying}.} 

We make the following assumptions, including Independence and Invariance:
\begin{enumerate}[label={}, leftmargin=*]
    \item Assumption 1: Consumer observes $\xi^u_{ij}, \xi^z_{ij}$ and the distribution of $\varepsilon_{ij}$ at the beginning of search. 
    \item Assumption 2: Consumer only knows the value of $\varepsilon_{ij}$ once product $j$ is inspected. 
    \item Assumption 3 (Independence): Inspecting a product $j$ reveals no information on $\varepsilon_{ik}, \forall k\not=j$. 
    \item Assumption 4 (Invariance): Products not inspected and not purchased in each step remain feasible for inspection in the next step with the same underlying purchase values and search costs. 
\end{enumerate}

To express the probability function, we divide the observed sequences into two cases. We first consider the case where the purchased product $h$ is not the last inspected product $J$. Let $\vec{\bm{z}}_i^k$ denote the reservation values of inspected products, $\vec{\bm{z}}_i^n$ the reservation values of uninspected products, and $\vec{\bm{u}}_i^{k^\prime}$ the purchase values of inspected but unpurchased products, ordered as follows; each vector is then decomposed into component vectors, with variables sharing the same superscript ($k, n, k^\prime$) arranged in the same order. 
\begin{gather*}
    \vec{\bm{z}}_{i}^k := (z_{i,J}, z_{i,J-1}, \cdots, z_{i,1})^\top, \ \vec{\bm{z}}_{i}^n := (z_{i,J+1}, \cdots, z_{i,|\mathcal{M}|})^\top, \
    \vec{\bm{u}}_{i}^{k^\prime} := (u_{i,1}, \cdots, u_{i,h-1}, u_{i,h+1}, \cdots u_{i,J})^\top; \\[1ex]
    \vec{\bm{z}}_{i}^k = \vec{\delta}_i^{z,k}(\vec{\bm{X}}_i^{z,k}) + \vec{\bm{\xi}}_{i}^{u,k} + \vec{\bm{\xi}}_i^{z,k}, \quad \vec{\bm{z}}_{i}^n = \vec{\delta}_i^{z,n}(\vec{\bm{X}}_i^{z,n}) + \vec{\bm{\xi}}_{i}^{u,n} + \vec{\bm{\xi}}_i^{z,n}, \quad \vec{\bm{u}}_{i}^{k^\prime} = \vec{\delta}_i^{u,k^\prime}(\vec{\bm{X}}_i^{u,{k^\prime}}) + \vec{\bm{\xi}}_{i}^{u,k^\prime} + \vec{\bm{\varepsilon}}_{i}^{k^\prime}. 
\end{gather*}

Following Equation \eqref{eq:pr_jp}, the probability of the sequence is: 
\begin{align*}
    \mathrm{Pr}(\{H, \mathcal{S}, \mathcal{R}, \mathcal{M}\}_i) & = \mathrm{Pr}\left( \underbrace{\hat{D}}_{(J + |\mathcal{M}| - 1) \times (J + |\mathcal{M}|)} \begin{pmatrix} u_{ih} \\ \vec{\bm{z}}^{k}_{i} \\ \vec{\bm{z}}^{n}_{i} \\ \vec{\bm{u}}^{k^\prime}_{i}\end{pmatrix}_{(J + |\mathcal{M}|) \times 1} < \vec{\bm{0}} \right), \ \mbox{where } \hat{D} = \begin{pmatrix}
        \hat{D}_1 & \hat{D}_2 \\
        \hat{D}_3 & \hat{D}_4 \\
    \end{pmatrix}
\end{align*}

The differencing matrix $\hat{D}$ consists of four blocks: 
{\small
\begin{align*}
    \hat{D}_1 = \begin{pmatrix} 1 & -1 & 0 & \cdots & 0 \\ 0 & 1 & -1 & \cdots & 0 \\ \vdots & \vdots & \ddots & \ddots & \vdots &  \\ 0 & 0 & \cdots & 1 & -1 \end{pmatrix}_{J \times (J+1)}, & \quad \hat{D}_2 = \{ 0 \}_{J \times (|\mathcal{M}| - 1)} \\
    \hat{D}_3 = \begin{pmatrix} -1 & 0 & \cdots & 0 \\ -1 & 0 & \cdots & 0 \\ \vdots & \vdots & \ddots & \vdots \\ -1 & 0 & \cdots & 0 \end{pmatrix}_{(|\mathcal{M}|-1) \times (J+1)}, & \quad \hat{D}_4 = I_{(|\mathcal{M}|-1) \times (|\mathcal{M}|-1)}. 
\end{align*}
}%

Hence, $\hat{D}$ has rank $J + |\mathcal{M}| - 1$, with its structure determined by the observed sequence.

For the case where the purchased product $h$ is the last inspected, we follow the vectorized form in the previous case and establish the probability function as below: 
\begin{align*}
    \mathrm{Pr}(\{H, \mathcal{S}, \mathcal{R}, \mathcal{M}\}_i) & = \mathrm{Pr}\left( \underbrace{\tilde{D}}_{(J + 2 \cdot |\mathcal{M}| - 3) \times (J + |\mathcal{M}|)} \begin{pmatrix} u_{ih} \\ \vec{\bm{z}}^{k}_{i} \\ \vec{\bm{z}}^{n}_{i} \\ \vec{\bm{u}}^{k^\prime}_{i}\end{pmatrix}_{(J + |\mathcal{M}|) \times 1} < \vec{\bm{0}} \right), \mbox{where } \tilde{D} = \begin{pmatrix}
        \tilde{D}_1 & \tilde{D}_2 \\
        \tilde{D}_3 & \tilde{D}_4 \\
        \tilde{D}_5 & \tilde{D}_6
    \end{pmatrix}
\end{align*}

The differencing matrix $\tilde{D}$ is also of rank $J + |\mathcal{M}| - 1$, which consists of six parts: 
{\small
\begin{align*}
    \tilde{D}_1 = \begin{pmatrix} 0 & 1 & -1 & 0 & \cdots & 0 \\ 0 & 0 & 1 & -1 & \cdots & 0 \\ \vdots & \vdots & \vdots & \ddots & \ddots & \vdots   \\ 0 & 0 & 0 & \cdots & 1 & -1  \end{pmatrix}_{(J-1) \times (J+1)}, & \quad D_2 = \{ 0 \}_{(J-1) \times (|\mathcal{M}|-1)},  \\
    \tilde{D}_3 = \begin{pmatrix} -1 & 0 & 0 & \cdots & 0 \\ \vdots & \vdots & \vdots & \ddots & \vdots \\ -1 & 0 & 0 & \cdots & 0 \end{pmatrix}_{(|\mathcal{M}|-1) \times (J+1)}, & \quad \tilde{D}_4 = I_{(|\mathcal{M}|-1) \times (|\mathcal{M}|-1)} \\
    \tilde{D}_5 = \begin{pmatrix} 0 & -1 & 0 & \cdots & 0 \\ \vdots & \vdots & \vdots & \ddots & \vdots \\ 0 & -1 & 0 & \cdots & 0 \end{pmatrix}_{(|\mathcal{M}|-1) \times (J+1)}, & \quad \tilde{D}_6 = I_{(|\mathcal{M}|-1) \times (|\mathcal{M}|-1)} 
\end{align*}
}%

In the remainder of this paper, we denote the differencing matrix consistently as $D$, an element in $\{\hat{D}, \tilde{D}\}$, based on $\mathcal{R}_i$. Accordingly, the model's probability is expressed as follows: 
\begin{equation}
    \mathrm{Pr}(\{\mathcal{H}, \mathcal{S}, \mathcal{R}, \mathcal{M}\}_i) = \mathrm{Pr}\left( D \begin{pmatrix} u_{ih} \\ \vec{\bm{z}}^{k}_{i} \\ \vec{\bm{z}}^{n}_{i} \\ \vec{\bm{u}}^{k^\prime}_{i}\end{pmatrix} < \bm{0} \right) = \mathrm{Pr}\left(D \begin{pmatrix} \xi_{ih}^u + \varepsilon_{ih} \\ \vec{\bm{\xi}}^{u,k}_{i} + \vec{\bm{\xi}}^{z,k}_{i} \\ \vec{\bm{\xi}}^{u,n}_{i} + \vec{\bm{\xi}}^{z,n}_{i} \\ \vec{\bm{\xi}}^{u,k^\prime}_{i} + \vec{\bm{\varepsilon}}^{k^\prime}_{i}\end{pmatrix} < - D \begin{pmatrix} \delta_i^u(X^u_{ih}) \\ \vec{\delta}_i^{z,k}(\vec{\bm{X}}^{z,k}_{i}) \\ \vec{\delta}_i^{z,n}(\vec{\bm{X}}_i^{z,n}) \\ \vec{\delta}_i^{u,k^\prime}(\vec{\bm{X}}_i^{u,{k^\prime}}) \end{pmatrix} \right) \label{eq:pr_jp_dec}
\end{equation}

Equation \eqref{eq:pr_jp_dec} explicitly expresses the probability in a value-differencing form. While the formulation follows the general structure of a discrete choice probability \citep{train2009discrete}, it employs a data-specified full-rank differencing matrix to account for the nature of sequential search. Equation \eqref{eq:pr_jp_dec} therefore bridges the entire sequential search process and the discrete choice model family. To our knowledge, this connection has not been formally established in previous studies. Compared to Equation \eqref{eq:osr_jp_int}, Equation \eqref{eq:pr_jp_dec} separates the variation explained by the data (the right-hand side) from the variation implied by the stochastic structure (the left-hand side), allowing the high-dimensional integral in the probability function to be decomposed in a tractable manner. As a result, researchers can estimate sequential search models using the empirical strategy of discrete choice models with flexibility and computational efficiency, without sacrificing their rich informational content.

\subsection{Identification} \label{subsec:identification}

While most empirical studies discuss identification of a Weitzman model heuristically by linking the data variations to model parameters, a formal discussion remains necessary for broader applicability. Contributions in this direction are provided by \cite{morozov2021estimation} and \cite{ursu2024sequential}.\footnote{Other work explores identification under partial specifications. For instance, \cite{abaluck2025method} studies the identification of preference parameters when it is uncertain which attributes are known to consumers before inspections, and \cite{onzo2025bayesian} investigates nonparametric identification of parameter distributions.} However, these arguments rely on conditional probabilities of a subset of decisions rather than the full decision set. This leaves ambiguity because variation in a single decision may be driven by multiple parameters, while a single parameter may affect multiple decisions. For example, the stopping decision is jointly determined by preferences, search costs, and the scale of uncertainty, while preferences may also be inferred from the inspection order.

Once the probability function is rewritten as in Equation \eqref{eq:pr_jp_dec}, many identification arguments under linear specifications become straightforward. Since the differencing matrix $D$ is always of full rank, preference parameters can be identified separately from search costs, following arguments similar to those in standard discrete choice models (e.g., \cite{berry2014identification}). However, because the stochastic components in reservation values and purchase values are dependent, it is essential to reconsider whether specific distributional assumptions are required for normalization. While the necessity of such assumptions depends on the specification, our discussion focuses on two fundamental principles: ``only differences in utility matter,'' and ``the scale of utility is arbitrary.'' These principles determine location and scale normalizations in discrete choice models, and remain critical for identification in sequential search models.

Let us examine the first principle. In the linear specification, the absolute levels of reservation and purchase values are irrelevant. As long as the probability function can be reformulated as Equation \eqref{eq:pr_jp_dec}, any constant added to all values cancels out with the differencing matrix. 

We point out that the value-differencing form holds only if both reservation and purchase values include a conditionally independent stochastic component, which ensures the decomposition in the second equality of Equation \eqref{eq:pr_jp_dec}. Such randomness in purchase values is guaranteed by $\varepsilon_{ij}$, whereas reservation values require additional stochasticity from $\xi^u_{ij}$ or $\xi^z_{ij}$. In the full absence of these terms, the search process is determined entirely by observed heterogeneity, and consumers with identical preferences and search costs always follow the same process. This leads to degenerate likelihood functions and a perfect separation issue, resulting in estimation failure for likelihood-based methods.\footnote{An alternative is to increase the dimensionality of preference heterogeneity, but this typically introduces high-dimensional heteroskedasticity and additional identification difficulties.} Hence, in empirical applications, incorporating a conditionally independent stochastic component in action values is therefore essential.

We now turn to the second principle. In standard discrete choice models, scale is irrelevant under linear specifications, and identification is typically achieved by normalizing the variance of the error term. This property, however, does not extend naturally to a Weitzman model. The key reason is that the distribution of $\varepsilon_{ij}$ influences not only the scale of purchase values but also the magnitude of reservation values through $m_\varepsilon(c_{ij})$, which is generally nonlinear in $c_{ij}$. Consequently, $c_{ij}$ and the distribution of $\varepsilon_{ij}$ must be identified jointly based on observed actions. In the absence of exclusive search cost shifters, identifying the distributional parameters of $\varepsilon_{ij}$ requires two conditions. First, $m_\varepsilon(\cdot)$ must be homogeneous of degree one; otherwise, the estimated intercept for search costs becomes sensitive to scale normalization. Second, if $c_{ij}$ is stochastic, its distribution must satisfy specific conditions to ensure that the associated parameters are scale invariant. Moreover, while satisfying these conditions may enable identification in theory, heteroskedasticity often complicates estimation in practice. 

We illustrate this point with a stylized specification from \cite{kim2010online}, in which the consumer has a pre-search taste for each product before inspections. The taste is denoted by $\xi_{ij}$ and affects both reservation and purchase values. The specification is stated as: 
\begin{align}
    u_{ij} \ & = X_{ij} \beta_i + \xi_{ij} + \varepsilon_{ij} \label{eq:kim2010_start} \\ 
    z_{ij} \ & = X_{ij} \beta_i + \xi_{ij} + m_\varepsilon(c_0) \label{eq:kim2010_end}
\end{align}

Here, $\xi^u_{ij} = \xi_{ij}$ captures the unobserved stochasticity in reservation values, while the standard deviation of $\xi^z_{ij} = m_\varepsilon(c_0)$ is assumed to be zero. Based on this specification, we decompose the probability function in the stacked vectorized form as follows:
\begin{align*}
    \mathrm{Pr}(\{\mathcal{H},\mathcal{S}, \mathcal{R}, \mathcal{M}\}_i) & = \mathrm{Pr}\left(D \begin{pmatrix} \xi_{ih} + \varepsilon_{ih} \\ \bm{\xi}^{k}_{i} \\ \bm{\xi}^{n}_{i} \\ \bm{\xi}^{k^\prime}_{i} + \bm{\varepsilon}^{k^\prime}_{i}\end{pmatrix} < - D \begin{pmatrix} X_{ih} \beta_i \\ \bm{X}_i^{k} \beta_i + \vec{m_\varepsilon}(c_0) \\ \bm{X}_i^{n} \beta_i + \vec{m_\varepsilon}(c_0) \\ \bm{X}_i^{k^\prime} \beta_i \end{pmatrix} \right). 
\end{align*}

We focus on whether $\sigma_\varepsilon$ and $\sigma_\xi$ can be identified without additional assumptions. In this setup, $\xi_{ij}$ is a shared component that enters both the reservation and purchase values. Rescaling $\xi_{ij}$ does not affect the relative scale of the preference parameters. Hence, the central problem lies in the search cost parameter, as well as the absolute or relative magnitudes of $\sigma_\varepsilon$ and $\sigma_\xi$.

We first note that normalization based on $\sigma_\xi$ is not generally applicable, as $m_\varepsilon(c_0)$ may not be a homogeneous function of degree one. \cite{kim2010online} provides a sufficient condition under which this normalization is valid: when $\varepsilon_{ij}$ is independently normally distributed, $m_\varepsilon(c_0)$ can be expressed in the form $m_\varepsilon(c_0) = \sigma_\varepsilon \cdot m(c_0/\sigma_\varepsilon)$, where the function $m(\cdot)$ is irrelevant to the distribution of $\varepsilon_{ij}$. Hence, we can normalize the model as follows: 
\begin{align}
    & \tilde{u}_{ij} = u_{ij}/\sigma_\xi =  X_{ij}(\beta_i/\sigma_\xi) + \xi_{ij}/\sigma_\xi + (\sigma_\varepsilon/\sigma_{\xi}) \cdot (\varepsilon_{ij} / \sigma_{\varepsilon}); \\
    & \tilde{z}_{ij} = z_{ij}/\sigma_\xi = X_{ij}(\beta_i/\sigma_\xi) + \xi_{ij}/\sigma_\xi + (\sigma_\varepsilon/\sigma_{\xi}) \cdot m\left(\frac{c_0/\sigma_{\xi}}{\sigma_\varepsilon/\sigma_{\xi}}\right). \label{eq:kim2010_end_variation}
\end{align}

For search costs, since they enter the probability function only through the intercept term given by $m(c_0/\sigma_\varepsilon)$, identification of $c_0$ can be achieved if the ratio $\sigma_\varepsilon/\sigma_\xi$ can be identified from heteroskedasticity. Although this ratio is, in principle, directly identifiable, empirical work documents substantial practical difficulties in doing so \citep{jiang2021consumer, chung2025simulated}. The source of this difficulty is that $\xi_{ij}$, as unobserved heterogeneity, induces correlation between the purchase value and the reservation value for the same product. When such a correlation is present, stable identification typically requires additional exclusion restrictions, namely, observable variables that affect only purchase values or only reservation values; otherwise, identification may become fragile \citep{keane1992note}. However, these exclusion restrictions are often difficult to justify from an economics perspective or infeasible due to data limitations.\footnote{\cite{yavorsky2021consumer} provides a successful example of applying exclusion restrictions. They use distance to automobile dealers as an extra variable that affects only reservation values, thereby achieving identification of $\sigma_\varepsilon$.} In light of this reality, existing empirical studies commonly impose the assumption $\sigma_\varepsilon/\sigma_\xi = 1$ to obtain robust estimation performance, at the cost of making search cost estimates dependent on this assumption.\footnote{A key implication is that if the identification of the search cost relies on distributional assumptions, the resulting estimates should not be directly monetized \citep{ursu2024sequential, compiani2024online, greminger2024heterogeneous}.} Appendix \ref{subapp:iden_chung} discusses the identification in a different specification \citep{chung2025simulated}, which excludes $\xi_{ij}$ and introduces randomness into reservation values through a stochastic search cost $c_{ij}$. This specification removes the correlation between reservation values and purchase values, thereby yielding more stable identification.

The identification arguments discussed in this subsection apply broadly to sequential search settings beyond the Weitzman model. The principle that ``only utility differences matter'' requires the value indices of actions to contain conditionally independent randomness, typically implemented through univariate independent shocks. By contrast, the principle that ``utility scales are arbitrary'' does not always hold. Even when scale invariance is satisfied, the resulting heteroskedasticity may weaken identification in practice, making additional distributional assumptions a practical compromise. In later sections, we extend the analysis to extended models whose underlying processes still admit a ranking representation. We point out that the identification arguments also apply to the identification of extended models. 

\subsection{Estimation Method} \label{subsec:estimation}

Rewriting the probability function as in Equation \eqref{eq:pr_jp_dec} allows the use of established methods from ranking models to estimate a sequential search model. Among these methods, the GHK-style simulator is a natural choice for computing the simulated likelihood.
This simulator has been used in prior studies \citep{jiang2021consumer, chung2025simulated} to compute the likelihood function given by Equation \eqref{eq:osr_jp}, whereas we apply it to compute Equation \eqref{eq:pr_jp_dec}. The following description provides a general guideline for the GHK-style simulator that applies not only to the Weitzman setting but also to all sequential search models discussed in the remainder of the paper.

\begin{enumerate}
    \item Draw $y_i$ from its distribution. Conditional on $y_i^d$, use the GHK sampling technique to sequentially draw values for the selected actions consistent with the part of the ranking derived from the observed action sequence, and compute its conditional probability.
    \item Compute the probability that all unselected actions fall below $y_i^d$, and multiply it by the result of Step 1 to obtain the simulated likelihood.
\end{enumerate}

Taking the model with specifications in Equations \eqref{eq:kim2010_start} to \eqref{eq:kim2010_end} as an example, we describe the implementation procedure, and show its application to the example in Figure \ref{fig:OSRrepresentation} in Figure \ref{fig:implementation}:

\begin{enumerate}
    \item If preferences are heterogeneous, make draws to determine $\beta_i^d$ for each individual $i$. Draw $\xi_{iJ}$ randomly to determine the latent variable $z_{iJ}^d$.
    \item Sequentially draw $\xi_{ij}$ for $j = J-1, \dots, 1$ conditional on $z_{ij} > z_{i,j+1}^d$ to determine $z_{ij}^d$, and compute $p_{i,a}^d = \prod_{j=1}^{J-1} \Pr(z_{ij} > z_{i,j+1}^d)$.
    \item If $h \neq J$, draw $\varepsilon_{ih}$ conditional on $u_{ih} < z_{iJ}^d$ to determine $u_{ih}^d$ and compute $p_{i,b}^d = \Pr(u_{ih} < z_{iJ}^d)$. Otherwise, draw $\varepsilon_{ih}$ randomly and set $p_{i,b}^d = 1$.
    \item Compute $p_{i,c}^d = \prod_{J < k \leq |\mathcal{M}_i|} \Pr(z_{ik} < y_i^d)$ and $p_{i,d}^d = \prod_{1 \leq j \leq J, j \neq h} \Pr(u_{ij} < y_i^d)$.
    \item Define the likelihood contribution $L_i^d = p_{i,a}^d \cdot p_{i,b}^d \cdot p_{i,c}^d \cdot p_{i,d}^d$ and average across $D$ draws to obtain the simulated likelihood $\hat{L}_i = \frac{1}{D} \sum_{d=1}^D L_i^d$.\footnote{We do not consider the outside option here because its incorporation typically depends on extra assumptions. We treat this as a case of a sequential search process beyond the Weitzman-style in Section \ref{sec:extension}.}
\end{enumerate}

\begin{figure}[h]
    \centering
    \begin{minipage}{\linewidth} 
    {\centering
        \input{TikzPics/figure_implementation.tex}
        \par}
        \caption{Rank-Based GHK Implementation for the Search Process in Figure \ref{fig:OSRrepresentation}}
        \label{fig:implementation}
        \vspace{5pt}
        {\footnotesize \emph{Notes:} This figure illustrates the implementation of the rank-based GHK simulator for computing the likelihood of the process in Figure \ref{fig:OSRrepresentation}, based on the ranking representation in Figure \ref{fig:optimality}. At each step, newly drawn values are shown in blue, previously drawn values are shown in green, and probabilities computed analytically are shown in gray.\par}
    \end{minipage}
\end{figure}

We assess the performance of our proposed simulator through a Monte Carlo experiment, comparing it with the GHK-style simulator presented in \cite{ursu2024sequential}, a simplified version of the one applied in \cite{jiang2021consumer}. For clarity, we refer to the existing simulator as the policy-based GHK simulator and to our proposed method as the rank-based GHK simulator.\footnote{We do not compare the rank-based GHK simulator with other methods, such as the Crude or Kernel-smoothed Frequency Simulators, because prior studies have shown that the policy-based GHK method outperforms these approaches. We acknowledge the contributions of \cite{ursu2024sequential} and \cite{chung2025simulated}, who provide extensive simulation-based comparisons between policy-based GHK and alternative methods. More recent, targeted approaches, such as importance sampling \citep{morozov2021estimation} and Bayesian MCMC methods \citep{morozov2023measuring, onzo2025bayesian}, could, in principle, also benefit from the simplified likelihood function derived from the ranking representation. Given the limited use of these methods in conventional ranking models, we do not pursue this discussion further in this paper.}

To ensure comparability, we adopt the benchmark specification from \cite{ursu2024sequential}, constructing the policy-based GHK likelihood strictly following their publicly available MATLAB code, with modifications only to the number of products and the product attribute details.
\begin{align} 
    & u_{ij} = \sum_s \boldsymbol{\gamma}_s \cdot x^s_{j} + \beta \cdot p_{ij} + \zeta_{ij} + \varepsilon_{ij}, \quad \mbox{where } \zeta_{ij} \sim N(0,1) \mbox{ and } \varepsilon_{ij} \sim N(0, \sigma_\varepsilon); \label{eq:spec_est_start} \\
    & z_{ij} = \sum_s \boldsymbol{\gamma}_s \cdot x^s_{j} + \beta \cdot p_{ij} + \zeta_{ij} + m_\varepsilon(\exp(\bar{c})); \\
    & u_{i}^{outside} = \gamma^{outside} + \varepsilon_{i0}, \quad \mbox{where } \varepsilon_{i0} \sim N(0, \sigma_\varepsilon). \label{eq:spec_est_end} 
\end{align} 

Here, the purchase value of product $j$ consists of a series of attributes, $x^s_j$, and a consumer-specific price, $p_j$. All preference parameters are assumed to be homogeneous across consumers.\footnote{In theory, preference heterogeneity is parametrically identifiable in cross-sectional settings, while nonparametric identification is generally considered infeasible \citep{morozov2023measuring}. In practice, the preference heterogeneity is difficult to identify even parametrically because of $\zeta_{ij}$. Appendix \ref{subapp:hete_pref} provides a detailed investigation. } Search costs are set to be constant. As discussed in Section \ref{subsec:identification}, while it is theoretically possible to estimate $\sigma_\varepsilon$, doing so without cost shifters is empirically challenging. We therefore fix $\sigma_\varepsilon = 1$. To ensure comparability independent of the number of simulation draws, we fix the number of groups of draws at 500. The results are reported in Table \ref{tab:ghk}.

\begin{table}[h]\footnotesize
    \centering
    \caption{Monte Carlo Simulation Results between the GHK-style simulators} \label{tab:ghk}
    \input{Tables/tab_ghk.tex}
\end{table}

We compare the estimation performance of the two simulators across two market sizes. The first corresponds to a minimal market or experimental environment with 8 products. The purchase value of each product is specified as a linear function of three binary attributes and price, encompassing all possible attribute combinations. The product with attributes $[0,0,0]$ is used as the normalized base alternative. The second corresponds to a typical medium-to-large market with 32 products. The product attributes are given by five binary attributes and price, and the product with attributes $[0, 0, 0, 0, 0]$ serves as the base alternative. In both settings, the search cost is held constant. The results are summarized in Columns (1) to (4) of Table \ref{tab:ghk}.

We find that in the small market (with an average of 1.4 products inspected), the two simulators yield very similar estimates for both search costs and attributes preference parameters, with negligible differences in RMSE.\footnote{RMSE is calculated following \cite{ursu2024sequential} as $\sqrt{1/N_\theta \times \sum [\hat{\theta - \theta^{true}}]^2}$, where the summation covers all parameters, including attribute preferences and the search cost.} In the larger market (with an average of 3.3 products inspected), differences become pronounced: policy-based GHK performs better in estimating preference parameters, while rank-based GHK produces better estimates of search costs. Overall, rank-based GHK achieves a lower RMSE and slightly outperforms policy-based GHK. 

Beyond estimation performances, the primary advantage of rank-based GHK over policy-based GHK lies in its simplicity and computational efficiency, which directly address the most demanding limitations of the latter.

First, in terms of implementation simplicity, the rank-based GHK is considerably more streamlined than policy-based GHK, requiring less than 40\% of the latter's coding effort. This simplification arises because policy-based GHK must separately construct likelihood functions for different types of observed sequences, depending on whether the purchased product is the outside option, the last inspected, or an earlier inspected alternative. While in rank-based GHK, such distinctions only determine whether the distributional restriction applies when sampling $y_i$.

Second, rank-based GHK is computationally more efficient. In a market with 32 products, it requires less than 60\% of the computation time of policy-based GHK. This efficiency stems from the lower simulation dimensionality: policy-based GHK typically draws unobserved components for all unselected purchase actions, requiring up to $J + |\mathcal{M}_i|$ values to be simulated per sequence. By contrast, rank-based GHK makes draws only for the values of selected actions, with a total of $J + 1$ values to be simulated.\footnote{The minimum simulation dimensionality is $J$. Under the current specification, $\zeta_{i1}$ is additionally drawn to compute the probability associated with $u_{i1}$, which may be unnecessary under alternative specifications.} On average, computing the simulated likelihood for a sequence requires 33.7 draws under policy-based GHK, but only 4.3 draws under rank-based GHK. For a fair comparison, we match the total number of draws of the two methods, with the results reported in Column (5) of Table \ref{tab:ghk}. When the total number of draws is held constant, rank-based GHK produces preference parameter estimates nearly identical to those from policy-based GHK while delivering substantially more accurate estimates of the outside option value and the search cost, with an RMSE approximately half that of the policy-based GHK.

In conclusion, compared with policy-based GHK, rank-based GHK achieves higher estimation accuracy while substantially reducing computational and implementation costs. Moreover, it fully retains the advantages of policy-based GHK over the Crude Frequency Simulator and the Kernel-smoothed Frequency Simulator, two other commonly used empirical methods.\footnote{The Crude Frequency Simulator is mainly suitable for short search sequences in small markets, as it requires a large number of draws to avoid zero-probability issues. The Kernel-smoothed Frequency Simulator alleviates this problem by introducing external smoothing, but its performance depends critically on the choice of smoothing parameters, which lack theoretical guidance and often require calibration through simulation, thereby increasing implementation complexity and limiting comparability across models. In contrast, the policy-based GHK method avoids these issues, with its primary drawback being implementation complexity \citep{ursu2024sequential}.}

Another method for estimating ranking models is the exploded logit approach \citep{beggs1981assessing, chapman1982exploiting}, which \cite{compiani2024online} adapts to the Weitzman model through a ``double logit'' framework. Essentially, this approach also follows the probability function described in Equation \eqref{eq:pr_jp_dec}. By imposing a Gumbel distribution on both the randomness in reservation and purchase values \footnote{Technically, this can be achieved through a Gumbel-preserving distribution for random search costs\citep{moraga2023consumer}, which ensures that when the post-inspection taste shock $\varepsilon_{ij}$ follows a Gumbel distribution, the inspection propensity $\xi^z_{ij}$ derived from Equation \eqref{eq:rv} also follows a Gumbel distribution.}, the approach delivers a closed-form solution for the likelihood function. Its main advantage lies in the strong identification performance induced by the Gumbel assumption. Its limitation, however, is that conditional probabilities in later steps still depend on previously revealed purchase values. Thus, sequential dependence is addressed but not eliminated. As a result, implementation requires the consumer's complete action sequence to be directly observed or, when unobserved, enumerated. We show in the next section and in Appendix \ref{app:simulator_scope} that our strategy does not rely on such path-level observability and is more flexible for extended models in practice.

\section{The Ranking Representation in Extended Models} \label{sec:extension}

We now highlight the other main advantage of the ranking representation: its broad empirical applicability. Since Lemma \ref{lem:equivalence} applies to any branching project that satisfies Independence and Invariance, any sequential search process that can be regarded as a branching project satisfying these assumptions can be fully captured by an equivalent ranking, rendering it static and free of sequence dependence. Therefore, the ranking representation is particularly useful for handling many variants of sequential search models that would otherwise be difficult to analyze.

In this section, we first provide a broad definition of sequential search processes and, through several examples, illustrate that many commonly studied settings beyond the classical Weitzman-style can also be accommodated within this framework. We then focus on two representative examples to demonstrate how they admit a ranking representation and can be estimated by directly applying the rank-based GHK method. Finally, we characterize the application boundary of the ranking representation: its existence requires only an identified topological ordering among feasible actions. That is, the representation is well-defined provided that the set of feasible actions forms a unique directed acyclic graph up to transitive equivalence. 

\subsection{The Broad Class of Sequential Search Models} \label{subsec:extension_examples}

A broad definition of the sequential search process describes how a decision-maker gradually performs actions across multiple alternatives and ultimately selects one. The utility of each alternative may be fully known at the outset or may need to be revealed through a sequence of actions with immediate payoffs. For each alternative, the associated actions must be conducted in a fixed order, with subsequent actions becoming feasible only after the preceding ones are completed. Selecting an action may reveal partial utility information about one or more alternatives or may simply serve as a prerequisite for subsequent actions. The decision process unfolds over multiple stages, with the decision-maker selecting one feasible action at each stage. An alternative becomes eligible for eventual selection only once its utility is fully revealed. The process terminates upon making a final choice. 

Under this definition, the Weitzman model constitutes a special case of sequential search models. We provide several additional examples to illustrate the range of encompassed models.
\begin{enumerate}[label={}, leftmargin=0pt]
    \item \textbf{Example 1}, standard discrete choice. A standard discrete choice problem can be viewed as a degenerate sequential search process, in which all product utilities are fully known at the outset, and the decision-maker chooses only the final purchase, with no further actions required.
    \item \textbf{Example 2}, Weitzman-style sequential search with a restricted outside option. In many empirical applications, researchers observe only consumers who have conducted at least one action. To align with the data structure, we typically assume that consumers can select the action of leaving the market after the first inspection. 
    \item \textbf{Example 3}, partially observable Weitzman-style sequential search. Researchers may observe only part of the consumer’s action sequence, and at each step, selectable alternatives and feasible actions may be unclear. As we show in the next subsection, a partially observed search process can itself be viewed as a sequential search process involving constructed composite actions.
    \item \textbf{Example 4}, two-stage product search \citep{gibbard2022model}. In this model, each product in the market entails two sources of uncertainty that must be resolved sequentially with different actions: first browse to learn some product attributes, then inspect the remainder. 
    \item \textbf{Example 5}, sequential search with discovery \citep{greminger2022optimal}. In this model, the consumer initially knows a subset of products in the market. In addition to inspecting and purchasing known products, she must undertake discovery actions to become aware of additional products; only discovered products can be inspected. Since products may undergo one or more discovery actions before being discovered, the process introduces additional action types and implies that a product may require an indeterminate number of stages before it can be purchased.
\end{enumerate}

Despite differences in their specific settings, these examples share a key feature: they can all be mapped into multi-stage discrete choice processes over actions structured as a branching project. As in previous sections, we focus on sequential search processes that satisfy the Independence and Invariance assumptions. Under these conditions, the optimal strategies conform to the Gittins index policy \citep{keller2003branching}, and Lemma \ref{lem:equivalence} applies, providing a unified foundation for the ranking representation and its associated empirical strategy.

\subsection{Partially Observed Search Process} \label{subsec:extension_incomplete}

In many empirical works using the Weitzman model, the inability to observe the complete search process poses a significant challenge. Under optimal rules, sequence dependence implies that missing information about the early search process distorts the conditional probabilities of subsequent decisions. Hence, empirical works typically require full observation of the entire search process. Otherwise, researchers must impute missing actions through assumptions or enumeration \citep[e.g.,][]{ursu2018power, compiani2024online, chung2025simulated}, or reconstruct optimality in some less data-intensive settings based on revised optimal rules, such as the Eventual Purchase Theorem \citep{armstrong2017ordered, choi2018consumer, kleinberg2016descending}. The former approach may introduce bias through restrictive assumptions or entail substantial computational costs from full enumeration, whereas the latter leads to a tough trade-off between retaining useful information that falls outside the revised optimal rules and the model's tractability.

Due to partial observability, insufficient information prevents the construction of a branching project corresponding to the original search process. This can be categorized into two cases.

The first case arises when some selected actions or their order are unobserved in some steps, while information on other steps remains fully observed. This allows us to directly recover the ranking relations implied in the underlying branching project. For example, in Figure \ref{fig:OSRrepresentation}, suppose the order in which products 2 and 3 are inspected is unobserved, while it is known that product 1 is inspected in the first step and product 4 in the fourth step. This implies that inspecting products 2 and 3 both rank above inspecting product 4. Accordingly, in the branching project, these two steps can be treated as additional restrictions relative to inspecting product 4. The resulting branching project and the corresponding ranking representation are illustrated in Figure \ref{fig:missing_steps}. 

\begin{figure}[h]
    \centering
    \begin{minipage}{\linewidth}
        {\centering
        \input{TikzPics/figure_recover.tex}
        \par}
        \caption{Branching Project and Ranking Representations of Missing Steps}
        \label{fig:missing_steps}
        \vspace{5pt}
        {\footnotesize \emph{Notes:} The figure presents the branching project and ranking representations corresponding to the search process in Figure \ref{fig:OSRrepresentation} when the second and third steps are unobserved. In the left panel, the observed selected actions $\{I_1, I_4, P_3\}$ are highlighted in blue. At each step, actions selected in previous steps are indicated by dashed circles, actions selected after the previous step but before the current step are shown in gray, and feasible but unselected actions are shown in green. In the right panel, selected actions are marked in blue, unselected actions in green, and red arrows indicate dominance relations.\par}
    \end{minipage}
\end{figure}

The second case arises when missing information in some steps prevents the identification of feasible action sets in other steps. Hence, although the selected actions in the remaining steps are observed, the researcher cannot determine which actions are dominated by these selections, and therefore cannot construct the ranking representation. To address this issue, we construct a new branching project whose fully observed action sequence corresponds to the original partially observed sequence. The following lemma formalizes how the observed information is consistently mapped into this reconstructed process.

\begin{lemma} \label{lem:composite}
Consider a branching project that satisfies Independence and Invariance, and focus on two consecutive actions associated with an alternative $j$, the parent action $b_j$, and its child action $a_j$, with index values $V_{b_j}$ and $V_{a_j}$. If the feasibility of $a_j$ cannot be determined at the time of decision-making, the pair $(b_j, a_j)$ can be treated as a single composite action $e_j$, which is considered feasible whenever $b_j$ is feasible and is regarded as selected only when $a_j$ is selected. Moreover, its effective index value is: 
\begin{align*}
    V_{e_j} = \min \{ V_{b_j}, V_{a_j} \}
\end{align*}
\end{lemma}
\begin{proof}
See Appendix \ref{app:proof_effective}.
\end{proof}

Composite actions offer an effective solution to partial observability. If a consumer inspects an alternative $j$ ($I_j$), then for any other alternative $k$, two scenarios arise. In the first, purchasing $k$ ($P_k$) is infeasible because its parent action, inspecting $k$ ($I_k$), has not been selected, implying that $I_j$ dominates $I_k$. In the second, $P_k$ is feasible but not selected, implying that $I_j$ dominates $P_k$. A composite action combining $I_k$ and $P_k$ unifies these cases, enabling comparison when $k$’s associated action is undetermined. Crucially, if the original process with individual actions satisfies Independence and Invariance, the process with composite actions also satisfies these assumptions. Therefore, Lemma \ref{lem:equivalence} remains applicable, and the Gittins index policy retains its optimality. Composite actions can additionally be combined with other individual actions.

Lemma \ref{lem:composite} can be flexibly applied to transform a partially observed sequential search process into a new branching project. As an example, consider the process illustrated in Figure \ref{fig:OSRrepresentation}, but only observe that Product 3 is purchased. This implies that the observable action sequence is $\{I_3, P_3\}$. Denoting the composite action for inspecting and purchasing the same product $j$ as $E_j$, we transform this sequence into the branching project as in the left panel of Figure \ref{fig:single_branching_partial}. 

\begin{figure}[htbp]
    \centering
    \begin{minipage}{\linewidth}
        {\centering
        \input{TikzPics/figure_ept.tex}
        \par}
        \caption{Branching Project and Ranking Representations under Partial Observability}
        \label{fig:single_branching_partial}
        \vspace{5pt}
        {\footnotesize \emph{Notes:} The figure illustrates the branching project and ranking representations corresponding to the search process in Figure \ref{fig:OSRrepresentation} when only the purchased product is observed. In the left panel, the observed actions $\{I_3, P_3\}$ are highlighted in blue at respective steps, other feasible actions are shown in green, with selected actions marked by dashed circles. In the right panel, selected actions are marked in blue, unselected actions in green, and red arrows indicate dominance relations.\par}
    \end{minipage}
\end{figure}
For other cases involving partially observed action sequences, Appendix \ref{app:transformation} provides an algorithm for implementing this transformation with Lemma \ref{lem:composite}. Based on the transformation, the next theorem delivers the ranking representation of the reconstructed sequential search process.

\begin{theorem}\label{theorem:incomplete}
Consider a Weitzman-style sequential search process that satisfies Independence and Invariance, with a potentially partially observed action sequence. Suppose the consumer ranks all feasible actions (or composite actions) throughout the sequential search process. Denote the action selected in step $j$ by $a_j$, the process is optimal only if the resulting ranking satisfies:
\begin{enumerate}
    \item $a_{j} \succ a_{j+t}$ for all $j$ and $t \in \mathbb{Z}_{>0}$, if $a_{j+t}$ is feasible at step $j$.
    \item All unselected actions rank lower than the preventive action $\bar{a}$. 
\end{enumerate}
\end{theorem}

We formally define the preventive action as follows: 
\begin{definition}[Forward-Accessibility]
A step in the search process is \textbf{forward-accessible} if any action feasible at that step is selected in a subsequent step.
\end{definition}

\begin{definition}[Preventive Action]
The \textbf{preventive action} is defined as the lowest-ranked action among all actions selected in forward-inaccessible steps. Its value is denoted by $y_i$, which equals the minimum value among these selected actions.
\end{definition}

The definition of the preventive action guarantees that no feasible but unselected action dominates any observed selected action. This principle applies to both individual and composite actions. In a Weitzman-style sequential search process, if the purchased product is inspected before the final step, only the purchase step is forward-inaccessible; if the product is inspected at the final step, then both the final inspection and the purchase steps are forward-inaccessible. Accordingly, the definition of the preventive action in Theorem \ref{theorem:main} is consistent with that in Theorem \ref{theorem:incomplete}: as illustrated in Figure \ref{fig:single_branching_partial}, both Step 1 and Step 2 are forward inaccessible, and thus $\bar{a}$ is the lower-ranked between $I_J$ and $P_h$, and $y_i = \min\{u_{ih}, z_{ih}\}$, as in Proposition \ref{Prop:weitzman}.

The proof of Theorem \ref{theorem:incomplete} is provided in Appendix \ref{app:proof_incomplete}. The theorem establishes that, even without complete observation of a Weitzman-style sequential search process, a reconstruction process based on a partially observed action sequence still admits a ranking representation. This implication is illustrated in Figure \ref{fig:single_branching_partial}: applying Theorem \ref{theorem:incomplete} to the branching project in the left panel yields the ranking representation in the right panel. Notice that Theorem \ref{theorem:incomplete} classifies all observed feasible actions into three categories: selected actions in forward-accessible steps, selected actions in forward-inaccessible steps, and unselected actions. Incorporating all these categories ensures that the ranking representation entails no loss of information relative to the reconstructed sequential search process.

It can be seen that Theorem \ref{theorem:main} is a special case of Theorem \ref{theorem:incomplete} under full observability. Moreover, existing propositions in the literature concerning various forms of partial observation can be directly derived from Lemma \ref{lem:composite} and Theorem \ref{theorem:incomplete}. We illustrate with two examples. First, consider the case where no search information is observed and only the purchase is known:

\begin{proposition} \label{Prop:ept}
(Eventual Purchase) For any Weitzman-style sequential search process satisfying Independence and Invariance with a purchased product $j$, it must be that $\min\{u_{ij}, z_{ij}\} > \min\{u_{ik}, z_{ik}\}, \forall k \neq j$. 
\end{proposition}

\begin{proof}
Figure \ref{fig:single_branching_partial} illustrates the sequential search process reconstructed using Lemma \ref{lem:composite}. In the reconstructed process, the two steps of inspecting and purchasing product $j$ are both forward-inaccessible. Therefore, by Theorem \ref{theorem:incomplete}, for all $k \neq j$, we have $\min\{u_{ij}, z_{ij}\} > \min\{u_{ik}, z_{ik}\}$. 
\end{proof}

Proposition \ref{Prop:ept} shows that Theorem \ref{theorem:incomplete} subsumes the famous Eventual Purchase Theorem, a result widely used for allowing researchers to infer demand directly from a discrete choice model without requiring detailed search process data.\footnote{For example, \cite{moraga2023consumer} derives a closed-form solution for consumers’ final purchase in a Weitzman model following Proposition \ref{Prop:ept}, under a Gumbel-preserving distribution for random search costs, embeds it within a BLP style framework, and applies it to the Dutch car market.}

Unobservable inspection order is a common limitation in empirical studies, also present in widely used sources such as Expedia hotel booking data \citep[e.g.,][]{ursu2018power, compiani2024online, greminger2024heterogeneous, kaye2024personalization}. Based on Theorem \ref{theorem:incomplete}, the process can be immediately reformulated to yield three sets of inequalities that fully characterize optimality in this setting.\footnote{These inequalities provide an alternative formulation of Proposition 2.1 in \citet{jolivet2019consumer} and Appendix F in \cite{greminger2024heterogeneous}.}

\begin{proposition} \label{Prop:nopath}
(No Inspection Order) Consider a Weitzman-style sequential search process of consumer $i$ that satisfies Independence and Invariance, with the inspection order entirely unobserved. The process can only be optimal if the following conditions are satisfied:
\begin{align*}
        w_{ih} > u_{ij}, \ \forall j \in \mathcal{S}_i\backslash \mathcal{H}_i;  
        \quad \quad w_{ih} < z_{ij},  \ \forall j \in \mathcal{S}_i\backslash \mathcal{H}_i; 
        \quad \quad w_{ih} > z_{ik},  \ \forall k \in \mathcal{M}_i\backslash \mathcal{S}_i. 
    \end{align*}
\end{proposition}

\begin{proof}
    See Appendix \ref{app:proof_nopath}. 
\end{proof}

For the remainder of this section, we conduct Monte Carlo simulations to evaluate the model’s estimation performance under varying degrees of partial observability using the rank-based GHK simulator. Specifically, we raise five illustrative scenarios of partial observability:
\begin{enumerate}[label={}, leftmargin=1.5em]
    \item \textbf{Scenario 1}: Only the final purchase is observed; 
    \item \textbf{Scenario 2}: Only the set of inspected products and the final purchase are observed; 
    \item \textbf{Scenario 3}: Only the inspection actions and their order are observed; 
    \item \textbf{Scenario 4}: Only the first inspection and the final purchase are observed; 
    \item \textbf{Scenario 5}: Only the search process over a subset of products is observed. 
\end{enumerate}

In all these scenarios, we do not rely on revisions to the optimal rules, external assumptions, or full enumerations. Instead, we apply Lemma \ref{lem:composite} and Theorem \ref{theorem:incomplete} to establish a ranking representation for these partially observed search processes, and carry out estimation using modified rank-based GHK simulators. Appendix \ref{app:incomplete_likelihood} provides the implementation details for each case.\footnote{In Scenarios 3 and 5, the ranking representation can only be established for part of the search process, in which case the rank-based GHK method is applied locally and combined with the crude frequency method. }

The model specification used in the assessment is as follows:
\begin{align} 
    & u_{ij} = \sum_{s=1}^3 \gamma_s x^s_{j} + \beta p_{ij} + \zeta_{ij} + \varepsilon_{ij}, \quad \mbox{where } \zeta_{ij} \sim N(0,\sigma_\zeta^2); \label{eq:spec_est_start} \\
    & z_{ij} = \sum_{s=1}^3 \gamma_s x^s_{j} + \beta p_{ij} + \zeta_{ij} + m_\varepsilon(\bar{c}). \label{eq:spec_est_end}
\end{align} 

We simulate 50 datasets, each consisting of 2,000 consumers and 8 products, without an outside option. Estimation relies exclusively on partial information from each dataset. The purchase value of each product is modeled as a linear function of price and three binary attributes, with the eight products covering all possible combinations of these attributes. The product with attributes $[0,0,0]$ is taken as the base alternative for location normalization. Table \ref{table:firstview} reports the estimation results. The first five columns present estimates for five partial-observability scenarios, while the final column reports estimates based on full observability. In Scenario 5, the observed action sequences cover six of eight products, forming a proper subset of the full choice set.

\begin{table}[h]\footnotesize
\centering
\caption{Monte Carlo Simulation Results with Partially Observed Action Sequences} \label{table:firstview}
\input{Tables/tab_firstview_main}
\end{table}

Table \ref{table:firstview} shows that linear preference parameters are well estimated across all partial observability scenarios. However, when only final purchase data are observed, the estimation of search costs deteriorates substantially, with both the standard deviation and RMSE much higher than in other cases. This is not surprising, as in this case the model collapses to a standard discrete-choice setting, in which identification of search costs relies only on variation in irregularly-distributed inspection propensities, making it weak and sensitive to sampling variation. As information on inspections increases, the precision of search cost estimates improves markedly.

These results highlight two core advantages of our empirical strategy in handling partial observability. First, we allow partially observed information to be fully exploited for estimation in a unified and parsimonious manner. Our strategy does not rely on additional or revised optimal rules, nor does it require extra assumptions or data processing tailored to a specific empirical setting. Instead, once partially observed action sequences are viewed as realizations of a constructed branching project, whose ranking representation can be directly used for implementation. Second, we allow partial observations to be used independently for estimation. When datasets are excessively large, the approach provides substantial flexibility in data usage. For example, when the main objective is to estimate consumer preferences, researchers can rely solely on the set of inspected products (Scenario 2). When a strategy of interest affects only the early stages of search, the analysis can be restricted to those stages (Scenario 4).

Note that correctly identifying the observability of actions is crucial for estimation. For any unselected action, the researcher must determine whether it was unobserved or observed but not selected. When this distinction is unavailable, separate likelihood contributions must be specified for both cases and summed, weighted by their observability probabilities. 

In Appendix \ref{app:expedia}, we apply the approaches above to the Expedia dataset to examine whether consumers reveal consistent preferences within the search process and in purchase decisions. Following the specification in \cite{ursu2018power}, our results show systematic differences between preference estimates based on the searched set data and searched set and purchase data, suggesting that treating search and purchase decisions as arising from the same preference environment may be misspecified. In particular, consumers place greater weight on hotel star ratings at the purchase stage, while review scores, location scores, and chain affiliation tend to matter more at the search stage. These findings suggest that search and purchase decisions may reflect different preference tradeoffs, with practical implications for recommendation system design. 

\subsection{Multi-Stage Information Acquisition} \label{subsec:extension_discovery}

In increasingly rich clickstream data from online platforms, researchers can observe a wide range of consumer behaviors related to information acquisition. For example, consumers may scroll or navigate to the next page to discover additional products, open supplementary tabs to view detailed specifications, proceed to the checkout page to review shipping and tax costs, or decide to return a product based on the post-purchase experience. These actions suggest that information acquisition in the search process may occur not in a single step but gradually through multiple stages involving different types of actions. Characterizing and understanding the general sequential information-acquisition process in a sequential search model is a natural approach for studying the effects of marketing strategies or policy interventions.

However, incorporating richer types of information-acquisition actions into sequential search, as represented by optimal policies, poses substantial empirical challenges. Accommodating additional stages and action types often requires re-deriving the optimal rules, which is technically demanding and increases estimation complexity. In practice, researchers therefore often rely on partial information from the data, trading informational richness for tractability.

This subsection extends the ranking representation and the associated empirical strategy to sequential search with multi-stage information acquisition. The theory is straightforward: whenever the process can be expressed as a branching project that satisfies Independence and Invariance, (i) Lemma \ref{lem:equivalence} continues to apply, and (ii) the Gittins index policy remains optimal. The following theorem establishes the ranking representation for the extension case.

\begin{theorem} \label{theorem:general}
    Consider an action sequence describing a sequential search process with multi-stage information acquisition that satisfies Independence and Invariance. Suppose the consumer ranks all feasible actions throughout the process. Denote the action selected at step $j$ by $a_j$, the process is optimal if and only if, in the corresponding ranking: 
    \begin{enumerate}
        \item $a_{j} \succ a_{j+t}$ for all $j$ and $t \in \mathbb{Z}_{>0}$, if $a_{j+t}$ is feasible at step $j$.
        \item Every unselected action $\ell$ ranks lower than its current preventive action $\bar{a}_{\ell}$.
    \end{enumerate}
    where $\bar{a}_{\ell}$ is the $\ell$-conditional preventive action.
\end{theorem}

\begin{definition}[Conditional Preventive Action]
The \textbf{$\ell$-conditional preventive action} is defined as the lowest-ranked selected action among all forward-inaccessible steps at which action $\ell$ is feasible. Its value is denoted by $y_{i\ell}$, equal to the minimum value among these selected actions.
\end{definition}

The proof of Theorem \ref{theorem:general} is provided in Appendix \ref{app:proof_general}. The theorem extends the main result from a structural perspective. Even when the assumptions on information acquisition stages and action types are relaxed, an optimal sequential search process that satisfies Independence and Invariance still admits a ranking representation. Theorem \ref{theorem:incomplete} therefore arises as a special case of Theorem \ref{theorem:general}, corresponding to a setting with two action types, a two-stage structure, and a common conditional preventive action for all unselected actions. 

Theorem \ref{theorem:general} further broadens the applicability of the ranking representation. We use Example 5 from Section \ref{subsec:extension_examples}, the sequential search with discovery model \citep{greminger2022optimal}, to illustrate how the ranking representation can be applied to such settings.\footnote{Another example in Section \ref{subsec:extension_examples} is the two-stage sequential search model \citep{gibbard2022model}, a brief discussion of which is offered in Appendix \ref{app:ts_implementation}. Note that some multi-stage information acquisition models, such as the search-purchase-return model in \cite{ibragimov2024clicks}, assume that once a purchase is made, all other feasible actions are eliminated, leaving only the decision of whether to return the product, which violates the Invariance assumption. Such settings are typically addressed by re-deriving action values that incorporate the possibility of returns.} In this model, a product must be discovered before it can be inspected or purchased. While the consumer is initially aware of only a subset of products, they can expand this set via discovery actions through different channels like product lists or promotion pages. Within each channel, a subsequent discovery becomes feasible only after the preceding one is completed. Hence, products outside the initial feasible set must undergo one or more discovery steps before they become feasible for inspection.

To extend the ranking representation to discovery actions, these actions must also admit well-defined and comparable values. This requires that discovery actions satisfy the Independence and Invariance assumptions. Independence ensures that the expected payoff of any discovery action is unaffected by inspection outcomes or other discoveries, so its value is predetermined and does not involve belief updating. Invariance ensures that once a discovery becomes feasible, it remains feasible until selected, which guarantees a stable feasible set. In addition, \cite{greminger2022optimal} imposes a weak monotonicity condition to ensure traceability. Along the same route, the value of any later discovery cannot exceed that of earlier ones. This prevents a discovery from being selected solely to access more valuable future discoveries.

Under this condition, similar to Equation \eqref{eq:rv}, \cite{greminger2022optimal} defines the Gittins index for the $t$-th discovery on route $r$, referred to as the \textit{discovery value}. Let
\begin{align*}
w_{irt} = \max_{j \text{ to be discovered at the } t\text{-th discovery on route } r} \min \{u_{ijr}, z_{ijr}\} ,
\end{align*}
which is unknown, while its cumulative distribution function $G_{irt}(w)$ can be derived. The discovery value $q_{irt}$ is then defined implicitly as the solution to
\begin{align*}
c_{irt}^{dis} = \int_{q_{irt}}^\infty (w - q_{irt}) \ dG_{irt}(w),
\end{align*}
where $c_{irt}^{dis}$ denotes the discovery cost of the $t$-th discovery on route $r$ for consumer $i$.

We illustrate how a sequential search process with discovery can be represented as a ranking through a simple example. Suppose the consumer initially knows only two products, 1 and 2, but can select $D_\mathrm{I}$ to discover product $3$ through a unique discovery route, which also generates a subsequent discovery opportunity $D_\mathrm{II}$. In this example, the consumer takes four sequential actions: inspecting product $1$ ($I_1$), performing the first discovery ($D_\mathrm{I}$), inspecting product $3$ ($I_3$), and purchasing product $3$ ($P_3$).

\begin{figure}[h]
    \centering
    \begin{minipage}{\linewidth}
        {\centering
        \input{TikzPics/figure_spd.tex}
        \par}
        \caption{\small \textls[-20] Branching Project and Ranking Representations of Sequential Search with Discovery}
        \label{fig:spd_illustration}
        \vspace{5pt}
        {\footnotesize \emph{Notes:} The figure presents the branching project and ranking representations for the example of sequential search with discovery. In the left panel, the selected actions $\{I_1, D_\mathrm{I}, I_3, P_3\}$ are highlighted in blue, while other feasible actions at each step appear in green, with selected actions marked by dashed circles. In the right panel, selected actions are highlighted in blue and unselected actions in green. Red arrows denote dominance relations.\par}
    \end{minipage}
\end{figure}

The branching project and ranking representations of this example process are shown in Figure \ref{fig:spd_illustration}. The mapping between them is established by Theorem \ref{theorem:general}. Among the four steps, the second, third, and fourth are forward-inaccessible, and the values of three conditional preventive actions can be computed from the values of selected actions $D_\mathrm{I}$, $I_3$, and $P_3$ in these steps. Among the unselected actions, $I_2$ and $P_1$ are feasible in all three forward-inaccessible steps, whereas $D_\mathrm{II}$ is infeasible in the second step (when $D_\mathrm{I}$ is selected).

We apply a modified rank-based GHK simulator for estimation based on the ranking representation. Its performance is demonstrated via a Monte Carlo study with the following specifications:
\begin{align*}
u_{ijr} &= \sum_{s=1}^3 \gamma_s \cdot x^s_{jr} + \beta \cdot p_{ijr} + \zeta_{ijr} + \varepsilon_{ijr}, \\
\log(c^{ins}_{ijr}) &= c^0, \\
\log(c^{dis}_{irt}) &\sim N(c^1,\sigma_c^2), \\
\zeta_{ijr},\,\varepsilon_{ijr} &\sim N(0,1), \\
\sigma_c &= 0.25,\quad r \in \{1,2\},\quad n_r = 2.
\end{align*}
where $c^{ins}_{ijr}$ denotes the search cost for product $j$ on route $r$, $n_r$ denotes the number of products revealed per discovery on route $r$. Notice that we assume that $c^{dis}_{irt}$ is stochastic and follows a given distribution, mirroring the arguments in Section \ref{subsec:identification}: variation in discovery costs provides an independent source of randomness for discovery values. We further assume that the variances of the pre- and post-inspection taste shocks are equal to 1 to strengthen identification. 

We perform the Monte Carlo simulation using 100 simulated datasets. Each dataset consists of 2000 consumers searching in a market of 1000 products. These products are split into two distinct routes: Route 1 (600 products) and Route 2 (400 products), where the latter offers lower average prices at the cost of reduced attribute variation. Each consumer begins with an initial set of 1 product plus an outside option and is randomly assigned 14 products to discover across the two routes. Throughout the process, each discovery action reveals $n_r = 2$ new products, unless only one remains undiscovered along the route. Importantly, consumers operate without knowing the total product count, such that they always believe that a discovery yields 2 products. As shown in Table \ref{table:discovery}, the results confirm the effectiveness of our modified simulator.

\begin{table}[h]\footnotesize
\centering
\caption{Monte Carlo Simulation Results with Sequential Search with Discovery\label{table:discovery}}
\input{Tables/tab_spd.tex}
\end{table}

The sequential search with discovery model has been empirically applied in \cite{zhang2024product}. Building on the optimal rules proposed in \cite{greminger2022optimal}, the paper implements estimation using a Kernel-smoothed Frequency Simulator. However, with a more complex search environment and a wider range of action types, implementation becomes considerably more difficult, and performance can be much weaker than that in the Weitzman model. By contrast, the rank-based GHK simulator is not subject to these limitations. Its implementation continues to follow the general guideline outlined in Section \ref{subsec:estimation}, and its estimation performance shows no evident deterioration relative to that in the Weitzman model. 

Notice that a multi-stage sequential search process can be partially observed. For example, \cite{greminger2024heterogeneous} provides a set of implications characterizing the optimality of a sequential search with discovery process when the inspection order is unobserved, and constructs the likelihood function based on these implications. Appendix \ref{app:spd_incomplete_implementation} shows that these implications follow directly from a combination of Theorem \ref{theorem:general} and \ref{theorem:incomplete} proposed in this paper.

\subsection{Applicability of the Ranking Representation} \label{subsec:extension_applicability} 
The examples in the preceding sections demonstrate that the ranking representation remains valid across a broad class of sequential search settings. However, the necessary conditions for its applicability must be formally established: given an observed action sequence, under what constraints can the underlying sequential search process be represented as a ranking?

The validity of the ranking representation hinges on structural conditions that go beyond the assumptions of Independence and Invariance. As in Section \ref{subsec:formalization}, we require that the search process can be decomposed into a sequence of discrete choices. Moreover, this sequence of discrete choices must satisfy the structural properties of a branching project.

The first condition requires that at each step, both the selected action and the alternative set are well-defined. A typical violation in which the feasible action set is unobserved occurs when the market choice set is unknown, preventing the researcher from determining the feasible alternatives against which the selected action is compared. Failure to observe the selected action occurs when an unobserved action is selected within the sequence. For example, if the data indicate only that a product is the second to be inspected, while the identity of the first inspected product remains unknown. In all scenarios discussed in Section \ref{subsec:extension_incomplete}, partial observability prevents the original search process from being decomposed into a sequence of discrete choices. As a result, Lemma \ref{lem:composite} is required to construct composite actions, thereby restoring representability in the form of a branching project. 

When the requirements for identifying the corresponding discrete choice sequence are satisfied, whether the sequence admits a branching project depends on the presence of sufficient and consistent dominance relations revealed in these choices. Drawing on concepts from graph theory, we establish the following theorem:

\begin{theorem} \label{theorem:DAG}
    Let $\mathfrak{G} = (V, E)$ be a directed graph. The node set, $V$, denotes the set of all feasible actions in a sequential search process satisfying Independence and Invariance. The edge set $E$ corresponds to the binary dominance relations implied by the sequence of discrete choices in the sequential search process. This sequential search process admits a branching project if and only if $\mathfrak{G}$ is a \textbf{unique Directed Acyclic Graph (DAG)} up to transitive equivalence.\footnote{A directed graph consists of a set of nodes and a set of edges. A DAG is a directed graph that contains no directed cycles. Two DAGs are transitively equivalent if they reach the same set of nodes from any given node. For any DAG, there exists a unique transitive reduction, which is the graph with the fewest possible edges that maintains the same reachability relation.}

    Furthermore, it follows from Lemma \ref{lem:equivalence} that $\mathfrak{G}$ corresponds to the ranking representation of the sequential search process up to a transitive equivalence.
\end{theorem}

The proof of Theorem \ref{theorem:DAG} is stated in Appendix \ref{app:proof_DAG}. It rests on the following fact: a topological sort exists only when the underlying graph is a DAG \citep[Proposition 1.4.3]{bang2008digraphs}, which allows all feasible actions to be represented as a ranking. When the discrete choices fail to form a DAG, cycles arise. For example, if we observe a consumer inspecting product A twice before and after inspecting a different product B, this implies a cycle between inspecting products A and B. This violates the structural requirement of a branching project that selected actions are removed from the feasible action set, and therefore no valid branching project can be established. If the discrete choices correspond to multiple DAGs, this suggests that the observed information is insufficient to determine the feasible action set or to establish relations among actions; thus, the ranking representation cannot be identified. 

Theorem \ref{theorem:DAG} generalizes the ranking representation of sequential search. Under the assumptions of Invariance and Independence, it extends the representation from specific settings to a broader framework. Its primary structural requirement is that the feasible action set forms a unique directed acyclic graph, ensuring the existence and consistency of a topological ranking. From an empirical perspective, this property allows researchers to bypass ad hoc derivations of optimal policies for varying environments and instead construct likelihood functions directly from the ranking representation, providing a unified empirical approach across diverse market settings.

Theorem \ref{theorem:DAG} also helps identify potential redundant conditions that may arise in implementing sequential search models. Although the ranking representation in Theorem \ref{theorem:main} contains no redundant conditions, the representations constructed for extended models in Theorems \ref{theorem:incomplete} and \ref{theorem:general} may still introduce redundant constraints. In such cases, one can construct the corresponding directed acyclic graph based on Theorem \ref{theorem:DAG} and apply transitive reduction to eliminate these redundancies, thereby avoiding the additional complexity induced by redundant constraints when implementing the model, in the same spirit as simplifying the optimal rules in Proposition \ref{Prop:weitzman}.

Finally, whenever a ranking representation applies under Theorem \ref{theorem:DAG}, the rank-based GHK simulator remains a natural tool for estimation, though its applicability is not universal. Appendix \ref{app:simulator_scope} discusses the scope and limitations of the simulator in greater detail. Yet, note that the GHK-style simulator is not the only method for performing the estimation. Generally speaking, once a ranking representation is established, estimation can draw on the full set of methods developed for ranking models. Improving computational strategies, particularly for large markets and long action sequences, remains an important direction for future research, for which the ranking representation provides a streamlined empirical foundation.\footnote{Recent progress includes \cite{wei2025pre}, who develops a neural network-based approach for estimating sequential search models under a given specification.} 

\section{Discussion} \label{sec:discussion}

Before concluding, we briefly discuss violations of the fundamental assumptions of Independence and Invariance, which clarify both the implications and boundaries of optimal sequential search.

Although the ranking representation established in this paper is primarily intended for empirical application, its economic interpretation should not be overlooked. The central insight is that, given all action values and under Independence and Invariance, a consumer’s optimal sequential search can be equivalently represented as a ranking over all feasible actions. Put differently, an optimal sequential search can be viewed as a discrete choice at the level of action rankings, in which consumers choose among all possible rankings according to their preferences and search costs. Under this interpretation, the sequential search setting serves as a framework that extends the standard discrete choice problem from a single final purchase to a ranking of actions determined by their Gittins indices. The realized search process then serves only to reveal which category of partial ranking the consumer’s selections belong to. From the researcher’s perspective, once optimality is imposed, the realized search path is fully determined by ex ante draws, regardless of the complexity of the underlying model specification: it neither generates nor alters any source of randomness in the environment. Therefore, even without simplification, as in Proposition \ref{Prop:ept}, the sequential search problem can still be interpreted as a fully static discrete-choice problem defined over the space of possible rankings.

Therefore, under these assumptions, optimal sequential search is fundamentally static, and dynamics emerge only when these assumptions fail. Violations of Independence or Invariance can be taken as introducing dynamics into an otherwise static discrete choice problem through endogenous and exogenous channels, respectively.

First, relaxing Invariance introduces dynamics driven by exogenous variation. In this case, action values evolve over time or in response to other observable state variables, in ways that can be parameterized and empirically identified, leading to changes in consumers' rankings of actions across sessions. In a simple case in which such changes are not anticipated by consumers, the problem remains quasi-static: the demand structure within each session can still be characterized as an optimal sequential search, while cross-session differences reflect unanticipated ranking shifts induced by exogenous factors.\footnote{Closely related settings have been studied in discrete choice setups. For example, \cite{conlon2013demand} considers unanticipated stockouts across sessions that generate sudden variations in observed choice sets.} Under this interpretation, researchers can characterize each session separately using the optimal sequential search framework, while identifying the parameters governing exogenous transitions from variation in rankings across sessions.\footnote{\cite{klein2024Do} apply this method to study preference discovery in a multi-session Weitzman setting.}

In more complex environments, consumers may anticipate exogenous changes and adjust their search behavior accordingly. Search decisions are then made under forward-looking expectations. For example, consumers may understand that current search actions affect the future status of certain alternatives. The setting is then a dynamic decision problem with a given state-transition structure. For tractability, researchers may first estimate the transition matrix using frequency-based methods or Markov estimation, and then estimate the remaining structural parameters.\footnote{\cite{ursu2023search} proposes a setting in which pausing search releases search fatigue and lowers subsequent search costs, with the search cost reduction effect being parameterized and estimated; \cite{elberg2019dynamic} develops a belief-updating framework in repeated purchase settings, where beliefs evolve according to an empirically estimated transition matrix that is then used to identify search parameters; \cite{gardete2024multiattribute} constructs a decision tree over attribute spaces, enabling joint estimation of belief-driven value transitions and search parameters.} In these settings, the Gittins index strategy is generally no longer optimal, and approximate index rules \citep[e.g.][]{lin2015learning, compiani2024online} are often required to quantify non-purchase actions.

Finally, relaxing Independence implies that the outcome of a search action affects the rankings of all other alternatives, which evolve through endogenous state transitions driven by realized search outcomes, with the transition structure itself also depending on prior stochastic realizations. A key example of Independence failure is cross-product learning. In this case, forward-looking optimal strategies must jointly account for current payoffs, the evolution of state variables, and the endogenous impact of current outcomes on future transitions. Given the complexity of the resulting intertemporal objective, empirical implementations often rely on approximate solutions to optimal strategies.\footnote{An example is \cite{ursu2020search}, which departs from the sequential search framework of \cite{weitzman1979optimal} and applies the framework of \cite{chick2012sequential}.}

\section{Conclusion} \label{sec:conclusion}

This paper establishes a theoretical equivalence between the optimal sequential search process and a partial ranking over feasible actions throughout the search process under the two fundamental assumptions of Independence and Invariance. Exploiting this equivalence, we strip out the sequential dependence issue and yield two empirical advantages. First, we develop a simple empirical strategy for the baseline Weitzman model, which delivers lower computational burden, improved performance, and reduced implementation complexity. Second, we show that the ranking representation and its associated empirical strategy extend to a broader class of sequential search settings, thereby allowing the approach to be applied to more realistic settings, such as partially observed search data or multi-stage information acquisition processes.

The implications of our findings are relevant to two primary research communities. For applied economists, this paper bridges the gap between optimal sequential search and ranking models, providing a rigorous microfoundation for incorporating granular search data into demand analysis. This perspective can be used to address a range of empirical challenges in demand estimation. For example, existing studies \citep[e.g.,][]{goeree2008limited} often model demand under limited consideration as a two-stage process involving consideration set formation and purchase, whereas these two stages can be directly unified into a single ranking problem under the sequential search setting. The identification problem arising from zero market shares can be alleviated by exploiting non-zero search observations. Moreover, search sequences can serve as auxiliary data that enriches the identification of product complementarities and substitutabilities.

For marketing researchers, in response to the call in \cite{honka2024consumer}, this paper provides a robust and flexible toolkit for analyzing consumer behavior in data-rich environments. As online platforms generate increasingly granular clickstream data, our methodology offers an efficient way to integrate such high-frequency observations into structural analysis. This not only deepens our understanding of how consumers interact with search intermediaries and retail interfaces but also provides an actionable framework for evaluating the effectiveness of marketing strategies deployed throughout the consumer search journey.

\newpage

\phantomsection\addcontentsline{toc}{section}{\refname}
\bibliography{bibliography}
\clearpage{}

\appendix

\setcounter{table}{0}
\setcounter{figure}{0}
\setcounter{page}{0}
\setcounter{section}{0}
\renewcommand{\thetable}{\thesection.\arabic{table}}
\renewcommand{\thefigure}{\thesection.\arabic{figure}}
\renewcommand{\thepage}{Online Appendix \arabic{page}}

\pagenumbering{arabic}\renewcommand{\thepage}{A\arabic{page}}
\setcounter{footnote}{0} \setcounter{section}{0}

\thispagestyle{empty} \null

\begin{center}
{\Huge{}\vspace{5cm}
Appendix}{\Huge\par}
\par\end{center}

\vspace{5cm}

\pagebreak{}

\onehalfspacing

\begin{landscape}
\clearpage
\section{Related Empirical Studies} \label{app:literature}
\vspace{-0.2cm}
\noindent
\resizebox{\linewidth}{!}{
\begin{minipage}{1.04\linewidth} 
\input{Tables/tab_literature}
\end{minipage}
}
\end{landscape}
\clearpage

\onehalfspacing

\section{Omitted Proofs}

\subsection{Proof of Theorem \ref{theorem:main} and Lemma \ref{lem:equivalence}} \label{app:proof_optimality}

This section proves Theorem \ref{theorem:main}. The argument proceeds as follows. First, we represent the Weitzman-style sequential search process as an equivalent branching project over actions \citep{keller2003branching}. Second, under Independence and Invariance, we prove Lemma \ref{lem:equivalence}, which establishes a probabilistic equivalence between adjacent stage choices with stage-specific feasible sets and a single-stage ranking defined on the union of these sets. Third, we iterate this equivalence backward through the entire search process to obtain a ranking representation whose probability matches that of the optimal sequential search process. Finally, we invoke the optimality of the Gittins index policy for branching projects \citep{keller2003branching} to map the action ranking to a value ranking and complete the proof.

\mask{

\begin{figure}[h]
    \centering
    \resizebox{\linewidth}{!}{
        \input{TikzPics/figure_branching.tex}
    }
    \caption{The Sequential Search, The Branching Project, and The Ranking Representations}
    \label{fig:branching}
\end{figure}
}

A risk-neutral agent faces a multi-stage action-selection problem over an infinite horizon and maximizes the total discounted payoff. Each action yields an instantaneous payoff, which may be a reward or a cost. At the initial stage, the subject observes a set of feasible actions and knows their true payoffs. Each feasible action is linked to a finite set of initially infeasible first-order child actions. Child actions may themselves have further child actions, generating higher-order descendants of the original feasible action. Every infeasible action has exactly one parent, so the underlying linkage forms an out-tree, and each action is the root of a subtree consisting of all its descendants. The subject has only partial knowledge about the payoffs of infeasible actions but knows the out-tree structure \textit{ex ante}. 

Within the feasible set, each action can be selected at most once. When an action is selected, its instantaneous payoff is realized, the action is removed from the feasible set, and its first-order children become feasible in the next stage with their payoffs fully revealed. At the same time, the subject may receive deterministic or stochastic signals and update beliefs about other actions. Independence requires that any such signals concern only actions within the selected action’s subtree. Invariance requires that the feasibility status and instantaneous payoffs are unaffected by any other external factors. Under these assumptions, the set of actions and payoff primitives can be treated as fixed throughout the process, which makes alternatives across stages comparable.

\mask{
Figure \ref{fig:branching} (left and center) illustrates the mapping from sequential search to a branching project. The left panel depicts the process as a decision tree, where at each stage the consumer decides whether to stop searching and which product to inspect or purchase. This process is equivalently recast as a branching project with five initially feasible inspections and five infeasible purchases, where each purchase is the first-order child of its corresponding inspection. An inspection yields a cost, while a purchase yields the corresponding purchase value.\footnote{Since a purchase is a terminal action that renders all other actions infeasible, this seems to violate the Independence assumption. To resolve this, we recast each terminal action with an immediate payoff $u$ as an equivalent infinite absorbing subtree, where each action has a single child with the same payoff $(1 - \beta)u$, with $\beta$ being the discount factor.}
}

We now prove Lemma \ref{lem:equivalence}. The lemma states that in a branching project, when two consecutive stages have different choice sets, the joint probability of selecting an optimal action in the first stage and producing a full ranking in the second stage equals the probability of a ranking defined on the union of the two stage specific sets that is consistent with these events. The proof follows \citet{luce1959individual}. Let $P_\mathcal{A}(x)$ denote the probability of selecting element $x$ from the choice set $\mathcal{A}$, and let $P_\mathcal{A}(\mathcal{B})$ denote the probability that the selected element lies in $\mathcal{B} \subset \mathcal{A}$. Let $R_\mathcal{A}(\rho)$ denote the probability of a full ranking $\rho$ over all alternatives in $\mathcal{A}$.

\begin{enumerate}
    \item \textit{The Choice Axioms} \cite[p. 6]{luce1959individual}: Let $\mathcal{A}$ be a finite choice set. For every $\mathcal{A} \subset \mathcal{B}$, $P_\mathcal{A}$ is defined. 
    \begin{itemize}
        \item [(i)] If $P_{\{x,y\}}(x) \not= 0,1$ for all $x, y \in \mathcal{B}$, then for $\mathcal{C} \subset \mathcal{A} \subset \mathcal{B}$, $P_{\mathcal{B}}(\mathcal{C}) = P_{\mathcal{A}}(\mathcal{C}) \cdot P_{\mathcal{B}}({\mathcal{A}})$; 
        \item [(ii)] If $P_{\{x,y\}}(x) = 0$ for some $x, y \in {\mathcal{B}}$, then for every ${\mathcal{A}} \subset {\mathcal{B}}$, $P_{\mathcal{B}}(\mathcal{A}) = P_{\mathcal{B} - \{x\}}(\mathcal{A} - \{x\})$. 
    \end{itemize}
    \item \textit{The Ranking Postulates} \cite[p. 72]{luce1959individual}: The alternatives are ranked by sequentially deciding the alternative that is superior to the remaining alternatives. It leads to the following ranking postulate: 
    \begin{itemize}
        \item[(i)] $R_{\{x,y\}}(x \succ y) = P_{\{x,y\}}(x)$; 
        \item[(ii)] $R_{\mathcal{A}}(x \succ \rho) = P_\mathcal{A}(x)R_{\mathcal{A}/\{x\}}(\rho)$. 
    \end{itemize}
\end{enumerate}

Notably, Ranking Postulate (ii) implies that the selection of the top alternative from $\mathcal{A}$ is independent of the relative ranking among the remaining alternatives. Equivalently,
\begin{align*}
   \text{(ii)'} \quad P_\mathcal{A}(x) = P_\mathcal{A}(x \mid  \rho ), \text{where } \rho \text{ is a ranking defined on } \mathcal{B} \setminus \{x\} \text{ for all } \mathcal{B} \supseteq \mathcal{A}. 
\end{align*}
We allow the ranking to be defined over any proper superset $\mathcal{B} \setminus \{x\}$ of $\mathcal{A} \setminus \{x\}$ because any ranking $\rho$ over $\mathcal{A} \setminus \{x\}$ can be interpreted as the sum of probabilities over all rankings of $\mathcal{B} \setminus \{x\}$ in which the elements in $\mathcal{A} \setminus \{x\}$ are ranked according to $\rho$, and the remaining elements are ranked arbitrarily.\footnote{This is a generalized result of Theorem 9 in \citet[p. 72]{luce1959individual}. } Since these additional elements do not appear in the choice set $\mathcal{A}$, they are irrelevant to the selection probabilities within $\mathcal{A}$.

We now prove Equation \eqref{eq:lemma}. Let $\rho_1$ be a full ranking on $\mathcal{A}_2 \setminus \{a_1\}$, written as $\rho_1 = \{a_2 \succ a_3 \succ \cdots\}$, and let $\rho_\ell$ denote the tail ranking of $\rho_1$ by removing its first $\ell-1$ elements. We show:
\begin{align*}
    P_{\mathcal{A}_1}(a_1 \mid \rho_1) \cdot & R_{\mathcal{A}_2 / \{a_1\}}(\rho_1) = R_{\mathcal{A}_2}(a_1 \succ \rho_1) + \sum_{\ell=2}^{N-1} R_{\mathcal{A}_2} (a_2 \succ \cdots \succ a_{\ell} \succ a_1 \succ \rho_{\ell})
\end{align*}

Case 1: $a_2\in \mathcal{A}_1$, indicating that $a_2$ is feasible in the first stage. When $P_{\mathcal{A}_1}(a_1\mid \rho_1) = 0$, the identity holds trivailly. When $P_{\mathcal{A}_1}(a_1\mid \rho_1) > 0$, we have:  
\begin{align*}
    P_{\mathcal{A}_1}(a_1\mid \rho_1) \cdot R_{{\mathcal{A}_2}/\{a_1\}}(\rho_1) & = P_{\mathcal{A}_1}(a_1) \cdot R_{\mathcal{A}_2/\{a_1\}}(\rho_1) \\
    & = \frac{P_{\mathcal{A}_2}(a_1)}{P_{\mathcal{A}_2} (\mathcal{A}_1)} \cdot R_{\mathcal{A}_2/\{a_1\}}(\rho_1) \\
    & = P_{\mathcal{A}_2} (a_1) \cdot R_{\mathcal{A}_2/\{a_1\}}(\rho_1) \\
    & = R_{\mathcal{A}_2}(a_1 \succ \rho_1)
\end{align*}
The first equality uses the Ranking Postulate (ii)'. The second equality applies the Choice Axiom (ii) and the fact that $a_2 \in \mathcal{A}_1$. The third equality applies the Ranking Postulate (ii). 

Case 2: $a_2 \notin \mathcal{A}_1$. Here $a_2$ is revealed only after selecting $a_1$, so the relative order between $a_1$ and $a_2$ is not observed directly. Assume $P{\{a_1,a_2\}}(a_1) \not= 0, 1$, we have: 
\begin{align*}
    P_{\mathcal{A}_1}(a_1 \mid \rho_1) \cdot R_{\mathcal{A}_2/\{a_1\}}(\rho_1) & = \frac{P_{\mathcal{A}_2} (a_1)}{P_{\mathcal{A}_2} (\mathcal{A}_1)} \cdot R_{\mathcal{A}_2/\{a_1\}}(\rho_1) \\
    & = \frac{1}{P_{\mathcal{A}_2} (\mathcal{A}_1)} \cdot R_{\mathcal{A}_2}(a_1 \succ \rho_1) \\
    & = R_{\mathcal{A}_2}(a_1 \succ \rho_1) + \frac{P_{\mathcal{A}_2} ({\mathcal{A}_2}/{\mathcal{A}_1})}{P_{\mathcal{A}_2}(\mathcal{A}_1)} \cdot R_{\mathcal{A}_2}(a_1 \succ \rho_1) \\
    & = R_{\mathcal{A}_2}(a_1 \succ \rho_1) + \frac{P_{\mathcal{A}_2} (a_2)}{P_{\mathcal{A}_2}(\mathcal{A}_1)} P_{\mathcal{A}_2} (a_1) P_{{\mathcal{A}_2}/\{a_1\}} (a_2) R_{{\mathcal{A}_2}/\{a_1, a_2\}}(\rho_2) \\
    & = R_{\mathcal{A}_2}(a_1 \succ \rho_1) + P_{\mathcal{A}_2} (a_2) \cdot \underline{P_{\mathcal{A}_1}(a_1) R_{{\mathcal{A}_2}/\{a_1, a_2\}}(\rho_2)}
\end{align*}
The first equality follows from the Choice Axiom (i). The second equality applies the Ranking Postulate (ii). The third equality combines the Choice Axiom (ii) and the fact that $a_2$ is selected in the second step, which means $P_{\mathcal{A}_2/\mathcal{A}_1}(a_2) = P_{\mathcal{A}_2/a_1}(a_2) = 1$. The fourth equality again applies the Ranking Postulate (ii). The last equality follows $P_{\mathcal{A}_2/a_1}(a_2) = 1$ and the Choice Axiom (i). 

Notice that the underlined part corresponds to the left-hand side of the equation with $\{a_2\}$ removed from $\rho_1$ and the choice set $\mathcal{A}_2$. We can therefore continue the derivation: 
\begin{align*}
    & \ R_{\mathcal{A}_2}(a_1 \succ \rho_1) + P_{\mathcal{A}_2} (a_2) \cdot P_{\mathcal{A}_1} (a_1) R_{\mathcal{A}_2/\{a_1, a_2\}}(\rho_2) \\
    = & \ R_{\mathcal{A}_2}(a_1 \succ \rho_1) + P_{\mathcal{A}_2} (a_2) \cdot (R_{\mathcal{A}_2/\{a_2\}}(a_1 \succ \rho_2) + P_{\mathcal{A}_2/\{a_2\}} (a_3) \cdot P_{\mathcal{A}_1}(a_1) R_{\mathcal{A}_2/\{a_1, a_2, a_3\}}(\rho_3))  \\
    = & \ R_{\mathcal{A}_2}(a_1 \succ \rho_1) + R_{\mathcal{A}_2}(a_2 \succ a_1 \succ \rho_2) + P_{\mathcal{A}_2}(a_2) P_{\mathcal{A}_2/\{a_2\}} (a_3) \cdot \underline{P_{\mathcal{A}_1} (a_1) R_{\mathcal{A}_2/\{a_1, a_2, a_3\}}(\rho_3))}
\end{align*}

We can further decompose the underlined segment until identifying the first action in $\rho_1$ that belongs to $\mathcal{A}_1$, denoted by $a_N$. Then we have $P_{\mathcal{A}_1} (a_1) R_{\mathcal{A}_2/\{a_1, a_2 \cdots, a_{N-1}\}}(a_N \succ \rho_N) = R_{\mathcal{A}_2/\{a_2, \cdots, a_{N-1}\}}(a_1 \succ a_N \succ \rho_{N}) = R_{\mathcal{A}_2/\{a_2, \cdots, a_{N-1}\}}(a_1 \succ \rho_{N-1})$. Taking it back into the derivation, we obtain: 
\begin{align*}
    & P_{\mathcal{A}_1}(a_1 \mid \rho_1) R_{\mathcal{A}_2/\{a_1\}}(\rho_1) \\
    = & \ R_{\mathcal{A}_2}(a_1 \succ \rho_1) + R_{\mathcal{A}_2}(a_2 \succ a_1 \succ \rho_2) + R_{\mathcal{A}_2}(a_2 \succ a_3 \succ a_1 \succ \rho_3) + \cdots \\
    & \ + P_{\mathcal{A}_2} (a_2) P_{\mathcal{A}_2/\{a_3\}} (a_3) \cdots P_{\mathcal{A}_2/\{a_2, \cdots, a_{N-2}\}} (a_{N-1}) \cdot R_{\mathcal{A}_2/\{a_2, \cdots, a_{N-1}\}}(a_1 \succ \rho_{N-1}) \\
    = & \ R_{\mathcal{A}_2}(a_1 \succ \rho_1) + \sum_{\ell=2}^{N-1} R_{\mathcal{A}_2} (a_2 \succ \cdots \succ a_{\ell} \succ a_1 \succ \rho_{\ell})
\end{align*}
with the second equality following from the Ranking Postulate (ii). Hence, the proof of Lemma \ref{lem:equivalence} is complete.

We next extend the local equivalence in Lemma \ref{lem:equivalence} to the entire branching project associated with the Weitzman-style sequential search process. Suppose the consumer purchases product $h$ in the final step. Then all actions feasible but unselected in the final step must rank below ``purchase $h$'' in any ranking consistent with the observed terminal choice. Denote a ranking of these unselected actions by $\rho_0$, and consider the ranking event $\{\text{purchase } h \succ \rho_0\}$.

If $h < J$, then the action ``purchase $h$'' is already feasible in the penultimate step. Lemma \ref{lem:equivalence} implies that the joint probability of the last two step choices equals the probability of rankings on the union of the two step choice sets that satisfy $\{\text{inspect } J \succ \text{purchase } h \succ \rho_0\}$. Iterating this argument backward yields a class of rankings on the union of choice sets across all steps that satisfy $\{\text{inspect } 1 \succ \cdots \succ \text{inspect } J \succ \text{purchase } h \succ \rho_0\}$, whose probability equals that of the corresponding optimal sequential search process.

If $h = J$, then the final step choice implies two admissible local ranking configurations over the last two step choice sets, namely $\{\text{inspect } J \succ \text{purchase } J \succ \rho_0\}$ and $\{\text{purchase } J \succ \text{inspect } J \succ \rho_0\}$. The first configuration is covered by the argument above. For the second configuration, backward expansion generates partial rankings in which ``purchase $J$'' is ordered above all unselected actions, while the inspection actions satisfy $\{\text{inspect } 1 \succ \cdots \succ \text{inspect } J \}$. Combining the two configurations yields the ranking class $\{\text{inspect } 1 \succ \cdots \succ \text{inspect } J \succ \rho_0, \text{purchase } J \succ \rho_0\}$, whose probability equals that of the optimal sequential search process.

Hence, Theorem \ref{theorem:main} is proved. 

\mask{
Figure \ref{fig:branching} (center and right) illustrates the resulting ranking representation. The red arrows in the right panel indicate ranking relations; the blue circles denote selected actions, and the green circles denote feasible but unselected actions at the last stage. The purchase ranks above all unselected actions, and earlier rankings imply the successive relations among selected inspections encoded by the branching project.
}
~\\
\textit{Remark}: In principle, the optimality of the Gittins index policy established by \citet{keller2003branching} can be invoked at any stage of the proof to map the relations among actions to the relations among Gittins indices. However, in the proof of Theorem \ref{theorem:main}, we deliberately refrain from making this correspondence in order to emphasize that the probabilistic equivalence between the optimal sequential search process and the partial ranking of actions does not rely on any quantitative relationship. This probabilistic equivalence rests solely on Lemma \ref{lem:equivalence}, which depends only on the Independence and Invariance assumptions. Although both the optimality of the Gittins index policy and the validity of Lemma \ref{lem:equivalence} rely on these assumptions, they are logically distinct and should not be conflated. In subsequent proofs, we directly invoke the optimality of the Gittins index policy, without repeatedly distinguishing between action rankings and value rankings.

\subsection{Proof of Proposition \ref{Prop:weitzman}}\label{app:proof_weitzman}

We first prove the necessity that violating conditions in the proposition always violates the optimal rules. 
    \begin{itemize}
        \item When the Distribution Condition is violated, Optimal Continuing is violated. 
        \item When the Ranking Condition is violated, Optimal Ranking is violated. 
        \item When the Purchase Choice Condition is violated, there would be two cases. When $\exists j, s.t. z_{iJ} < u_{ij}$, Optimal Continuing is violated; Optimal Purchasing is violated when $\exists j, s.t. u_{ih} < u_{ij}$. 
        \item When the Inspection Choice Condition is violated, there would be two cases. In the case of $\exists k, s.t. z_{iJ} < z_{ik}$, it violates the Optimal Ranking. If $\exists k, s.t. u_{ih} < z_{ik}$, given Optimal Purchasing is not violated, i.e., $u_{ih} > \max_{j \leq J, j \not= h} u_{ij}$, Optimal Stopping is violated. 
    \end{itemize}
    Next, we prove sufficiency in showing that violating the optimal rules also violates conditions in the proposition. 
    \begin{itemize}
        \item When Optimal Ranking is violated. If $\exists j_1 < j_2 < J$ such that $z_{ij_1} < z_{ij_2}$, the Ranking Condition is violated; if $\exists j < J$ such that $z_{ij} < \max_{k > J} z_{ik}$, the Inspection Choice Condition is violated. 
        \item When Optimal Stopping is violated. If the Purchase Choice Condition holds, we have $u_{ih} = \max_{j \leq J} u_{ij} < \max_{k > J} z_{ik}$. Hence, $\exists k > J, s.t. z_{ik} > u_{ih}$, which violates the Inspection Choice Condition. 
        \item When Optimal Continuing is violated, it is to say that $\exists \ell < j \leq J, s.t. z_{ij} < u_{i\ell}$. With the Ranking Condition holds, we have $z_{iJ} < z_{ij} < u_{i \ell}$. If $\ell \not= h$, the Purchase Choice Condition is violated; if $\ell = h$, the Distribution Condition is violated. 
        \item When Optimal Purchasing is violated, the Purchase Choice Condition is violated. 
    \end{itemize}
    
\subsection{Proof of Lemma \ref{lem:composite}} \label{app:proof_effective}

Following the idea of defining the reservation value, we introduce a fallback option with utility $\bar{u}$. If the fallback option is not selected before $a_j$, the consumer first selects $b_j$ rather than the fallback option and then selects $a_j$. By the optimality of the Gittins index policy, these selections respectively imply $V_{b_j} > \bar{u}$ and $V_{a_j} > \bar{u}$. Therefore, when $\min\{V_{b_j}, V_{a_j}\} > \bar{u}$, both $b_j$ and $a_j$ are selected, meaning that the composite action $e_j$ is selected. Conversely, when $\min\{V_{b_j}, V_{a_j}\} < \bar{u}$, the fallback option is selected, implying that at least one of $b_j$ or $a_j$ is not selected, and thus $e_j$ is not selected. Thus, $\min\{V_{b_j}, V_{a_j}\}$ can be regarded as the effective index value of $e_j$. 

\subsection{Proof of Proposition \ref{Prop:nopath}} \label{app:proof_nopath}

The recast sequential search process obtained using Algorithm \ref{alg:backward} reduces to a single-step discrete choice (with an illustration in Figure \ref{fig:nopath_illustration}). Although we observe both the inspection and purchase of the purchased product, we treat them as a composite action. In this discrete choice, the consumer selects this composite action against purchasing any other inspected product or inspecting any other uninspected product, while all inspections of previously inspected products occur before this selection. The results then follow directly from Theorem \ref{theorem:incomplete}. 

\subsection{Proof of Theorem \ref{theorem:incomplete}}\label{app:proof_incomplete}

Since the recast sequential search process still satisfies the Independence and Invariance assumptions, the optimality of the Gittins index policy remains valid: at each step, the consumer selects the highest-valued feasible action, and her ranking of feasible actions is based on these values.

The first part of the theorem follows directly. Suppose an action is feasible at step $j$ but is not selected until step $j + t$. If its value satisfies $v_{i,j+t} > v_{ij}$, then the consumer fails to choose the highest-valued action at step $j$, contradicting the optimality of the Gittins index policy.

It remains to prove the second part of the theorem. Suppose there is an unselected action $k$ such that $v_{ik} > y_i$. Then $v_{ik}$ must exceed the value of some action selected at a forward-inaccessible step—say, step $j$, with selected action value $v_{ij} < v_{ik}$. To avoid violating the Gittins index policy, action $k$ must have been infeasible at step $j$.

Since all inspection actions are feasible from the start of the search process, if $k$ is either an inspection action or a composite action combining inspection and purchase of the same product, the Invariance assumption implies it would also have been feasible at step $j$, which contradicts the Gittins index policy. Therefore, $k$ must be a purchase action.

In this case, the infeasibility of $k$ at step $j$ implies that its corresponding product had not yet been inspected. But since all inspections are feasible from the outset, the inspection of that product must have been feasible at step $j$, again contradicting the assumption that step $j$ is forward-inaccessible.

Therefore, $v_{ik} < y_i$, completing the proof of the second part. 

\subsection{Proof of Theorem \ref{theorem:general}}\label{app:proof_general}

The necessity parallels the proof of Theorem \ref{theorem:incomplete}. First, since the sequence still satisfies the Independence and Invariance assumptions, the Gittins index policy remains optimal, and the first part of the theorem follows directly. Second, suppose there exists an unselected action $k$ such that $v_{ik} > y_{i\ell}$. Then $v_{ik}$ must exceed the value of some action selected in a forward-inaccessible step, say step $j$, at which $k$ was feasible and $v_{ij} < v_{ik}$. This contradicts the optimality of the Gittins index policy.

We now prove sufficiency. Suppose that at some step in the sequence, the consumer does not follow the Gittins index policy. That is, at a step where action $j$ is selected, there exists some other feasible action $j'$ such that $v_{ij'} > v_{ij}$. Three cases can arise:
\begin{enumerate}
    \item If this step is forward-inaccessible, then $v_{ij'}$ is never selected later. Since $v_{ij'} > v_{ij}$, it must also hold that $v_{ij'} > y_{ij'}$ because $y_{ij'} < v_{ij}$. This contradicts condition (ii) in the theorem.

    \item If this step is forward-accessible and $v_{ij'}$ is not selected later, then by the Invariance assumption, there exists a forward-inaccessible step after $j$ where some action $k$ is selected, $v_{ij'}$ is feasible, but not selected. If condition (i) holds, then $v_{ik} < v_{ij} < v_{ij'}$, which again contradicts condition (ii), as shown in the first case.

    \item If this step is forward-accessible and $v_{ij'}$ is selected later, then $v_{ij} < v_{ij'}$ implies that at least one intermediate step between the selections of $j$ and $j'$ violates condition (i).
\end{enumerate}

In all three cases, we arrive at a contradiction. Hence, Theorem \ref{theorem:general} is proven.

\subsection{Proof of Theorem \ref{theorem:DAG}}\label{app:proof_DAG}

The proof is organized into three parts: the construction and uniqueness of the induced graph $\mathfrak{G}$, the necessary and sufficient conditions for a branching project, and the role of transitive reduction in establishing a ranking representation.

The induction of the graph $\mathfrak{G}$ is grounded in the structural prerequisites of the model. The search process is identified by a discrete choice $\{(a^*_j, \mathcal{A}_j)\}_{j=1}^J$, where $a^*_j$ is the action selected from the feasible set $\mathcal{A}_j$ at stage $j$. Given the observed data, the construction of the edge set $E$ follows a deterministic rule: for each $j$, the subject's choice of $a^*_j$ over the remaining actions in $\mathcal{A}_j \setminus \{a^*_j\}$ uniquely induces a set of binary dominance relations edges $E_j = \{ (a^*_j, a): a \in \mathcal{A}_j, a \neq a^*_j \}$, and $E = \bigcup_j E_j$ defines the complete edge set. Since the prerequisites guarantee that both the feasible set $\mathcal{A}_j$ and the selected action $a^*_j$ are observable, the mapping from the search sequence to $\mathfrak{G}= (V, E)$ is injective. Hence, for any given search sequence, $\mathfrak{G}$, as well as its transitive reduction, is uniquely determined.

For the existence of a branching project representation, we first prove the necessity of the DAG condition. If a sequence admits a branching project representation, then each action is selected at most once and is permanently removed from the feasible set upon selection. Hence, no path can logically return to a previously selected action. Should a cycle exist in $\mathfrak{G}$, it would imply that action became feasible and preferred again after its removal, which violates the setting of a branching project. Thus, $\mathfrak{G}$ must be a DAG. 

We next prove the sufficiency. If $\mathfrak{G}$ is a DAG, there exists at least one topological sort for $\mathfrak{G}$. Based on this sorting, we can define a total preference order that is consistent with all directed relations in $\mathfrak{G}$. This ensures that in every discrete choice, the selected action is strictly preferred over all other feasible actions. On this basis, the out-tree structure is identified by the entry of actions into the feasible sets: for any action, its parent is the previously selected action that triggered its feasibility. This construction confirms the existence of a branching project representation. Furthermore, the requirement that $\mathfrak{G}$ be unique up to transitive equivalence is essential. Suppose two transitively inequivalent graphs $\mathfrak{G}_1$ and $\mathfrak{G}_2$ could be constructed from the same observed sequence. Th
is would imply a contradiction in the observational facts: there would exist at least one pair of nodes $(u, v)$ such that $v$ is revealed to be feasible when $u$ is selected in one graph ($\mathfrak{G}_1$), but absent in the other ($\mathfrak{G}_2$). Since the feasible sets are observable, the reachability relation, and thus the DAG up to transitive equivalence, must be unique.

Lastly, we explain the role of transitive reduction in defining the ranking representation. Specifically, if the sequence reveals that $a \to b$ and $b \to c$, the direct edge $a \to c$ constitutes redundant information that can be derived through transitivity. Given that $\mathfrak{G}$ is a unique DAG, its transitive reduction $\mathfrak{G}^R$ is also uniquely defined. This minimal structure eliminates all implied transitive edges while preserving the essential binary relations. Therefore, $\mathfrak{G}^R$ serves as the ranking representation of the sequential search process.

\section{Verification of Information Completeness} \label{app:bruteforce}

\begin{figure}[h]
    \centering
        \includegraphics[width=0.65\linewidth]{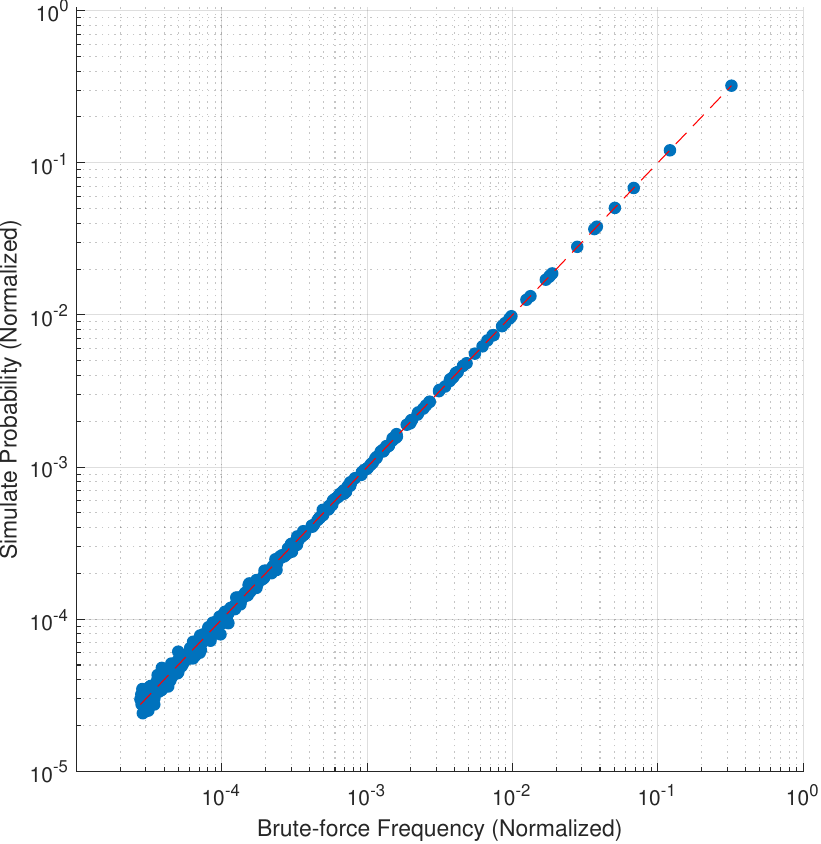} 
        \caption{Comparison between Brute-force Frequency and Simulated Probability} 
        \label{fig:bruteforce}
\end{figure}

We simulate 50 million pseudo consumers in a 5-product market who follow the optimal rules in their search processes and obtain the brute-force frequencies of 1240 distinct action sequences. For the 400 most frequent sequences, we construct their rankings and compute their simulated probabilities using the GHK method with 200000 draws across 20 runs, and take the average of the estimates. Table \ref{table:bruteforce} shows the similarity between the brute-force frequencies and the computed probabilities. The corresponding scatter plot is shown in Figure \ref{fig:bruteforce}.

\begin{table}[h]\footnotesize
\centering
\caption{Similarity between Brute-force Frequency and Simulated Probability} \label{table:bruteforce}
\input{Tables/tab_bruteforce.tex}
\end{table}

\section{Inspection Propensity in an Additive Specification} \label{app:linear_and_monotonicity}

Without loss of generality, consider the case where $\delta_{i}^z(X_{ij}^z) = \delta_{i}^u(X_{ij}^u)$. Denote $\delta_{i}^u(X_{ij}^u) + \xi^u_{ij}$ in Equation \eqref{eq:base_model_start} by $v_{ij}$, which represents the value of product $j$ that is identifiable from the observable product attributes before inspections. Denote the cumulative density function of $\varepsilon_{ij}$ by $F^\varepsilon(\cdot)$ and the probability density function by $f^\varepsilon(\cdot)$. Taking it into Equation \eqref{eq:rv}: 
\begin{align*}
    c_{ij} = \ & \int_{\varepsilon > \bar{u} - v_{ij}} (\varepsilon - (\bar{u} - v_{ij})) \ d F^\varepsilon(\varepsilon) \\
    = \ & \left(1 - F^\varepsilon\left(\frac{\bar{u} - v_{ij}}{\sigma_\varepsilon}\right) \right) \int_{\varepsilon > \bar{u} - v_{ijr}}^{\infty}(\varepsilon - (\bar{u} - v_{ij})) \frac{f(\varepsilon)}{1 - F\left(\frac{\bar{u} - v_{ij}}{\sigma_\varepsilon}\right) } d \varepsilon \\
    = \ & \left(1 - F^\varepsilon\left(\frac{\bar{u} - v_{ij}}{\sigma_\varepsilon}\right) \right) \cdot \mathrm{E}(\varepsilon - (\bar{u} - v_{ij}) \mid \varepsilon > \bar{u} - v_{ij}) \\
    = \ & \left(1 - F^\varepsilon\left(\frac{\bar{u} - v_{ij}}{\sigma_\varepsilon}\right) \right) \cdot \left[ \sigma_\varepsilon \cdot  \frac{f^\varepsilon\left(\frac{\bar{u} - v_{ij}}{\sigma_\varepsilon} \right)}{1 - F^\varepsilon\left(\frac{\bar{u} - v_{ij}}{\sigma_\varepsilon} \right)} - \sigma_\varepsilon \cdot \frac{\bar{u} - v_{ij}}{\sigma_\varepsilon}\right] \\ 
    = \ & \sigma_\varepsilon \left[ f^\varepsilon\left(\frac{\bar{u} - v_{ij}}{\sigma_\varepsilon} \right) - \frac{\bar{u} - v_{ij}}{\sigma_\varepsilon} \left(1 - F^\varepsilon\left(\frac{\bar{u} - v_{ij}}{\sigma_\varepsilon} \right)\right)\right]
\end{align*}
Dividing both sides by $\sigma_{\varepsilon}$, the above equation is only about $\frac{c_{ij}}{\sigma_\varepsilon}$ at the left-hand side and a function of $\frac{\bar{u} - v_{ij}}{\sigma_\varepsilon}$ at the right-hand side. In addition, \cite{kim2010online} point out that
\begin{align*}
    \frac{\partial \left[ f^\varepsilon\left(x \right) - x \left(1 - F^\varepsilon\left(x \right)\right)\right]}{\partial x} = -\left(1 - F^\varepsilon\left(x \right)\right) < 0
\end{align*}
which is always negative with a finite $x$. Hence, the equality implies a bijection between $\frac{c_{ij}}{\sigma_\varepsilon}$ and $\frac{\bar{u} - v_{ij}}{\sigma_\varepsilon}$, and thus between $\bar{u}$ and $c_{ij}$. This yields a unique solution for $\bar{u}$, denoted $z_{ij}$. Defining the inspection propensity as $m_\varepsilon(x) = \sigma_\varepsilon \bigl[f^\varepsilon(x) - x \bigl(1 - F^\varepsilon(x)\bigr)\bigr]^{-1} = \sigma_\varepsilon \cdot m(x)$, the reservation value in Equation \eqref{eq:base_model_end} can be written as $z_{ij} = v_{ij} + \sigma_\varepsilon \cdot m(c_{ij})$, consistent with that in Equation \eqref{eq:kim2010_end_variation}.

\section{Identification under Alternative Model Specifications} \label{app:iden_alter}
\setcounter{table}{0}

\subsection{Identification of \cite{chung2025simulated}'s Specification} \label{subapp:iden_chung}
An alternative specification is presented in \cite{chung2025simulated}, which introduces stochasticity into the reservation value by assuming a random search cost and, consequently, a heterogeneous $\xi^z_{ij}$. We consider the following specifications for our analysis in this appendix section:
\begin{align}
    u_{ij} \ & = \sum_{s=1}^3 \gamma_s x^s_{j} + \beta_{i} p_{ij} + \varepsilon_{ij} \label{eq:chung2025_start} \\
    z_{ij} \ & = \sum_{s=1}^3 \gamma_s x^s_{j} + \beta_{i} p_{ij} + \xi^z_{ij}, \quad \mbox{where } \xi^z_{ij} = m_{\varepsilon}(c_{ij})\label{eq:chung2025_end}
\end{align}

Its joint probability function can be expressed as: 
\begin{align*}
    \mathrm{Pr}(\{\mathcal{H},\mathcal{S}, \mathcal{R}, \mathcal{M}\}_i) & = \mathrm{Pr}\left( D \begin{pmatrix} \varepsilon_{ih} \\ \bm{\xi}^{z,k}_{i} \\ \bm{\xi}^{z,n}_{i} \\ \bm{\varepsilon}^{k^\prime}_{i}\end{pmatrix} < - D \begin{pmatrix} X_{ih} \beta_i \\ \bm{X}_i^{k} \beta_i \\ \bm{X}_i^{n} \beta_i \\ \bm{X}_i^{{k^\prime}} \beta_i \end{pmatrix} \right)
\end{align*}

The joint probability incorporates two sources of stochasticity: $\varepsilon_{ij}$ and $\xi^z_{ij} = m_\varepsilon(c_{ij})$. Following \cite{kim2010online}, we assume that $\varepsilon_{ij}$ follows a normal distribution, ensuring that $m_\varepsilon(\cdot)$ remains a homogeneous function of degree one. Hence, the model simplifies to:
\begin{align*}
    & \tilde{u}_{ij} = x_{ij} (\beta_i / \sigma_\varepsilon) + (\sigma_\varepsilon / \sigma_\varepsilon ) \cdot (\varepsilon_{ij} / \sigma_\varepsilon)\\
    & \tilde{z}_i = x_{ij} (\beta_i / \sigma_\varepsilon) + (\sigma_\varepsilon / \sigma_\varepsilon ) \cdot m\left(\frac{c_{ij}/\sigma_\varepsilon}{\sigma_\varepsilon/\sigma_\varepsilon}\right)
\end{align*}

Unlike \cite{kim2010online}'s specification, $c_{ij}$ is set as a random variable with an estimable distribution. To ensure stable identification under scale transformations, the distribution of $c_{ij}$ must be closed under scaling: if $c_{ij} \sim T(a_1, a_2, ..., a_K)$, then for any positive scalar $n$, it must hold that $n \cdot c_{ij} \sim T(n a_1, n a_2, ..., n a_K)$.\footnote{The exponential distribution is a standard example of a scale-closed family. \cite{chung2025simulated} found it to yield more stable estimates than the log-normal distribution, which lacks this property. Uniform distributions with parameterized bounds also satisfy scale closure.} When this condition holds, it suffices to normalize $\sigma_\varepsilon$ for model identification. Otherwise, $\sigma_\varepsilon$ and the parameters of $c_{ij}$ must be jointly estimated, which increases complexity and typically requires additional assumptions.

\begin{table}[h]\footnotesize
\centering
\caption{Monte Carlo Simulation Results under \cite{chung2025simulated}'s Specification} \label{table:iden_Chung}
\input{Tables/tab_iden_Chung}
\end{table}

We illustrate this issue in Table \ref{table:iden_Chung} by estimating two simulated datasets under alternative specifications. The first follows Equations \eqref{eq:chung2025_start}-\eqref{eq:chung2025_end}, while the second replaces the cost distribution by specifying $c_{ij}$ as log-normally distributed with known variance. We set $\sigma_\varepsilon = 1$ and estimate the model under a range of preset values $\hat{\sigma}_\varepsilon$. Across specifications, the log-likelihood values and linear preference parameters are nearly identical and scale-invariant, indicating that these parameters are not identified without normalization. When $\hat{\sigma}_\varepsilon = \sigma_\varepsilon$, the search cost is accurately recovered in both cases. However, as reported in parentheses, the implied cost parameter behaves differently across distributions: under the exponential specification, the estimate of $1/\lambda$ remains proportional to $\hat{\sigma}_\varepsilon$, whereas under the log-normal specification, the estimate of $c_0$ varies non-proportionally. Thus, the scale property of the search cost identification depends on its distributional specification. 

\subsection{Identification of Heterogeneous Preferences} \label{subapp:hete_pref}

In Appendix \ref{subapp:iden_chung}, we extend the Monte Carlo simulations in Section \ref{subsec:estimation} to incorporate heterogeneity in consumers' price preference. While the identification of this randomness appears robust (as shown in Table \ref{table:iden_Chung}), this performance is largely contingent on \cite{chung2025simulated}'s specification, where the stochasticity governing reservation values and purchase values is determined by two entirely independent sets of random terms. In contrast, under the specification of \cite{kim2010online}, the presence of $\zeta_{ij}$ induces an inherent correlation between the reservation and purchase values for any inspected product, thereby complicating empirical identification. 

\begin{table}[h]\footnotesize
\centering
\caption{Monte Carlo Simulation Results with Preference Heterogeneity} \label{table:iden_hete}
\input{Tables/tab_iden_hete}
\end{table}

\begin{figure}[h]
    \centering
        \includegraphics[width=0.65\linewidth]{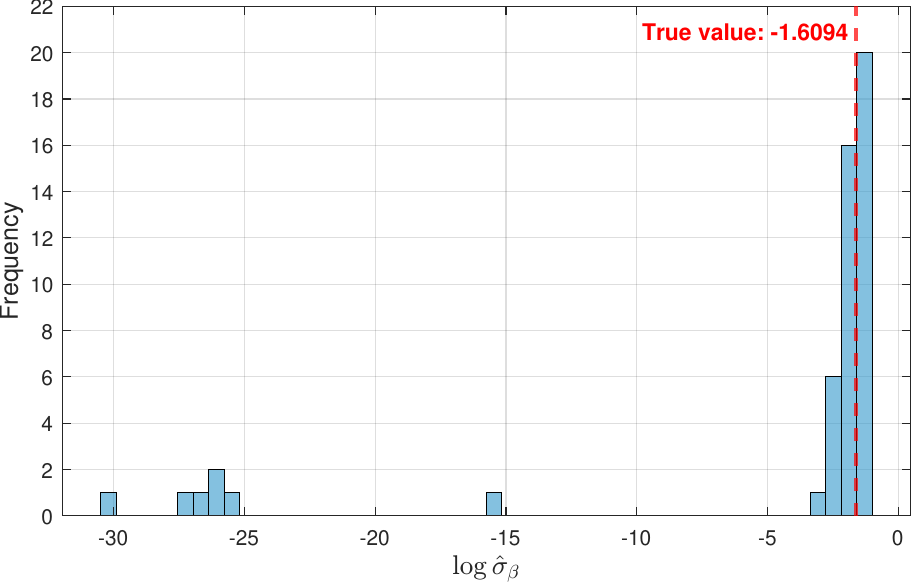}
        \caption{Empirical Distribution of Monte Carlo Simulation Results for $\log \sigma_\beta$} \label{fig:counterfactual_1}
\end{figure}

Table \ref{table:iden_hete} presents the Monte Carlo estimation results for the \cite{kim2010online}'s specification with the inclusion of price preference heterogeneity, where each consumer's price preference $\beta_i$ is assumed to follow the independent and identical distribution $\mathcal{N}(\bar\beta, \sigma_\beta^2)$. The empirical distribution of the estimates for $\log \sigma_\beta$ across 50 simulation runs is illustrated in Figure \ref{fig:counterfactual_1}. 

As shown in the figure, while most estimates cluster around the true value, a subset of simulations yields results that deviate substantially, with the estimated preference heterogeneity collapsing toward zero. This suggests that, in the absence of additional information, identifying $\sigma_\beta^2$ solely from distributional assumptions about $\zeta_{ij}$ and $\beta_i$ via price variation is feasible but highly fragile. A common empirical remedy for this fragility is to leverage panel-level action sequence data \citep{morozov2021estimation}.

\section{Transformation Algorithm for Partial Observability} \label{app:transformation}

We present an algorithm that enables researchers to construct a new sequential search process from an original one with partially observed action sequences. The algorithm requires knowledge of the market choice set and the underlying parent-child relationships among actions.

\vspace{0.2cm}

\begin{algorithm}[H] 
\footnotesize
\onehalfspacing
\SetAlgoLined
\caption{Backward Transformation for Partially Observed Branching Search} \label{alg:backward}
\SetKwInOut{Input}{Input}
\SetKwInOut{Output}{Result}
\Input{Observed sequence $\{a_j\}_{j=1}^J$, alternative set of final selection actions $\mathcal{K}$.}
\Output{A sequence of discrete choices $\{(a^*_j, \mathcal{K}_j)\}_{j=1}^J$ with dominance relations $\succ$.}
\Begin{
    \tcp{Stage 1: Initialization at Step $J$}
    $a^*_J = a_J$; $\mathcal{K}_J = \{a \in \mathcal{K} \mid a \text{ is not confirmed infeasible in Step } J\}$\;
    \ForEach{$\tilde{a} \in \mathcal{K}_J \setminus \{a^*_J\}$}{
        \While{feasibility of $\tilde{a}$ is unknown in Step $J$}{
            $\tilde{a} \leftarrow \text{composite of } \tilde{a} \text{ and its parent action}$\;
        }
        $a^*_J \succ \tilde{a}$; \quad Update $\tilde{a}$ in $\mathcal{K}_J$\; 
    }

    \tcp{Stage 2: Backward induction from Step $J-1$ to $1$}
    \For{$j = J-1$ \KwTo $1$}{
        $a^*_j = a_j$; $\mathcal{K}_j = \emptyset$\;
        \If{$a_j$ is a parent action of $a_{j+1}$}{
            $a^*_j \leftarrow \text{composite of } a_{j+1} \text{ and } a^*_j$\;
        }
        \ForEach{$\tilde{a} \in \mathcal{K}_{j+1} \setminus (\{a_j\} \cup \text{descendants of } a_j)$}{
            \If{$\tilde{a}$ is confirmed infeasible in Step $j$}{
                $\tilde{a}^0 \leftarrow \text{parent of } \tilde{a} \text{ in Step } j+1$\; 
                \While{$\tilde{a}^0$ is not confirmed feasible in Step $j$}{
                    $\tilde{a}^0 \leftarrow \text{composite of } \tilde{a}^0 \text{ and its parent action}$\;
                }
                $\tilde{a}^0 \succ a^*_{j+1}$; \quad $\tilde{a} = \tilde{a}^0$\;
            }
            \While{feasibility of $\tilde{a}$ is unknown in Step $j$}{
                $\tilde{a} \leftarrow \text{composite of } \tilde{a} \text{ and its parent action}$\;
            }
            $a^*_j \succ \tilde{a}$; \quad $\mathcal{K}_j = \mathcal{K}_j \cup \{\tilde{a}\}$\; 
        }
        $\mathcal{K}_j = \mathcal{K}_j \cup \{a^*_j\}$\;
    }

    \tcp{Stage 3: Additional dominance relation for the initial state}
    \ForEach{$\tilde{a} \in \mathcal{K}_1$}{
        \If{$\tilde{a}$ is confirmed infeasible initially}{
            $\tilde{a}^0 \leftarrow \text{parent of } \tilde{a} \text{ in Step } 1$\; 
            \While{$\tilde{a}^0$ is not confirmed feasible initially}{
                $\tilde{a}^0 \leftarrow \text{composite of } \tilde{a}^0 \text{ and its parent action}$\;
            }
            $\tilde{a}^0 \succ a^*_1$\;
        }
    }
}
\end{algorithm}

\section{Rank-based GHK Simulator under Partial Observability} \label{app:incomplete_likelihood}
\setcounter{figure}{0}

In this section, we show how to transform a partially observed action sequence into a branching project of a new sequential search process, establish the ranking representation following Theorem \ref{theorem:incomplete}, and implement the rank-based GHK Simulator under the following five scenarios: 
\begin{enumerate}
    \item Only the final purchase is observed; 
    \item Only the set of inspected products and the final purchase are observed; 
    \item Only the inspection actions and their order are observed; 
    \item Only the first inspection and the final purchase are observed; 
    \item Only the search process over a subset of products is observed. 
\end{enumerate}

We adopt the specification defined by Equations \eqref{eq:spec_est_start} to \eqref{eq:spec_est_end}, imposing the normalization $\sigma_\zeta = 1$ and excluding the outside option. To maintain clarity, we continue to use the 5-product example in Figure \ref{fig:OSRrepresentation} as the illustrative example, where the consumer's fully observed action sequence is given by the market $\mathcal{M}_i = \{1, 2, 3, 4, 5\}$, the set of inspected products $\mathcal{S}_i = \{1, 2, 3, 4\}$, the observed search order $\mathcal{R}_i = \{1 \succ 2 \succ 3 \succ 4\}$, and the purchase singleton $\mathcal{H}_i = \{3\}$. 

\subsection{Scenario 1}
In this case, the branching project and ranking representations are shown in Figure \ref{fig:single_branching_partial}. The probability function can be established based on Proposition \ref{Prop:ept}: 
\begin{align*}
    \Pr(\{\mathcal{H}, \mathcal{S}, \mathcal{R}, \mathcal{M}\}_i) = \int_{\mathbf{z}_i, \mathbf{u}_i^{insp}} \mathbb{I}(& w_{i3} > w_{i1},\ w_{i3} > w_{i2},\ w_{i3} > w_{i4}, w_{i3} > w_{i5} )  \ d\mathcal{F}(\mathbf{z}_i, \mathbf{u}_i^{insp}) 
\end{align*}
where $w_{ij} \equiv \min\{u_{ij}, z_{ij}\}$. The rank-based GHK simulator can be implemented using the following procedure: 
\begin{enumerate}
    \item If preferences are heterogeneous, make draws to determine $\beta_i^d$ for each individual $i$.
    \item Draw $\zeta_{i3}$ and $\varepsilon_{i3}$ randomly to determine $z_{i3}^d$, $u_{i3}^d$, and $w_{i3}^d$.
    \item Draw $\zeta_{ij}$ for $j \in \{1, 2, 4, 5\}$ randomly to determine $z_{ij}^d$.
    \item For $j \in \{1, 2, 4, 5\}$, compute $p_{ij}^d = \Pr(u_{ij} < w_{i3}^d \mid z_{ij}^d)$ if $z_{ij}^d > w_{i3}^d$, and assign $p_{ij}^d = 1$ otherwise. Compute the joint probability of the draw $p_i^d = \prod_{j \neq 3} p_{ij}^d$. 
    \item Define the likelihood contribution of the draw as $L_i^d = p_i^d$ and average across $D$ draws to obtain the simulated likelihood $\hat{L}_i = \frac{1}{D} \sum_{d=1}^D L_i^d$.
\end{enumerate}

\subsection{Scenario 2}
In this case, the observed selected actions are the inspection and purchase of the purchased product. Following Algorithm \ref{alg:backward}, the two actions should be combined as a composite action, and the branching project and ranking representations should be as in Figure \ref{fig:nopath_illustration}. 

\begin{figure}[h]
    \centering
    \begin{minipage}{\linewidth}
        {\centering
        \input{TikzPics/figure_nopath.tex}
        \par}
        \caption{Branching Project and Ranking Representations of Scenario 2}
        \label{fig:nopath_illustration}
        \vspace{5pt}
        {\footnotesize \emph{Notes:} The figure presents the branching project and ranking representations corresponding to the search process in Figure \ref{fig:OSRrepresentation} when only the inspected set and the final purchase are observed. In the left panel, the observed action $\{E_3, I_3\}$ is highlighted in blue, and actions selected prior to $E_3$ are shown in gray; in each step, the actions selected before the step are marked by dashed circles, and feasible but unselected actions appear in green. In the right panel, selected actions are marked in blue and unselected actions in green, with red arrows indicating dominance relations.\par}
    \end{minipage}
\end{figure}

Following Theorem \ref{theorem:incomplete}, the value of the preventive action is $y_i = w_{i3}$, as selecting $E_3$ and $P_3$ are both forward-inaccessible steps, yet $P_3$ is absorbed by $E_3$. The probability function can then be constructed from Proposition \ref{Prop:nopath}:
\begin{align*}
    \Pr(\{\mathcal{H}, \mathcal{S}, \mathcal{R}, \mathcal{M}\}_i) = \int_{\mathbf{z}_i, \mathbf{u}_i^{insp}} \mathbb{I}(& z_{i1} > w_{i3},\ z_{i2} > w_{i3},\ z_{i4} > w_{i3}, \\
    & w_{i3} > z_{i5}, \ w_{i3} > u_{i1}, \ w_{i3} > u_{i2}, \ w_{i3} > u_{i4} )  \ d\mathcal{F}(\mathbf{z}_i, \mathbf{u}_i^{insp}) 
\end{align*}

The rank-based GHK simulator implementation procedure is as follows: 
\begin{enumerate}
    \item If preferences are heterogeneous, make draws to determine $\beta_i^d$ for each individual $i$.
    \item Draw $\zeta_{i3}$ and $\varepsilon_{i3}$ randomly to determine $z_{i3}^d$, $u_{i3}^d$, and $w_{i3}^d$.
    \item For $j \in \{1, 2, 4\}$, draw $\zeta_{ij}$ conditional on $z_{ij} > w_{i3}^d$ to determine $z_{ij}^d$, and compute $p_{i,a}^d = \prod_{j \in \{1, 2, 4\}} \Pr(z_{ij} > w_{i3}^d)$.
    \item Compute $p_{i,b}^d = \left( \prod_{j \in \{1, 2, 4\}} \Pr(u_{ij} < w_{i3}^d \mid z_{ij}^d) \right) \cdot \Pr(z_{i5} < w_{i3}^d)$.
    \item Compute the likelihood contribution $L_i^d = p_{i,a}^d \cdot p_{i,b}^d$ and average across $D$ draws to obtain the simulated likelihood $\hat{L}_i = \frac{1}{D} \sum_{d=1}^D L_i^d$.
\end{enumerate}

\subsection{Scenario 3}

In this case, the observable selected actions include the ordered inspections of Products 1, 2, 3, and 4. According to Algorithm \ref{alg:backward}, we construct a sequential discrete choice process as illustrated in the upper panel of Figure \ref{fig:nopurchase_illustration}. The first four steps of this process form a branching project and are converted into a ranking representation, whereas the final step is completed by constructing a selected alternative to recover the missing final selection in which $I_5$ is not selected, thereby yielding the lower panel of Figure \ref{fig:nopurchase_illustration}. The equivalent value of the constructed final selection must ensure that $I_5$ is not selected.

\begin{figure}[h]
    \centering
    \begin{minipage}{\linewidth}
        {\centering
        \input{TikzPics/figure_nopurchase.tex}
        \par}
        \caption{Branching Project and Ranking Representations of Scenario 3}
        \label{fig:nopurchase_illustration}
        \vspace{5pt}
        {\footnotesize \emph{Notes:} The figure presents the branching project and ranking representations corresponding to the search process in Figure \ref{fig:OSRrepresentation} when only the inspections and their order are observed. In the upper panel, the observed actions $\{I_1, I_3, I_3, I_4\}$ are highlighted in blue, and the action $\{I_5\}$ that is not selected at the final step is shown in gray; in each step, the actions selected before the step are marked by dashed circles, and feasible but unselected actions appear in green. In the lower panel, selected actions are marked in blue, actions that may be selected at the final step are shown in green, and the known unselected actions appear in gray. Red arrows indicate dominance relations, while the blue rectangle indicates a maximum among the blocked actions.\par}
    \end{minipage}
\end{figure}

Accordingly, the equivalent value of the final selection is given by:
\begin{align*}
    e_i = \min\{z_{i4}, \max\{u_{i1}, u_{i2}, u_{i3}, u_{i4} \} \}
\end{align*}

Here, $u_{i1}$, $u_{i2}$, and $u_{i3}$ are in the choice set at the step of inspecting product $4$, while purchasing $4$ is not. Hence, the corresponding joint probability expression is as follows: 
\begin{align*}
    \Pr(\{\mathcal{H}, \mathcal{S}, \mathcal{R}, \mathcal{M}\}_i) = \int_{\mathbf{z}_i, \mathbf{u}_i^{insp}} \mathbb{I}(& z_{i1} > z_{i2}, z_{i2} > z_{i3}, z_{i3} > z_{i4}, \ z_{i4} > \max\{u_{i1}, u_{i2}, u_{i3}\}, \\
    & \min\{z_{i4}, \max\{u_{i1}, u_{i2}, u_{i3}, u_{i4} \} \} > z_{i5})  \ d\mathcal{F}(\mathbf{z}_i, \mathbf{u}_i^{insp}) 
\end{align*}

The rank-based GHK simulator implementation procedure is as follows: 
\begin{enumerate}
    \item If preferences are heterogeneous, make draws to determine $\beta_i^d$ for each individual $i$.
    \item Draw $\zeta_{i4}$ randomly to determine the baseline latent variable $z_{i4}^d$.
    \item For $j = 3, 2, 1$, sequentially draw $\zeta_{ij}$ conditional on $z_{ij} > z_{i,j+1}^d$ to determine $z_{ij}^d$, and compute $p_{i,a}^d = \prod_{j=1}^3 \Pr(z_{ij} > z_{i,j+1}^d)$.
    \item For $j \in \{1, 2, 3\}$, draw $\varepsilon_{ij}$ conditional on $u_{ij} < z_{i4}^d$ to determine $u_{ij}^d$, and compute $p_{i,b}^d = \prod_{j=1}^3 \Pr(u_{ij} < z_{i4}^d \mid z_{ij}^d)$.
    \item Draw $\varepsilon_{i4}$ randomly to obtain $u_{i4}^d$, calculate the threshold $e_i^d = \min \{ z_{i4}^d, \max_{k \in \{1, \dots, 4\}} \{u_{ik}^d\} \}$, and compute $p_{i,c}^d = \Pr(z_{i5} < e_i^d)$.
    \item Define the likelihood contribution $L_i^d = p_{i,a}^d \cdot p_{i,b}^d \cdot p_{i,c}^d$ and average across $D$ draws to obtain the simulated likelihood $\hat{L}_i = \frac{1}{D} \sum_{d=1}^D L_i^d$.
\end{enumerate}

\subsection{Scenario 4}

In this case, the observable actions include the inspection and purchase of the first inspected product, as well as the inspection and purchase of the ultimately purchased product. According to Algorithm \ref{alg:backward} and Theorem \ref{theorem:incomplete}, we construct the branching project and ranking representations as shown in Figure \ref{fig:firstview_illustration}.

\begin{figure}[h]
    \centering
    \begin{minipage}{\linewidth}
        {\centering
        \input{TikzPics/figure_firstview.tex}
        \par}
        \caption{Branching Project and Ranking Representations of Scenario 4}
        \label{fig:firstview_illustration}
        \vspace{5pt}
        {\footnotesize \emph{Notes:} The figure presents the branching project and ranking representations corresponding to the search process in Figure \ref{fig:OSRrepresentation} when only the first inspected product and the final purchase are observed (when they correspond to different products). In the left panel, the observed actions $\{I_1, I_3, P_3\}$ are highlighted in blue; in each step, the actions selected before the step are marked by dashed circles, and feasible but unselected actions appear in green. In the right panel, selected actions are marked in blue and unselected actions in green, with red arrows indicating dominance relations.\par}
    \end{minipage}
\end{figure}

As the purchased product is observed, the value of the preventive action is given by $y_i = w_{i3} = \min\{u_{i3}, z_{i3}\}$. For our example, the joint probability expression is as follows: 
\begin{align*}
    \Pr(\{\mathcal{H}, \mathcal{S}, \mathcal{R}, \mathcal{M}\}_i) = \int_{\mathbf{z}_i, \mathbf{u}_i^{insp}} \mathbb{I}(& z_{i1} > z_{i3}, \\
    & w_{i3} > u_{i1}, \ w_{i3} > w_{i2}, \ w_{i3} > w_{i4}, \ w_{i3} > w_{i5} )  \ d\mathcal{F}(\mathbf{z}_i, \mathbf{u}_i^{insp}) 
\end{align*}

If the purchased product is the first inspected product ($A$), the joint probability is similar to that in Scenario 1: 
\begin{align*}
    \Pr(\{\mathcal{H}, \mathcal{S}, \mathcal{R}, \mathcal{M}\}_i) = \int_{\mathbf{z}_i, \mathbf{u}_i^{insp}} \mathbb{I}(w_{i1} > u_{i2}, \ w_{i1} > w_{i3}, \ w_{i1} > w_{i4}, \ w_{i1} > w_{i5} ) \ d\mathcal{F}(\mathbf{z}_i, \mathbf{u}_i^{insp}) 
\end{align*}

The rank-based GHK simulator implementation procedure is as follows: 
\begin{enumerate}
    \item Draw the heterogeneities in preferences and determine $\beta_i^d$. 
    \item Draw $\zeta_{i3}$ and $\varepsilon_{i3}$ randomly to determine $z_{i3}^d$, $u_{i3}^d$ and $w_{i3}^d$. 
    \item Draw $\zeta_{i1}$ conditional on $z_{i1} > z_{i3}^d$ to determine $z_{i1}^d$. 
    Compute $p_{i,a}^d = \Pr(z_{i1} > z_{i3}^d)$. 
    \item Draw $\zeta_{i2}, \zeta_{i4}$ and $\zeta_{i5}$ randomly to obtain $z_{i2}^d, z_{i4}^d$ and $z_{i5}^d$. 
    \item For those draws with $z_{iB}^d > w_{iC}^d$, compute $p_{i2}^d = \Pr(u_{i2} < w_{i3}^d\mid z_{i2}^d)$. For the other draws, assign $p_{i2}^d = 1$. Do this also for products $4$ and $5$. Compute $p_{i,b}^d = p_{i2}^d \cdot p_{i4}^d \cdot p_{i5}^d$. 
    \item Compute the likelihood contribution of the draw $L_i^d = p_{i,a}^d \cdot p_{i,b}^d$. Take the average across draws to obtain the simulated likelihood. 
\end{enumerate}
For the case where the purchased product is the first inspected product, the implementation procedure follows Scenario 1. 

\subsection{Scenario 5}

We consider the case in Figure \ref{fig:OSRrepresentation} where information on products $3$ and $4$ is missing; that is, it is unknown whether the two products were inspected. In addition, the purchased product is also unobserved. In this case, the observable feasible actions include inspecting Products $1$ and $2$. Following Algorithm \ref{alg:backward}, we construct the sequential discrete choice process shown in the left panel of Figure \ref{fig:twocensored_illustration}. The first two steps form a branching project and are converted into a ranking representation, whereas the final step is completed by constructing a selected alternative to recover the missing final choice information, thereby yielding the right panel of Figure \ref{fig:twocensored_illustration}. During this process, the forward-inaccessible selection includes only $I_2$, yet the equivalent value of the constructed final selection must be larger than $z_5$ to prevent $I_5$ from being selected.

\begin{figure}[h]
    \centering
    \begin{minipage}{\linewidth}
        {\centering
        \input{TikzPics/figure_partial.tex}
        \par}
        \caption{Branching Project and Ranking Representations of Scenario 5}
        \label{fig:twocensored_illustration}
        \vspace{5pt}
        {\footnotesize \emph{Notes:} The figure presents the branching project and ranking representations corresponding to the search process in Figure \ref{fig:OSRrepresentation} when only a sequence of actions across part of the products in the market is observed. In the left panel, the observed actions $\{I_1, I_2\}$ are highlighted in blue, and the action $\{I_5\}$ that is not selected at the final step is shown in gray; in each step, the actions selected before the step are marked by dashed circles, and feasible but unselected actions appear in green. In the right panel, selected actions are marked in blue, actions that may be selected at the final step are shown in green, and unselected actions appear in gray. The red arrows indicate dominance relations, while the blue rectangle indicates a maximum among the blocked actions.\par}
    \end{minipage}
\end{figure}

Accordingly, the equivalent value of the final selection is given by:
\begin{align*}
    e_i = \min\{z_{i2}, \max\{u_{i1}, w_{i3}, w_{i4}, u_{i2} \} \}
\end{align*}
Here, $u_{i1}$, $w_{i3}$, and $w_{i4}$ are in the choice set at the step of inspecting $2$, while purchasing $2$ is not. Hence, the corresponding joint probability expression is as follows: 
\begin{align*}
    \Pr(\{\mathcal{H}, \mathcal{S}, \mathcal{R}, \mathcal{M}\}_i) = \int_{\mathbf{z}_i, \mathbf{u}_i^{insp}} \mathbb{I}(& z_{i1} > z_{i2}, \ z_{i2} > \max\{u_{i1}, w_{i3}, w_{i4}\}, \\
    & \min\{z_{i2}, \max\{u_{i1}, u_{i2}, w_{i3}, w_{i4} \} \} > z_{i5})  \ d\mathcal{F}(\mathbf{z}_i, \mathbf{u}_i^{insp}) 
\end{align*}

The rank-based GHK simulator implementation procedure is as follows: 
\begin{enumerate}
    \item If preferences are heterogeneous, make draws to determine $\beta_i^d$ for each individual $i$.
    \item Draw $\zeta_{ij}$ and $\varepsilon_{ij}$ for $j \in \{3, 4\}$ to determine $w_{ij}^d$, and let $w_i^d = \max \{w_{i3}^d, w_{i4}^d\}$. $w_i^d$ is defined as the maximum value across all products associated with their composite actions.
    \item Draw $\zeta_{i2}$ conditional on $z_{i2} > w_i^d$ to determine $z_{i2}^d$, and compute $p_{i,a}^d = \Pr(z_{i2} > w_i^d)$.
    \item Draw $\zeta_{i1}$ conditional on $z_{i1} > z_{i2}^d$ to determine $z_{i1}^d$, and compute $p_{i,b}^d = \Pr(z_{i1} > z_{i2}^d)$.
    \item Draw $\varepsilon_{i1}$ conditional on $u_{i1} < z_{i2}^d$ to determine $u_{i1}^d$, and compute $p_{i,c}^d = \Pr(u_{i1} < z_{i2}^d \mid z_{i1}^d)$.
    \item Draw $\varepsilon_{i2}$ randomly to obtain $u_{i2}^d$, and calculate the threshold $e_i^d = \min\{z_{i2}^d, \max\{u_{i1}^d, u_{i2}^d, w_i^d\} \}$.
    \item Compute $p_{i,d}^d = \Pr(z_{i5} < e_i^d)$.
    \item Define the likelihood contribution $L_i^d = p_{i,a}^d \cdot p_{i,b}^d \cdot p_{i,c}^d \cdot p_{i,d}^d$ and average across $D$ draws to obtain the simulated likelihood $\hat{L}_i = \frac{1}{D} \sum_{d=1}^D L_i^d$.
\end{enumerate}

\mask{
\setcounter{table}{0}

We examine the robustness and performance of the estimation method under both complete and incomplete search information using an alternative specification. Consider the specification following Equations \eqref{eq:chung2025_start} to \eqref{eq:chung2025_end}: 
\begin{align*}
\begin{split}
    & u_{ij} = \sum_{s=1}^3 \gamma_t x^s_{j} + \beta_{i} p_{ij} + \varepsilon_{ij}, \quad \mbox{where } \beta_{i} \sim N(\bar{\beta},\sigma_\beta^2) ;  \\
    & z_{ij} = \sum_{s=1}^3 \gamma_t x^s_{j} + \beta_{i} p_{ij} + m_\varepsilon(c_{ij}), \quad \mbox{where } c_{ij} \sim \mathrm{Exp}(\lambda). 
\end{split}
\end{align*}

Unlike the specification in \cite{kim2010online}, the reservation value and the purchase value of the same product are now conditionally independent. 

The Monte Carlo simulation results under the alternative specification are presented in Table \ref{table:firstview_app}. As in Table \ref{table:firstview}, we compare performances using complete search information and four different scenarios of limited search data: (1) final purchase only, (2) inspection set and purchase, (3) first inspected product and purchase, and (4) search data over six out of eight products. 

\begin{table}[h]\footnotesize
\centering
\caption{Monte Carlo Simulation Results of Table \ref{table:firstview} with Alternative Specifications} \label{table:firstview_app}
\input{Tables/tab_firstview_app}
\end{table}

It can be observed that, unlike the earlier case discussed in the main text, the estimator here exhibits a more pronounced sensitivity to the amount of available data. Specifically, Columns (2), (3), and (4) all deliver reasonably good estimation performance, with Column (2) yielding results closest to those based on complete search information. Columns (3) and (4) perform slightly worse, while Column (1), which uses only purchase data, performs very poorly.

This difference arises from the structural distinction between the two specifications. In the \cite{kim2010online} specification, the distribution of the effective value is defined as the minimum of two independent components: a Gaussian draw and a minimum between a fixed constant and a normally distributed random variable. In contrast, the \cite{chung2025simulated} specification defines the effective value as the minimum of a Gaussian draw and an exponentially distributed random variable. The latter is considerably more difficult to identify under limited information. Moreover, in the latter setting, difficulties in estimating the search cost tend to propagate into estimating linear preference parameters due to the lack of a scale normalization.

These results highlight that the consequences of incomplete search information vary considerably across model specifications. Missing ordering information has relatively minor effects, while the absence of inspection indicators can severely impair identification, leading to poor estimation outcomes. 


Suppose there exists an external mechanism that provides off-search ranking information about unselected actions. In that case, such information can be used in the estimation process in the same way as search data, without requiring modifications on Optimal Search Rules.\footnote{It is important that the information takes the form of a top-down ranking to satisfy the ranking postulate, which ensures that ranking probabilities can be aggregated. For example, among three products A, B, and C, if the information includes $u_{iA} > u_{iB}$ and $u_{iA} > u_{iC}$, it can be directly incorporated into estimation. However, if only $u_{iA} > u_{iC}$ and $u_{iB} > u_{iC}$ are observed, a top-down ranking must first be reconstructed before the associated probability can be computed.}

Consider a scenario in which a consumer, upon inspecting at least two products, is required to compare the second inspected product with the first and indicate her preference between the two. If this marking is costless and the information is fully truthful, it provides additional ranking data for consumers who eventually purchase the third or later inspected products.

We explore the contribution of the extra information to estimation with Monte Carlo simulations. For this purpose, we generate a smaller search dataset with 2,000 consumers following the specification in Equation \eqref{eq:spec_est_start} - \eqref{eq:spec_est_end} without the outside option. We consider two scenarios: incorporating and not incorporating the information from marking in model estimation. Incorporating this information requires slightly adjusting the rank-based GHK implementation in Step 4. The detailed process is as follows:

\begin{enumerate}
    \item Draw preference heterogeneity to obtain $\beta_i^d$. Draw $\zeta_{iJ}$ to determine $z_{iJ}^d$. 
    \item Sequentially draw $\zeta_{i,J-1}, \zeta_{i,J-2}, \cdots, \zeta_{i2}$ to determine $z_{i,J-1}^d, z_{i,J-2}^d, \cdots, z_{i2}^d$ conditional on $z_{ij} > z_{i,j+1}^d$. Compute $p_{i1}^d = \prod_{1 \leq j \leq J-1} \mathrm{Pr}(z_{ij} > z_{i,j+1} \mid z_{i,j+1} = z_{i,j+1}^d)$. 
    \item If $h \not= J$, draw $\varepsilon_{ih}$ conditional on $u_{ih} < z^d_{iJ}$ and compute $p_{i2}^d = \mathrm{Pr}(u_{ih} < z_{iJ} \mid z_{iJ} = z_{iJ}^d)$; \\ if $h = J$, draw $\varepsilon_{ih}$ randomly and assign $p_{i2}^d = 1$. Determine $y_i^d = \min\{u_{ih}^d, z_{iJ}^d\}$. 
    \item[4.1.] If the purchased product is inspected in the third place or later, denote $s = 1 \mbox{ or } 2$ as the preferred between the first two inspected products. Draw $\varepsilon_{is}$ to determine $u_{is}^d$ conditional on $u_{is} < y_{i}^d$ and compute $p_{i3}^d = \mathrm{Pr}(u_{is} < y_{i} \mid y_{i} = y_{i}^d) \cdot \mathrm{Pr}(u_{i,3-s} < u_{is} \mid u_{is} = u_{is}^d) \cdot \prod_{3 \leq j \leq J, j \not= h} \mathrm{Pr}(u_{ij} < y_i \mid y_i = y_i^d)$. \\ 
    Otherwise, compute $p_{i3}^d = \prod_{1 \leq j \leq J, j \not= h} \mathrm{Pr}(u_{ij} < y_i \mid y_i = y_i^d)$. 
    \item[4.2.] Compute $p_{i4}^d = \prod_{J < k \leq |\mathcal{M}_i|} \mathrm{Pr}(z_{ik} < y_i \mid y_i = y_i^d)$. 
    \item[5.] Compute the likelihood contribution of each draw $L_i^d = p_{i1}^d \cdot p_{i2}^d \cdot p_{i3}^d \cdot p_{i4}^d$. Take the average across draws to obtain the simulated likelihood. 
\end{enumerate}

The Monte Carlo simulation results are shown in Table \ref{table:additional}. 

\begin{table}[h]\footnotesize
\centering
\caption{Monte Carlo Simulation Results with Additional Ranking Information} \label{table:additional}
\input{Tables/tab_additional.tex}
\end{table}

\red{Based on our value settings, this marking information provides substantial supplemental ranking information for 9\% of the simulated consumers. The sampling processes of the two methods are identical in the first three steps, and the same random seed is used for the corresponding simulations. Consequently, the performance of the two methods is nearly identical when the true parameter values are used as starting points for estimation. However, when the starting values deviate significantly from the true values, incorporating the additional ranking information greatly improves the model's estimation performance. This enhancement reduces the likelihood of falling into local optima and yields average preference estimates closer to the true values. These results demonstrate that incorporating supplemental ranking information can significantly improve the estimation performance of the rank-based GHK simulator with only a minor increase in computational burden, offering practical value for estimating models using real-world datasets. }
}

\section{Empirical Application on Expedia Data} \label{app:expedia}

E-commerce platforms provide an important empirical setting for studying consumers’ search and purchase processes. In this setting, clickstream data can be viewed as records of consumers’ sequential actions and studied using sequential search models. Existing research uses such data to identify consumer preferences and search costs and to conduct counterfactual product policy evaluations. These studies typically rely on the premise that consumers’ decisions at both the search and purchase stages are governed by the same decision environment and preference parameters.

However, in actual markets, this premise is often suspected. It is common for consumers to exhibit preference tradeoffs and choice criteria at the purchase stage that differ from those in the search stage. For example, a consumer may complete a sequence of inspections and identify a target hotel on platforms such as Expedia or Booking, then use comparison sites such as Trivago to compare prices across platforms, and eventually complete the purchase on a different platform. Similarly, consumers may focus primarily on relative product comparisons during search while placing limited weight on budget constraints, but reconsider whether the selected product is worth the price at the final purchase stage. Relative to the information search stage, the purchase stage is often affected by additional factors, potentially leading to decision-making deviations.

This potential deviation poses a direct challenge for empirical research based on single-platform data. Because sequential search typically ends with a purchase as the final outcome after information collection, studying purchase decisions in isolation from the search process is difficult, as it is hard to determine whether purchase outcomes arise from changes in purchase-stage decision criteria or from additional private information formed during search. On the other hand, simply assuming that all purchase decisions are generated under the same conditions as search decisions may introduce systematic bias and further distort parameter estimates and policy analysis based on purchase data.

The empirical method proposed in this paper, based on partially observed data, provides an indirect way to identify this issue. The core idea is to compare estimates of preference parameters based on different subsets of search information and test whether systematic differences exist across them. We apply this method to data from the online hotel booking platform Expedia.

The dataset is a publicly available dataset\footnote{Data source: \hyperlink{www.kaggle.com/c/expedia-personalized-sort/data}{www.kaggle.com/c/expedia-personalized-sort/data
}.}. For each recorded consumer, we observe the set of hotels shown based on the search query, the set of hotels clicked, and the hotel booked, but not the order of clicks. The data is split into an Expedia-ranking sample (70\%), in which hotels are displayed according to the platform’s ranking algorithm, and a random-ranking sample (30\%), in which hotels are displayed in a random order. The resulting data structure corresponds to Scenario 3 in Section \ref{subsec:extension_incomplete}, with clicks interpreted as inspections and bookings as purchases.

Our strategy proceeds as follows. We first estimate the model under the three information scenarios introduced in Section \ref{subsec:extension_incomplete}: Scenario 1, using only purchase data; Scenario 2, using only the set of searched alternatives; and Scenario 3, combining both sources. We then compare the estimated preference parameters across these specifications. Consistency across specifications indicates no evidence of changes in decision-making criteria between search and purchase, whereas systematic differences suggest potential inconsistency.

The data processing follows \cite{ursu2018power} to retain comparability. Specifically, we retain the four largest destinations among the search queries and estimate the model separately for each. Price construction takes the average across consumers, excluding those with abnormal prices (nightly rates below 10 dollars or above 1000 dollars) and potentially mismeasured (total payment exceeding 130 percent of the nightly rate multiplied by the length of stay).

The model specification also follows \cite{ursu2018power}. Observable hotel attributes on the search results page include price, star rating, review score, chain affiliation, location score, and promotion indicators. In addition, unlike Section \ref{subsec:extension_incomplete}, we allow consumers to choose an outside option rather than booking any hotel. Search costs include a common intercept across hotels and a component that varies with a hotel’s position in the search results for a given query. The pre- and post-inspection taste shocks are both assumed to be distributed i.i.d. standard normal. 

A limitation of the dataset is that only consumers who have inspected at least once are recorded, implying that the outside option is feasible only after at least one inspection. This affects the informational content of the initial inspection. We therefore follow the literature and impose two additional assumptions: the feasibility of the outside option does not affect search decisions, and the outside option is homogeneous across all initial inspections. We adjust the rank-based GHK simulators for the three scenarios in Section \ref{subsec:extension_incomplete} to accommodate the presence of the outside option under these assumptions. The model setup is therefore as follows: 
\begin{align*}
    u_{ij} \ & = \underbrace{X_{ij} \gamma}_{v_{ij}} + \zeta_{ij} + \varepsilon_{ij} \label{eq:kim2010_start} \\ 
    z_{ij} \ & = X_{ij} \gamma + \zeta_{ij} + m_\varepsilon(\eta_0 + position \cdot \eta_1) \\
    u_{i}^{outside} \ & = \gamma^{outside} + \zeta_{i0} \\
    \zeta_{ij} \ & \sim N(0, 1), \ \forall j = 0, 1, 2, \cdots \\ 
    \varepsilon_{ij} \ & \sim N(0, 1), \ \forall j = 1, 2, \cdots 
\end{align*}
where $X_{ij}$ denotes the observable hotel attributes, $\xi_{ij}$ are the pre-inspection taste shocks and $\varepsilon_{ij}$ are the post-inspection taste shocks. $\gamma$ denotes the preference parameter vector for all observable hotel attributes, and $\gamma^{outside}$ is the constant for the outside option value. $\eta_0$ and $\eta_1$ are the two parameters for the search cost. 

It is important to note that we do not assume that the inspection order coincides with the order in which search results are displayed. This assumption is adopted in \cite{ursu2018power} to compensate for the lack of fully observed inspection orders, which is not required in our approach.

\begin{figure}[h]
    \centering
        \includegraphics[width=0.95\linewidth]{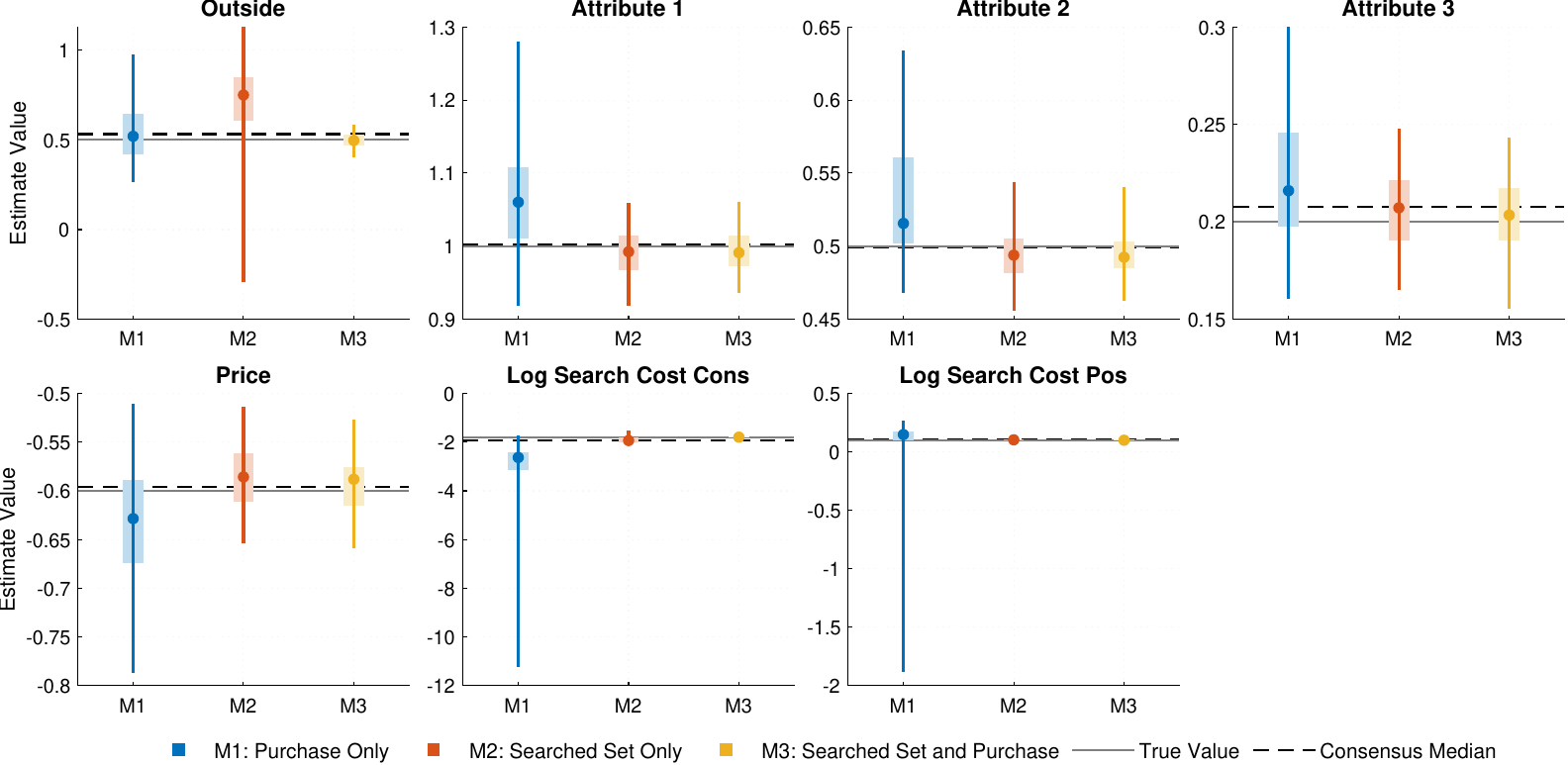}
        \caption{Distribution of Estimates with Partial Observability and Outside Option} \label{fig:est_comparison_montecarlo}
\end{figure}

We first conduct a Monte Carlo simulation to examine the performance of the estimators with the following specification: 
\begin{align*}
    v_{ij} = \sum_{s=1}^3 \gamma_s \cdot x^s_{j} + \beta \cdot p_{ij}
\end{align*}
where $p_{ij}$ denotes consumer-product-specific prices and three $x^s_{j}$ denote three binary attributes, with a total of eight products covering all possible combinations of these attributes. The results are reported in Table \ref{table:incomplete_outside} and the distribution of estimates is shown in Figure \ref{fig:est_comparison_montecarlo}. 
\begin{table}[h]\footnotesize
\centering
\caption{Monte Carlo Simulation Results with Partial Observability and Outside Option} \label{table:incomplete_outside}
\input{Tables/tab_incomplete_outside.tex}
\end{table}

The Monte Carlo results reveal several implications. First, when only purchase data are used, the estimation of search cost parameters performs poorly. As discussed in the main text, this is driven by the lack of direct comparisons between search actions and purchase decisions in the data, which creates identification difficulties unless specific assumptions are imposed on the distribution of post-inspection taste shocks. Second, when only search-set data are used, the estimation of the outside value is relatively weak because there are no observed selections in which the outside option is selected, leading to an effective market share of zero and, consequently, identification challenges. 

The comparison further shows that using only the searched set data does not affect the estimation of preference parameters for other attributes, whereas using only purchase data results in a significant deterioration in the quality of these estimates. Therefore, in the empirical application to the Expedia data, we compare only the preference parameter estimates from the latter two estimators. 

\begin{table}[h]\footnotesize
\centering
\caption{Expedia Data Application: Comparison of Estimates across Destinations} \label{table:estimator_comparison}
\input{Tables/tab_est_compare_expedia.tex}
\end{table}

Our results are reported in Table \ref{table:estimator_comparison}. Column (1) presents estimates using only information on products searched by consumers, while Column (2) reports joint estimates using both search information and final purchase outcomes. The first two columns are based on averages across 30 estimation runs, with 2000 draws per consumer in each run. Column (3) reports the estimates from \cite{ursu2018power}, obtained using the Kernel-smoothed Frequency Simulator with 50 runs, under the assumption that consumers search according to the displayed order of search results.

We first compare Columns (2) and (3) to assess the robustness of our estimates. Table \ref{table:estimator_comparison} shows that, except for Review Scores in Destinations 2 and 3, and Location Scores and Promotion in Destination 3, our estimates are broadly consistent with the benchmark study in both coefficient signs and significance. Given the use of a different simulation-based estimation method, this consistency suggests that the estimates in Column (2) provide a reliable basis for subsequent comparison.

\begin{landscape}
    \begin{figure}[htbp]
        \centering
        \includegraphics[width=0.95\linewidth]{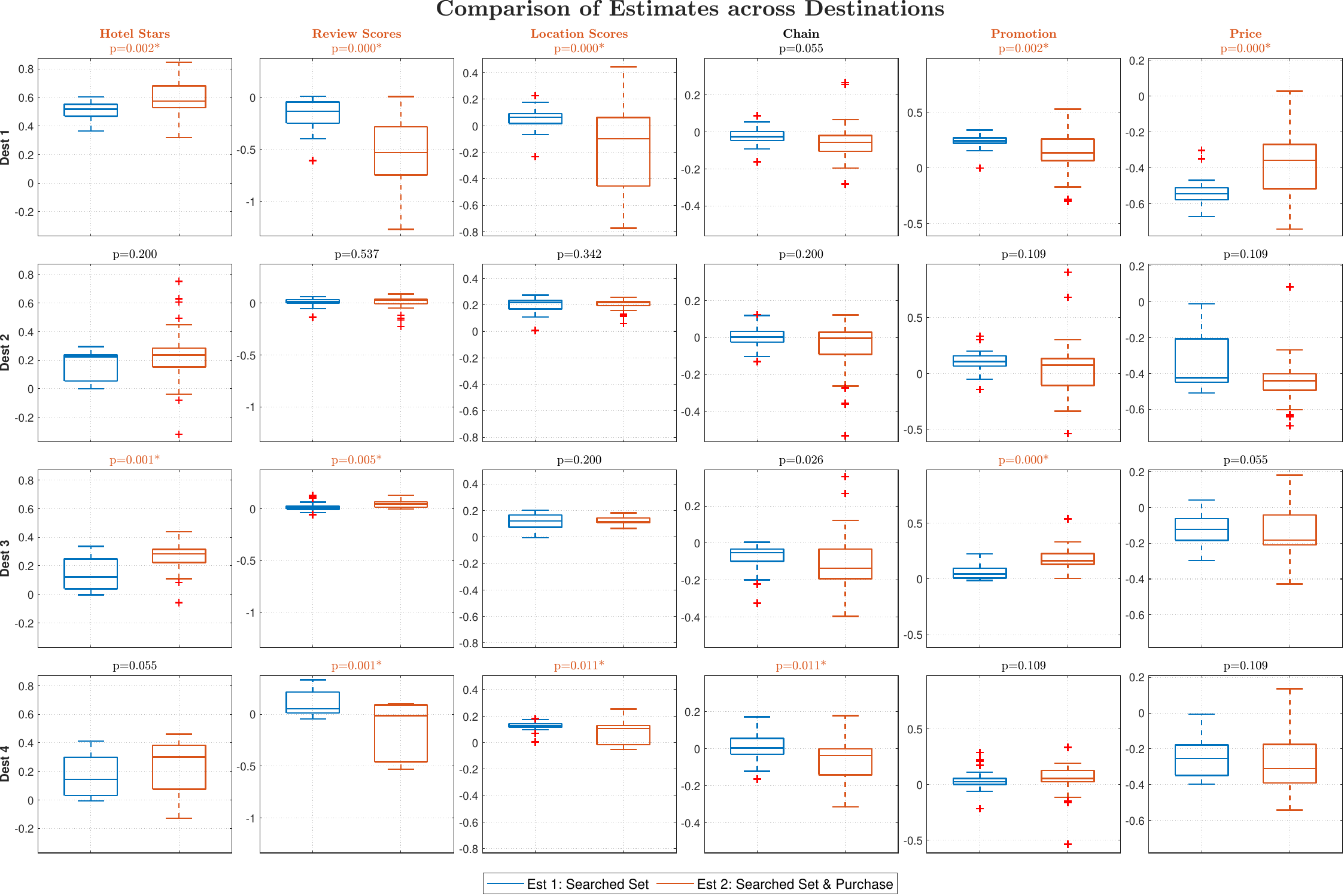}
        \caption{Expedia Data Application: Comparison of Estimates across Destinations}
        \label{fig:estimator_comparison}
        \vspace{2mm} 
    \raggedright 
    \footnotesize 
    \textit{Notes:} This figure compares the distribution of coefficient estimates for six hotel attributes across four different destinations. Blue boxes represent estimates based solely on the searched set, while orange boxes include both search and purchase information. The horizontal line within each box represents the median, and the whiskers denote the 1.5 interquartile range. The $p$-values indicate the results of a t-test for the difference in means between the two estimators, with asterisks (*) denoting significance at the 5\% level.
    \end{figure}
\end{landscape}

The comparison between Column (1), based only on searched set data, and Column (2), is further illustrated in Figure \ref{fig:estimator_comparison}, which reports the statistical differences across 30 estimation runs. The results show that, once purchase decisions are incorporated into the estimation, most parameters in Destination 1 and several in Destinations 3 and 4 shift significantly, while the remaining coefficients also exhibit varying degrees of adjustment. Specifically, consumers place greater importance on hotel star ratings at the purchase stage than during search. By contrast, the marginal importance of Review Scores, Location Score, and Chain affiliation declines in final purchase decisions. Preference estimates for Price and Promotion display heterogeneous changes across destinations. These differences suggest that treating search and purchase decisions as arising from the same revealed preference may be misspecified.

Our empirical findings indicate that search and purchase stages may reflect different decision tradeoffs. During search, attributes such as location and reviews appear to function primarily as screening signals that help consumers narrow the consideration set. At the final purchase stage, consumers place greater weight on the core attribute more directly tied to realized consumption value, such as hotel quality reflected by star ratings.

These results carry implications for hotel managers and platform design. For hotel operators, strong reviews and favorable location may increase the likelihood of entering consumers’ consideration sets, but final conversion still depends on the product's underlying value. For platforms, recommendation systems optimized on joint purchase data may prioritize products that align with final purchase preferences, but would be screened out early because of weak location or review attributes. A more effective recommendation system should therefore account for the distinct revealed preferences at different stages in the search process.

\section{Rank-based GHK Simulator for Sequential Search with Discovery} \label{app:spd_implementation}

To compute the probability, the observed action sequence is divided into segments defined by discoveries. Each segment comprises a sequence of inspections and concludes with a discovery, whereas the final segment concludes with a purchase. These segments are indexed in reverse order: the last segment is Segment 0, the second-to-last is Segment 1, and so on. Let $J(t)$ denote the index of the last product inspected before the end of Segment $t$. If no product is inspected in Segment 0, then $J(0) = J(1)$. The implementation procedure proceeds chronologically starting from Segment 0.

\begin{enumerate}
    \item Check if $J(0) > J(1)$. If not, set $p_{i,a,0}^d = p_{i,b,0}^d = 1$ and skip to Step 4. 
    \item Draw $\zeta^u_{i,J(0)}$ to determine $z_{i,J(0)}^d$ and set $p_{i,a,0}^d = 1$. 
    \item Sequentially draw $\zeta^u_{i,j}$ for $j = J(0)-1, \dots, J(1)+1$ conditional on $z_{i,j} > z_{i,j+1}^d$. Compute the associated probability $p_{i,b,0}^d = \prod_{j=J(1)+1}^{J(0)-1} \Pr(z_{i,j} > z_{i,j+1}^d)$.
\end{enumerate}
This completes the simulation for the last segment based on the ranking representation. For each subsequent segment $t \geq 1$, the procedure is as follows: 

\begin{enumerate}
\setcounter{enumi}{3}
    \item For the route $r(t)$ selected at the end of Segment $t$, draw $c_{i,r(t),t}^{dis}$ to determine the discovery value $q_{i,r(t),t}$. This value must exceed (i) the reservation values of products discovered by segment $t$ but inspected in segments $t' < t$, and (ii) the discovery values $q_{i,r',t}^d$ for all $r' \neq r(t)$. Compute $p_{i,o,t}^d$ as the probability that $q_{i,r(t),t}$ satisfies these conditions. For any route $r' \neq r(t)$ chosen in a previous segment $t' < t$, set $q_{i,r',t}^d = q_{i,r',t-1}^d$.
    \item Check if $J(t) > J(t+1)$. If not, set $p_{i,a,t}^d = p_{i,b,t}^d = 1$ and skip to Step 8.
    \item Draw $\zeta^u_{i,J(t)}$ conditional on $z_{i,J(t)} > q_{i,r(t),t}^d$ to determine $z_{i,J(t)}^d$, and compute $p_{i,a,t}^d = \Pr(z_{i,J(t)} > q_{i, r(t), t}^d \mid q_{i, r(t), t} = q_{i, r(t), t}^d)$.
    \item Sequentially draw $\zeta^u_{i,j}$ for $j = J(t)-1, \dots, J(t+1)+1$ conditional on $z_{i,j} > z_{i,j+1}^d$. Compute $p_{i,b,t}^d = \prod_{j=J(t+1)+1}^{J(t)-1} \Pr(z_{i,j} > z_{i,j+1}^d)$.
    \item Repeat Steps 4 through 7 for all segments until the final segment $T$ is reached.
\end{enumerate}
The ranking for the overall sequential search with discovery sequence is now simulated. The following steps determine the values of conditional preventive actions and the probabilities of unselected actions:

\begin{enumerate}
\setcounter{enumi}{8}
    \item If $J(t+1) < h \leq J(t)$, draw $\varepsilon_{ih}$ conditional on $u_{ih} < \min_{1 \leq t' \leq t}\{q_{i,r(t'),t'}^d\}$ to determine $u^d_{ih}$ and compute the probability $p_{i,c,t}^d = \Pr(u_{ih} < \min_{1 \leq t' \leq t}\{q_{i,r(t'),t'}^d\} \mid q_{i,r(1),1}^d, \dots, q_{i,r(t),t}^d)$. Otherwise, set $p_{i,c,t}^d = 1$.
    \item Define the values for conditional preventive actions: for $t = 0$, $y_{i0}^d = \min\{z_{i,J(0)}^d, u_{ih}^d \}$; for $t > 0$, $y_{it}^d = q_{i,r(t),t}^d$.
    \item For each route $r$, let $\bar{q}_{ir}^d$ be the drawn discovery value with the smallest $t$. If a route $r'$ was never discovered, set $\bar{q}_{i,r'}^d = + \infty$.
    \item Compute $p_{i,d}^d = \prod_{r'=1}^{N_R} \Pr(q_{i,r',0} < \min\{ y_{i0}^d, \min_{r'' \neq r'}\{\bar{q}_{i,r''}^d\} \})$ to account for the probabilities of unselected discovery actions given the values of conditional preventive actions.
    \item Compute $p_{i,e}^d = \prod_{h' \neq h, h' \leq J(t)} \Pr(u_{ih'} < \min_{{t: J(t) \geq h' } } { y_{it}^d}) $ to account for unselected purchase actions.
    \item Compute $p_{i,f}^d$ (analogous to $p_{i,5}^d$) for products that were discovered but not inspected, based on the value of the corresponding conditional preventive action in the segment following their discovery.
    \item The simulated likelihood contribution for the draw is the product $L_i^d = p_{i,d}^d \cdot p_{i,e}^d \cdot p_{i,f}^d \cdot \prod_t (p_{i,o,t}^d \cdot p_{i,a,t}^d \cdot p_{i,b,t}^d \cdot p_{i,c,t}^d)$. Average $L_i^d$ across $D$ draws to obtain the simulated likelihood $\hat{L}_i = \frac{1}{D} \sum_{d=1}^D L_i^d$.
\end{enumerate}

\section{The Ranking Representation of the Two-Stage Sequential Search Model} \label{app:ts_implementation}
\setcounter{figure}{0}

\cite{gibbard2022model} provides an example of a multi-stage sequential information acquisition process in which complete product information is gradually revealed over two stages. Unlike sequential search with discovery settings, consumers in this model can observe all products in the market at the outset, but the purchase value of each product consists of two sources of uncertainty that must be resolved sequentially before the product can be acquired.

Following the original notation, the first-stage action is referred to as browsing ($\mathcal{B}$), the second-stage action as clicking ($\mathcal{C}$), and the final purchase action as acquiring ($\mathcal{A}$). \cite{gibbard2022model} shows that, under certain conditions, a Gittins index can be computed for each types of action and used to characterize the optimal strategy for this process: regardless of the action type, the consumer selects at each step the action with the highest Gittins index value until the first time selecting an acquisition, consistent with the optimality of the Gittins index policy established in \cite{keller2003branching}.

We show that such a two-stage sequential search process can still be represented as a ranking under the assumptions of Independence and Invariance. As an example, consider a market with three products, $\mathcal{M}_i = \{1, 2, 3\}$, and suppose the consumer follows the sequence: browse product $1$, click product $1$, browse product $2$, consider product $2$, and acquire product $2$. The upper panel of Figure \ref{fig:twostage_illustration} presents the corresponding branching project representation, while the lower panel displays the ranking representation obtained from Theorem \ref{theorem:general}.

\begin{figure}[h]
    \centering
    \begin{minipage}{\linewidth}
        {\centering
        \input{TikzPics/figure_twostage.tex}
        \par}
        \caption{Branching Project and Ranking Representations of the Two-Stage Sequential Search Process}
        \label{fig:twostage_illustration}
        \vspace{5pt}
        {\footnotesize \emph{Notes:} The figure presents the branching project and ranking representations corresponding to the example of the two-stage sequential search process. In the upper panel, the observed actions $\{\mathcal{B}_1, \mathcal{C}_1, \mathcal{B}_2, \mathcal{C}_2, \mathcal{A}_2\}$ are highlighted in blue; in each step, the actions selected before the step are marked by dashed circles, and feasible but unselected actions appear in green. In the lower panel, selected actions are marked in blue, and unselected actions appear in green. The red arrows indicate dominance relations.\par}
    \end{minipage}
\end{figure}

In this example, the observed sequence contains three forward-inaccessible steps. For the two unselected actions in the search process, $\mathcal{B}_3$ and $\mathcal{A}_1$, their corresponding conditional preventive action is the lowest-ranked action among $\mathcal{B}_2$, $\mathcal{C}_2$, and $\mathcal{A}_2$, and its value equals the minimum of their Gittins indices.

\section{\mbox{Partially Observed Sequential Search with Discovery}} \label{app:spd_incomplete_implementation}
\setcounter{figure}{0}

A partially observed sequential search with discovery process remains within the scope of our approach. We focus on a concrete example based on the widely used Expedia dataset in empirical research. This dataset provides impression-level search data for each consumer, recording the ordered list of products, namely hotels, displayed on the search results page, the set of products actually inspected, and the final purchase outcome. However, the inspection order is unobserved, and discovery outcomes are observed only at the time of purchase.

Without loss of generality, we assume that the consumer initially knows only one product, and that each discovery reveals one additional product according to the display order. Because each discovery action becomes feasible only after the preceding discovery has been taken, selecting product $j$, which appears in position $j$ in the data, necessarily requires $j+1$ actions. These include purchase, with value $u_{ij}$; inspection, with value $z_{ij}$; potential discovery, with value $q_{ij}$; as well as all prior discovery actions leading to the product. For any consumer, the observed action sequence therefore contains all discovery actions preceding the purchased product, together with the discovery, inspection, and purchase of the product. We emphasize that reconstructing a branching project from this sequence remains possible by directly applying Algorithm \ref{alg:backward}, although the resulting redundancy in conditions makes the representation difficult to streamline.

\cite{greminger2024heterogeneous} introduce a simplifying assumption under which the value of discovery decreases monotonically with its position in the search display. This assumption implies that the decision to discover a product depends only on the value of that action itself, since all prior discovery actions have values no lower than that of discovery. Consequently, if the consumer is observed to discover a product, all preceding discovery actions must also be selected and therefore do not obstruct the observed discovery.

On this basis, for any product, we can restrict attention to three actions: discovering, inspecting, and purchasing the purchased product. With a slight abuse of notation, let $\mathcal{M}_i$ denote the set of all discovered products. Let $h$ denote the display order number of the purchased product. Let $\mathcal{S}_{i}^1$ be the set of products discovered and inspected before the display order $h$, and $\bar{\mathcal{S}}_{i}^1$ the set of products discovered but not inspected before the display order $h$. Similarly, let $\mathcal{S}_{i}^2$ be the set of products discovered and inspected after $h$, and $\bar{\mathcal{S}}_{i}^2$ the set of products discovered but not inspected after $h$. Applying Algorithm \ref{alg:backward}, we establish a three-step branching project and obtain the following sets of inequalities. 

The last selection is to select purchasing product $h$, therefore: 
\begin{align*}
    u_{ij} < u_{ih}, & \ \forall j \in \mathcal{S}_{i}^1 \cup \mathcal{S}_{i}^2 \\ 
    z_{ij} < u_{ih}, & \ \forall j \in \bar{\mathcal{S}}_{i}^1 \cup \bar{\mathcal{S}}_{i}^2 \\
    q_{i,|\mathcal{M}_i|+1} < u_{ih}. & 
\end{align*}

The second-to-last selection is the composite action between inspecting and purchasing product $h$, as inspecting $h$ is the parent action of purchasing $h$. The dominance relations among actions in this step, presented in their corresponding values, are as follows: 
\begin{align*}
    u_{ij} < \min\{z_{ih}, u_{ih}\}, & \ \forall j \in \mathcal{S}_{i}^1 \cup \mathcal{S}_{i}^2 \\
    z_{ij} < \min\{z_{ih}, u_{ih}\}, & \ \forall j \in \bar{\mathcal{S}}_{i}^1 \cup \bar{\mathcal{S}}_{i}^2 \\
    q_{i,|\mathcal{M}_i|+1} < \min\{z_{ih}, u_{ih}\}. &
\end{align*}

The third-to-last selection is the composite action between discovering, inspecting, and purchasing product $h$, as discovering $h$ is also the parent action of inspecting $h$. The dominance relations among actions in this step, presented in their corresponding values, are as follows: 
\begin{align*}
    \min\{q_{i,h+1}, \dots q_{ij}, z_{ij} \} = \min\{q_{ij}, z_{ij} \} > \min\{z_{ih}, u_{ih}\}, & \ \forall j \in \mathcal{S}_{i}^2 \\
    \min\{q_{i,h+1}, \dots q_{ij}\} = q_{ij} > \min\{z_{ih}, u_{ih}\}, & \ \forall j \in \bar{\mathcal{S}}_{i}^2 \\
    u_{ij} < \min\{q_{ih}, z_{ih}, u_{ih}\}, & \ \forall j \in \mathcal{S}_{i}^1 \\
    z_{ij} < \min\{q_{ih}, z_{ih}, u_{ih}\}, & \ \forall j \in \bar{\mathcal{S}}_{i}^1
\end{align*}

The additional condition following the third-to-last selection is: 
\begin{align*}
    z_{ij} > \min\{q_{ih}, z_{ih}, u_{ih}\}, & \ \forall j \in \mathcal{S}_{i}^1
\end{align*}

Remove all redundant inequalities to obtain the following: 
\begin{equation}
\begin{cases}
    q_{i,|\mathcal{M}_i|} > \min\{z_{ih}, u_{ih}\} > q_{i,|\mathcal{M}_i|+1} \\
    z_{ij} > \min\{z_{ih}, u_{ih}\}, & \forall j \in \mathcal{S}_{i}^2 \\
    u_{ij} < \min\{z_{ih}, u_{ih}\}, & \forall j \in \mathcal{S}_{i}^2 \\
    z_{ik} < \min\{z_{ih}, u_{ih}\}, & \forall k \in \bar{\mathcal{S}}_{i}^2 \\
    z_{ij} > \min\{q_{ih}, z_{ih}, u_{ih}\}, & \forall j \in \mathcal{S}_{i}^1 \\
    u_{ij} < \min\{q_{ih}, z_{ih}, u_{ih}\}, & \forall j \in \mathcal{S}_{i}^1 \\
    z_{ik} < \min\{q_{ih}, z_{ih}, u_{ih}\}, & \forall k \in \bar{\mathcal{S}}_{i}^1
\end{cases} \label{eq:greminger}
\end{equation}

These inequalities imply \cite{greminger2024heterogeneous}'s Proposition 2, which is reformulated as follows: 
\begin{itemize}
    \item \textbf{Stopping:} The consumer discovers all products up to the last one with a discovery value greater than or equal to $\tilde{w}_{ih}^2$.
    \item \textbf{Search and early discovery:} The consumer inspects all products in $\mathcal{S}_{i}^1$ and none in $\bar{\mathcal{S}}_{i}^1$.
    \item \textbf{Search and late discovery:} The consumer inspects all products in $\mathcal{S}_{i}^2$ and none in $\bar{\mathcal{S}}_{i}^2$.
    \item \textbf{Choice:} The consumer purchases product $h$ but no product in $\mathcal{S}_{i}^1$ or $\mathcal{S}_{i}^2$.
\end{itemize}

To calculate the likelihood function, because $q_{i,|\mathcal{M}_i|}$ and $q_{ih}$ are linked through a fixed function determined by parameters, \cite{greminger2024heterogeneous} simulate the distribution of $q_{ih}$ under the first condition of Equation \eqref{eq:greminger}. He then computes, within each region of the support of $q_{ih}$, the probabilities that the other six conditions in Equation \eqref{eq:greminger} are satisfied, which can be directly obtained when the reservation and purchase values contain independent stochastic components. The method is essentially a crude frequency simulator, with only the action values of product $j$ needing to be simulated. It can be combined with a GHK-style simulator if the purchase value relies on the stochastic components of the corresponding reservation value.

\section{Applicability of the Rank-Based GHK Simulator} \label{app:simulator_scope}

As established in Section \ref{theorem:DAG}, a ranking representation exists whenever the discrete choice process induced by the search sequence uniquely represents the relationships among feasible actions as a DAG. Accordingly, the probability function can be written in the value difference form shown in equation \eqref{eq:pr_jp_dec}. In addition, the general identification conditions presented in Section \ref{subsec:identification} must hold, namely that each action is associated with a conditionally independent random variable whose parametric distribution is known.

Yet, the conditions above remain insufficient to guarantee that the relevant probabilities can be computed using a GHK-style simulator, which relies on a chain decomposition of conditional probabilities: it sequentially draws values for action nodes and evaluates the conditional probabilities of subsequent nodes under truncation conditions determined by earlier draws. Hence, the simulator is inherently constrained by its sequential structure.

When simulation-based methods are used to evaluate the probability of the search process, the nodes that must be explicitly sampled in the DAG formed by feasible actions include all nodes with both positive in-degree and positive out-degree, as well as one node from each pairwise ranking relation. The values of these nodes serve as conditioning variables for ranking relations among feasible actions and must therefore be obtained through sampling. Under the assumption that the random components in action values are mutually independent, these nodes form a minimal set for simulation, while the probabilities of the remaining nodes can be computed directly from their conditional distributions given the sampled values.

Consider again the sequential-search process illustrated in Figure \ref{fig:OSRrepresentation}. Suppose the consumer cannot observe the order in which products 2 and 3 are drawn, yielding the ranking representation shown in Figure \ref{fig:ghk_sampling_routes}. The resulting graph is a DAG in which the nodes that require sampling are $z_{i2}, z_{i3}, z_{i4}$, along with the purchase value of the purchased product, $u_{i3}$. 

Under the benchmark case in which the random components across nodes are mutually independent (for example, specification in Equations \eqref{eq:chung2025_start} - \eqref{eq:chung2025_end}), the sampling order should avoid imposing truncation conditions from opposing directions on the same node, since such bidirectional constraints may eliminate the feasible support of that node and thereby block subsequent sampling. 

Figure \ref{fig:ghk_sampling_routes} presents four candidate sampling orders, three of which are implementable. The first two differ in their starting nodes and directions of expansion, yet both prioritize the four nodes with positive in-degree and out-degree, namely $z_{i2}, z_{i3}, z_{i4}$, and $u_{i3}$, allowing the conditional probabilities of remaining nodes to be calculated directly from truncated distributions. The third order corresponds to the policy-based GHK simulator. Although it also produces the correct probability, it starts with nodes with only positive in-degree, which entails a higher simulation dimension and computational cost. The fourth-order model also prioritizes these four core nodes, but the sampling order may cause the conditional support for $z_{i4}$ to collapse to the empty set under opposing restrictions, rendering effective sampling infeasible.

\begin{figure}[h]
    \centering
    \input{TikzPics/figure_route}
    \caption{The GHK Sampling Routes Example}
    \label{fig:ghk_sampling_routes}
\end{figure}

When the specification contains dependence among the random components across variables, the sampling order must respect this dependence structure. For example, if the specification follows Equations \eqref{eq:kim2010_start} to \eqref{eq:kim2010_end}, the purchase value of each product depends on the realization of the random component in its reservation value, which must therefore be sampled first. To determine whether a sampling route under such dependence can support a policy-based GHK simulator, we establish the following proposition.

\begin{proposition}
    Let $\mathfrak{G}' = (V, E)$ be a directed graph, where $V$ denotes the set of all feasible actions. The edge set $E = E_{dep} \cup E_{path}$ is composed of two classes of directed constraints: $E_{dep}$ represents the stochastic dependencies between action values, and $E_{path}$ represents a proposed sequential sampling order that does not involve truncation conditions from opposing directions at any node. The proposed order is implementable by a rank-based GHK simulator if and only if $\mathfrak{G}'$ is a DAG. 
\end{proposition}
\begin{proof}
    Necessity: Suppose $\mathfrak{G}'$ is not a DAG. Then there exists at least one directed cycle $\{v_1, v_2, \dots, v_k, v_1\}$. In the GHK sampling, the truncation interval $\mathcal{I}_{v_i}$ for any node $v_i$ is defined as a function of the realized values $x_u$ of all its predecessors $\{u \mid (u, v_i) \in E\}$, denoted as:
    $$\mathcal{I}_{v_i} = f(\{x_u\}_{(u, v_i) \in E})$$
    Due to the existence of the cycle, the interval $\mathcal{I}_{v_1}$ recursively depends on the realization $x_{v_k}$ for all $v_k$ in the cycle. This creates a circular definition in the computation, rendering it impossible to obtain a determinate numerical value for the integration limits of any node within the cycle. Thus, the sampling sequence is logically non-implementable.
    
    Sufficiency: If $\mathfrak{G}'$ is a DAG, then the graph has at least one topological order. When the GHK-style simulator executes in this order, the realized values of all subsequent nodes are determined by their predecessors. Thus, the truncation interval $I_{v_i}$ serves as a well-defined support for the conditional distribution at each step, ensuring that the truncation intervals are well defined and that the sampling sequence is logically implementable.
\end{proof}

Consider the proposed GHK sampling orders 1 and 2 in Figure \ref{fig:ghk_sampling_routes}. Their implementability is illustrated in Figure \ref{fig:ghk_sampling_routes_imple}. Order 1 induces a directed cycle among $I_3$, $I_4$, and $P_3$, whereas order 2 remains acyclic. Thus, a rank-based GHK simulator exists under order 2 but not under order 1.

\begin{figure}[h]
    \centering
    \input{TikzPics/figure_route_imple}
    \caption{The Implementability of the Sampling Routes}
    \label{fig:ghk_sampling_routes_imple}
\end{figure}

In some environments, no such sampling order exists. A practical alternative is to treat a cycle as a composite node and approximate its distribution using brute-force simulation. A detailed discussion of such hybrid approaches is beyond the scope of this paper.

\end{document}

%% file: TikzPics/tikz_setup.tex
\usetikzlibrary{graphs, shapes.geometric, arrows.meta, positioning, fit}
\tikzset{
  rect1/.style = {
    shape = rectangle,
    draw = black,
    text width = 2.8cm,
    align = center,
    minimum height = 0.8cm,
    font = \normalsize
  }
}

\tikzset{
  rect2/.style = {
    shape = rectangle,
    draw = black, 
    dashed, 
    text width = 2.8cm,
    align = center,
    minimum height = 0.8cm,
    font = \footnotesize
  }
}

\tikzset{
  rect3/.style = {
    shape = rectangle,
    draw = black, 
    dotted, 
    text width = 2.8cm,
    align = center,
    minimum height = 0.8cm,
    font = \footnotesize
  }
}

\tikzset{
  circ1/.style = {
    shape = circle,
    draw = black,
    text width = 0.6cm,
    align = center,
    minimum height = 0.5cm,
  }
}

\tikzset{
  circ2/.style = {
    shape = circle,
    draw = black,
    minimum size = 0.5cm,
    inner sep = 2pt, 
    align = center,
  }
}

\tikzset{
  circ3/.style = {
    shape = circle,
    draw = black,
    dashed,
    minimum size = 0.5cm,
    inner sep = 2pt, 
    align = center,
  }
}

%% file: TikzPics/figure_ssm.tex
\begin{tikzpicture}[scale = 1.05]
    \tikzstyle{every node}=[font=\small, scale = 0.83, fill = white]; 
    \tikzstyle{every path}=[thick]; 

    \draw (-2.2,0) node [rect1, text width = 1.8cm, fill = green!20!white](out){Start search \\ (Which one to inspect? )}; 
    \draw (0,0) node[rect1, text width = 1.4cm] (z_1){Inspect prod 1}; 
    \draw[dotted, ->] (out) -- (z_1); 
    \draw (0,1.4) node[rect1, text width = 1.4cm] (z_1_o){Inspect other prods}; 
    \draw[dotted, ->] (out) -- (z_1_o); 
    \draw (1.9,0) node[rect1, text width = 1.4cm, fill = green!20!white] (u_1){Stop search? };
    \draw[dotted, ->] (z_1) --  (u_1); 
    \draw (3.7,1.3) node[rect1, text width = 1.4cm, fill = green!20!white] (z_1_2){Which one to inspect? };
    \draw[dotted, ->] (u_1) -- (z_1_2); 
    \draw (2.2,0.8) node (){No}; 
    \draw (5.5,1.3) node[rect1, text width = 1.4cm] (z_2){Inspect prod 2}; 
    \draw[dotted, ->] (z_1_2) -- (z_2); 
    \draw (5.5,2.6) node[rect1, text width = 1.4cm] (z_2_o){Inspect other prods}; 
    \draw[dotted, ->] (z_1_2) -- (z_2_o); 
    \draw (7.3,1.3) node[rect1, text width = 1.4cm, fill = green!20!white] (u_2){Stop search? }; 
    \draw[dotted, ->] (z_2) -- (u_2); 
    \draw (9.1,2.6) node[rect1, text width = 1.4cm, fill = green!20!white] (z_2_3){Which one to inspect? };
    \draw[dotted, ->] (u_2) -- (z_2_3); 
    \draw (7.6,2.1) node (){No}; 
    \draw (10.9,2.6) node (end0){$\cdots$};
    \draw[dotted, ->] (z_2_3) -- (end0); 

    \draw (3.7,-1.3) node[rect1, text width = 1.4cm, fill = green!20!white] (buy1){Which one to buy? };
    \draw[dotted, ->] (u_1) -- (buy1); 
    \draw (2.1,-0.8) node (){Yes}; 
    \draw (5.5,-0.85) node[rect1, text width = 1.4cm] (end1_1){Buy prod 1}; 
    \draw[dotted, ->] (buy1) -- (end1_1); 
    \draw (5.6,-1.95) node[rect1, text width = 1.6cm] (end1_0){Leave (outside)}; 
    \draw[dotted, ->] (buy1) -- (end1_0); 
    \draw (9.1,0) node[rect1, text width = 1.4cm, fill = green!20!white] (buy2){Which one to buy? };
    \draw[dotted, ->] (u_2) -- (buy2); 
    \draw (7.6,0.5) node (){Yes}; 
    \draw (10.9,1.1) node[rect1, text width = 1.4cm] (end2_2){Buy prod 2}; 
    \draw[dotted, ->] (buy2) -- (end2_2); 
    \draw (10.9,0) node[rect1, text width = 1.4cm] (end2_1){Buy prod 1}; 
    \draw[dotted, ->] (buy2) -- (end2_1); 
    \draw (11.0,-1.1) node[rect1, text width = 1.6cm] (end2_0){Leave (outside)}; 
    \draw[dotted, ->] (buy2) -- (end2_0); 
    
    \end{tikzpicture}

%% file: TikzPics/figure_osr.tex
\begin{tikzpicture}[scale=0.75]
    \tikzstyle{every node}=[font=\normalsize, scale = 0.65]
    \draw (0,0) node[rect1, fill = green!10!white, text width = 1.6cm] (A_I_1){Which to inspect?}; 
    \draw (A_I_1) -- (1.6,2) node[rect1, fill = blue!20!white, text width = 1.1cm] (A_I_a_1){$I_1 (z_{i1})$}; 
    \draw (A_I_1) -- (1.6,1) node[rect1, fill = white, text width = 1.1cm] (A_I_b_1){$I_2 (z_{i2})$}; 
    \draw (A_I_1) -- (1.6,0) node[rect1, fill = white, text width = 1.1cm] (A_I_c_1){$I_3 (z_{i3})$}; 
    \draw (A_I_1) -- (1.6,-1) node[rect1, fill = white, text width = 1.1cm] (A_I_d_1){$I_4 (z_{i4})$}; 
    \draw (A_I_1) -- (1.6,-2) node[rect1, fill = white, text width = 1.1cm] (A_I_e_1){$I_5 (z_{i5})$}; 
    \draw[dashed] (2.3, -3.0) -- (2.3, 4.8); 
    \draw (1.0,-2.8) node {Step 1}; 
    
    \draw (A_I_a_1) -- (3,2) node[rect1, fill = green!10!white, text width = 1.2cm] (A_S_1){Stop?}; 
    \draw (3.85, 2.6) node {No}; 
    \draw (A_S_1) -- (4.9,2) node[rect1, fill = green!10!white, text width = 1.6cm] (A_I_2){Which to Inspect?}; 
    \draw (A_I_2) -- (6.5,3) node[rect1, fill = blue!20!white, text width = 1.1cm] (A_I_b_2){$I_2 (z_{i2})$}; 
    \draw (A_I_2) -- (6.5,2) node[rect1, fill = white, text width = 1.1cm] (A_I_c_2){$I_3 (z_{i3})$}; 
    \draw (A_I_2) -- (6.5,1) node[rect1, fill = white, text width = 1.1cm] (A_I_d_2){$I_4 (z_{i4})$}; 
    \draw (A_I_2) -- (6.5,0) node[rect1, fill = white, text width = 1.1cm] (A_I_e_2){$I_5 (z_{i5})$}; 
    \draw (3.5, -0.5) node {Yes}; 
    \draw (A_S_1) -- (4.5,-1.5) node[rect1, fill = green!10!white, text width = 1.8cm] (A_P_2){Which to Purchase?}; 
    \draw (A_P_2) -- (6.55,-1.5) node[rect1, fill = white, text width = 1.1cm] (A_P_a_2){$P_1 (u_{i1})$}; 
    \draw[dashed] (7.2, -3.0) -- (7.2, 4.8); 
    \draw (4.9,-2.8) node {Step 2}; 

    \draw (A_I_b_2) -- (7.9,3) node[rect1, fill = green!10!white, text width = 1.2cm] (A_S_2){Stop?}; 
    \draw (8.75, 3.6) node {No}; 
    \draw (A_S_2) -- (9.8,3.0) node[rect1, fill = green!10!white, text width = 1.6cm] (A_I_3){Which to Inspect?}; 
    \draw (A_I_3) -- (11.4,4.0) node[rect1, fill = blue!20!white, text width = 1.1cm] (A_I_c_3){$I_3 (z_{i3})$}; 
    \draw (A_I_3) -- (11.4,3.0) node[rect1, fill = white, text width = 1.1cm] (A_I_d_3){$I_4 (z_{i4})$}; 
    \draw (A_I_3) -- (11.4,2.0) node[rect1, fill = white, text width = 1.1cm] (A_I_e_3){$I_5 (z_{i5})$}; 
    \draw (8.2, 0.5) node {Yes}; 
    \draw (A_S_2) -- (9.4,-1.0) node[rect1, fill = green!10!white, text width = 1.8cm] (A_P_3){Which to Purchase?}; 
    \draw (A_P_3) -- (11.45,-0.5) node[rect1, fill = white, text width = 1.1cm] (A_P_a_3){$P_1 (u_{i1})$}; 
    \draw (A_P_3) -- (11.45,-1.5) node[rect1, fill = white, text width = 1.1cm] (A_P_b_3){$P_2 (u_{i2})$}; 
    \draw[dashed] (12.1, -3.0) -- (12.1, 4.8); 
    \draw (9.8,-2.8) node {Step 3}; 
    
    \draw (A_I_c_3) -- (12.8,4) node[rect1, fill = green!10!white, text width = 1.2cm] (A_S_3){Stop?}; 
    \draw (13.65, 4.6) node {No};
    \draw (A_S_3) -- (14.7,4) node[rect1, fill = green!10!white, text width = 1.6cm] (A_I_4){Which to Inspect?}; 
    \draw (A_I_4) -- (16.4,4) node[rect1, fill = blue!20!white, text width = 1.1cm] (A_I_d_4){$I_4 (z_{i4})$}; 
    \draw (A_I_4) -- (16.4,3) node[rect1, fill = white, text width = 1.1cm] (A_I_e_4){$I_5 (z_{i5})$}; 
    \draw (13, 1.5) node {Yes}; 
    \draw (A_S_3) -- (14.3,0) node[rect1, fill = green!10!white, text width = 1.8cm] (A_P_4){Which to Purchase?}; 
    \draw (A_P_4) -- (16.45,1) node[rect1, fill = white, text width = 1.1cm] (A_P_a_4){$P_1 (u_{i1})$}; 
    \draw (A_P_4) -- (16.45,0) node[rect1, fill = white, text width = 1.1cm] (A_P_b_4){$P_2 (u_{i2})$}; 
    \draw (A_P_4) -- (16.45,-1) node[rect1, fill = white, text width = 1.1cm] (A_P_c_4){$P_3 (u_{i3})$}; 
    \draw[dashed] (17.1, -3.0) -- (17.1, 4.8); 
    \draw (14.7,-2.8) node {Step 4}; 

    \draw (A_I_d_4) -- (17.8,4) node[rect1, fill = green!10!white, text width = 1.2cm] (A_S_4){Stop?}; 
    \draw (18.65, 4.6) node {No};
    \draw (A_S_4) -- (19.7,4) node[rect1, fill = green!10!white, text width = 1.6cm] (A_I_5){Which to Inspect?}; 
    \draw (A_I_5) -- (21.4,4) node[rect1, fill = white, text width = 1.1cm] (A_I_e_5){$I_5 (z_{i5})$}; 
    \draw (17.8, 2) node {Yes}; 
    \draw (A_S_4) -- (19.2,0.5) node[rect1, fill = green!10!white, text width = 1.8cm] (A_P_5){Which to Purchase?}; 
    \draw (A_P_5) -- (21.4,2) node[rect1, fill = white, text width = 1.1cm] (A_P_a_5){$P_1 (u_{i1})$}; 
    \draw (A_P_5) -- (21.4,1) node[rect1, fill = white, text width = 1.1cm] (A_P_b_5){$P_2 (u_{i2})$}; 
    \draw (A_P_5) -- (21.4,0) node[rect1, fill = blue!20!white, text width = 1.1cm] (A_P_c_5){$P_3 (u_{i3})$}; 
    \draw (A_P_5) -- (21.4,-1) node[rect1, fill = white, text width = 1.1cm] (A_P_d_5){$P_4 (u_{i4})$}; 
    \draw (19.7,-2.8) node {Step 5}; 

    \draw [red, very thick] (A_I_1) -- (A_I_a_1);
    
    \draw [blue, very thick] (A_S_1) -- (A_I_2); 
    \draw [red, very thick] (A_I_2) -- (A_I_b_2); 
    
    \draw [blue, very thick] (A_S_2) -- (A_I_3); 
    \draw [red, very thick] (A_I_3) -- (A_I_c_3); 
    
    \draw [blue, very thick] (A_S_3) -- (A_I_4); 
    \draw [red, very thick] (A_I_4) -- (A_I_d_4);
     
    \draw [blue, very thick] (A_S_4) -- (A_P_5); 
    \draw [red, very thick] (A_P_5) -- (A_P_c_5); 
    \draw [violet, very thick] (A_I_5) -- (A_I_e_5); 
    
\end{tikzpicture}

%% file: TikzPics/figure_example.tex
\begin{tikzpicture}[scale = 0.92]
    \tikzstyle{every node}=[font=\normalsize, scale = 0.9]; 
    \tikzstyle{every path}=[thick]; 
    \draw[black, thick, ->] (-7.5,-2.1) -- (-7.5, 2.1) node[left] (){Value}; 
    
    \draw (0,0) node [rect1, fill = blue!20!white, text width = 1.0cm, align=center] (u_h){$P_3$ \\ $(u_{i3})$}; 
    \draw (-1.6,0.4) node [rect1, fill = green!10!white, text width = 1.0cm] (z_J){$I_4$ \\ $(u_{i4})$};
    \draw[->, black] (z_J) -- (u_h); 
    
    \draw (-3.2,0.8) node [rect1, fill = green!10!white, text width = 0.8cm] (z_J-1){$I_3$ \\ $(z_{i3})$};
    \draw[black, ->] (z_J-1) -- (z_J); 
    \draw (-4.8,1.2) node [rect1, fill = green!10!white, text width = 0.8cm] (z_2){$I_2$ \\ $(z_{i2})$};
    \draw[black, ->] (z_2) -- (z_J-1); 
    \draw (-6.4,1.6) node [rect1, fill = green!10!white, text width = 0.8cm] (z_1){$I_1$ \\ $(z_{i1})$};
    \draw[black, ->] (z_1) -- (z_2); 

    \draw (2.6,-1.1) node [rect1, fill = white, text width = 0.8cm] (u_1){$P_1$ \\ $(u_{i1})$}; 
    \draw (4.2,-1.1) node [rect1, fill = white, text width = 0.8cm] (u_2){$P_2$ \\ $(u_{i2})$};
    \draw (5.8,-1.1) node [rect1, fill = white, text width = 0.9cm] (u_4){$P_4$ \\ $(u_{i4})$}; 
    \draw (7.4,-1.1) node [rect1, fill = white, text width = 0.9cm] (z_5){$I_5$ \\ $(z_{i5})$};

    \draw[black, ->] (u_h) -- (1.75,-0.4); 

    \draw[dotted] (1.75,-0.2) -- (8.35,-0.2) -- (8.35,-2) -- (1.75,-2) -- (1.75,-0.2); 
    \draw (5, -0.1) node [above = 0.1](){\begin{tabular}{l}
        Unselected actions with values $< u_{i3}$. 
    \end{tabular}}; 
    \draw[dotted] (0.8,-0.6) -- (0.8,2.2) -- (-7.05,2.2) -- (-7.05,-0.6) -- (0.8,-0.6); 
    \draw (-4, -0.7) node [below = 0.1](){\begin{tabular}{l}
        Selected actions with values $\geq u_{i3}$. 
    \end{tabular}}; 
\end{tikzpicture}

%% file: TikzPics/figure_single_branching.tex
\begin{tikzpicture}[scale=0.53]
    \tikzstyle{every node}=[font=\normalsize, scale = 0.85]
    
    \draw (0,0) -- (1.8,3.6) node[circ2, fill = blue!20!white, text width = 0.5cm] (I_a_1){$I_1$}; 
    \draw (0,0) -- (1.8,1.8) node[circ2, fill = green!10!white, text width = 0.5cm] (I_b_1){$I_2$}; 
    \draw (0,0) -- (1.8,0) node[circ2, fill = green!10!white, text width = 0.5cm] (I_c_1){$I_3$}; 
    \draw (0,0) -- (1.8,-1.8) node[circ2, fill = green!10!white, text width = 0.5cm] (I_d_1){$I_4$}; 
    \draw (0,0) -- (1.8,-3.6) node[circ2, fill = green!10!white, text width = 0.5cm] (I_e_1){$I_5$}; 

    \draw (I_a_1) -- (3.6,3.6) node[circ2, fill = white, text width = 0.5cm] (B_P_a_1){$P_1$}; 
    \draw (I_b_1) -- (3.6,1.8) node[circ2, fill = white, text width = 0.5cm] (B_P_b_1){$P_2$}; 
    \draw (I_c_1) -- (3.6,0) node[circ2, fill = white, text width = 0.5cm] (B_P_c_1){$P_3$}; 
    \draw (I_d_1) -- (3.6,-1.8) node[circ2, fill = white, text width = 0.5cm] (B_P_d_1){$P_4$}; 
    \draw (I_e_1) -- (3.6,-3.6) node[circ2, fill = white, text width = 0.5cm] (B_P_e_1){$P_5$}; 

    \draw (2.5, -5) node[font=\normalsize] (Step_1){Step 1};

    \draw (6,0) -- (7.8,3.6) node[circ3, fill = white, text width = 0.5cm] (I_a_2){}; 
    \draw (6,0) -- (7.8,1.8) node[circ2, fill = blue!20!white, text width = 0.5cm] (I_b_2){$I_2$}; 
    \draw (6,0) -- (7.8,0) node[circ2, fill = green!10!white, text width = 0.5cm] (I_c_2){$I_3$}; 
    \draw (6,0) -- (7.8,-1.8) node[circ2, fill = green!10!white, text width = 0.5cm] (I_d_2){$I_4$}; 
    \draw (6,0) -- (7.8,-3.6) node[circ2, fill = green!10!white, text width = 0.5cm] (I_e_2){$I_5$}; 

    \draw (I_a_2) -- (9.6,3.6) node[circ2, fill = green!10!white, text width = 0.5cm] (B_P_a_2){$P_1$}; 
    \draw (I_b_2) -- (9.6,1.8) node[circ2, fill = white, text width = 0.5cm] (B_P_b_2){$P_2$}; 
    \draw (I_c_2) -- (9.6,0) node[circ2, fill = white, text width = 0.5cm] (B_P_c_2){$P_3$}; 
    \draw (I_d_2) -- (9.6,-1.8) node[circ2, fill = white, text width = 0.5cm] (B_P_d_2){$P_4$}; 
    \draw (I_e_2) -- (9.6,-3.6) node[circ2, fill = white, text width = 0.5cm] (B_P_e_2){$P_5$};  

    \draw (8.5, -5) node[font=\normalsize] (Step_2){Step 2};

    \draw (12,0) -- (13.8,3.6) node[circ3, fill = white, text width = 0.5cm] (I_a_3){}; 
    \draw (12,0) -- (13.8,1.8) node[circ3, fill = white, text width = 0.5cm] (I_b_3){}; 
    \draw (12,0) -- (13.8,0) node[circ2, fill = blue!20!white, text width = 0.5cm] (I_c_3){$I_3$}; 
    \draw (12,0) -- (13.8,-1.8) node[circ2, fill = green!10!white, text width = 0.5cm] (I_d_3){$I_4$}; 
    \draw (12,0) -- (13.8,-3.6) node[circ2, fill = green!10!white, text width = 0.5cm] (I_e_3){$I_5$}; 

    \draw (I_a_3) -- (15.6,3.6) node[circ2, fill = green!10!white, text width = 0.5cm] (B_P_a_3){$P_1$}; 
    \draw (I_b_3) -- (15.6,1.8) node[circ2, fill = green!10!white, text width = 0.5cm] (B_P_b_3){$P_2$}; 
    \draw (I_c_3) -- (15.6,0) node[circ2, fill = white, text width = 0.5cm] (B_P_c_3){$P_3$}; 
    \draw (I_d_3) -- (15.6,-1.8) node[circ2, fill = white, text width = 0.5cm] (B_P_d_3){$P_4$}; 
    \draw (I_e_3) -- (15.6,-3.6) node[circ2, fill = white, text width = 0.5cm] (B_P_e_3){$P_5$};  

    \draw (14.5, -5) node[font=\normalsize] (Step_3){Step 3};

    \draw (18,0) -- (19.8,3.6) node[circ3, fill = white, text width = 0.5cm] (I_a_4){}; 
    \draw (18,0) -- (19.8,1.8) node[circ3, fill = white, text width = 0.5cm] (I_b_4){}; 
    \draw (18,0) -- (19.8,0) node[circ3, fill = white, text width = 0.5cm] (I_c_4){}; 
    \draw (18,0) -- (19.8,-1.8) node[circ2, fill = blue!20!white, text width = 0.5cm] (I_d_4){$I_4$}; 
    \draw (18,0) -- (19.8,-3.6) node[circ2, fill = green!10!white, text width = 0.5cm] (I_e_4){$I_5$}; 

    \draw (I_a_4) -- (21.6,3.6) node[circ2, fill = green!10!white, text width = 0.5cm] (B_P_a_4){$P_1$}; 
    \draw (I_b_4) -- (21.6,1.8) node[circ2, fill = green!10!white, text width = 0.5cm] (B_P_b_4){$P_2$}; 
    \draw (I_c_4) -- (21.6,0) node[circ2, fill = green!10!white, text width = 0.5cm] (B_P_c_4){$P_3$}; 
    \draw (I_d_4) -- (21.6,-1.8) node[circ2, fill = white, text width = 0.5cm] (B_P_d_4){$P_4$}; 
    \draw (I_e_4) -- (21.6,-3.6) node[circ2, fill = white, text width = 0.5cm] (B_P_e_4){$P_5$};  

    \draw (20.5, -5) node[font=\normalsize] (Step_4){Step 4}; 

    \draw (24,0) -- (25.8,3.6) node[circ3, fill = white, text width = 0.5cm] (I_a_5){}; 
    \draw (24,0) -- (25.8,1.8) node[circ3, fill = white, text width = 0.5cm] (I_b_5){}; 
    \draw (24,0) -- (25.8,0) node[circ3, fill = white, text width = 0.5cm] (I_c_5){}; 
    \draw (24,0) -- (25.8,-1.8) node[circ3, fill = white, text width = 0.5cm] (I_d_5){}; 
    \draw (24,0) -- (25.8,-3.6) node[circ2, fill = green!10!white, text width = 0.5cm] (I_e_5){$I_5$}; 

    \draw (I_a_5) -- (27.6,3.6) node[circ2, fill = green!10!white, text width = 0.5cm] (B_P_a_5){$P_1$}; 
    \draw (I_b_5) -- (27.6,1.8) node[circ2, fill = green!10!white, text width = 0.5cm] (B_P_b_5){$P_2$}; 
    \draw (I_c_5) -- (27.6,0) node[circ2, fill = blue!20!white, text width = 0.5cm] (B_P_c_5){$P_3$}; 
    \draw (I_d_5) -- (27.6,-1.8) node[circ2, fill = green!10!white, text width = 0.5cm] (B_P_d_5){$P_4$}; 
    \draw (I_e_5) -- (27.6,-3.6) node[circ2, fill = white, text width = 0.5cm] (B_P_e_5){$P_5$};  

    \draw (26.5, -5) node[font=\normalsize] (Step_5){Step 5}; 
\end{tikzpicture}

%% file: TikzPics/figure_implementation.tex
\begin{tikzpicture}[scale=0.85, every node/.style={scale=0.8}, >=stealth]

    \tikzset{
        step_label/.style={font=\bfseries\small, yshift=-1.5cm}
    }

    \begin{scope}[xshift=0cm, yshift=0cm]
        \draw (0,-0.2) node [circ1, fill = white] (u_3A){$u_{i3}$}; 
        \draw (-1.2,0.4) node [circ1, fill = blue!20!white] (z_4A){$z_{i4}^d$};
        \draw (-2.4,1) node [circ1, fill = white] (z_3A){$z_{i3}$};
        \draw (-3.6,1.6) node [circ1, fill = white] (z_2A){$z_{i2}$};
        \draw (-4.8,2.2) node [circ1, fill = white] (z_1A){$z_{i1}$};
        \draw (-4.8,-1.6) node [circ1, fill = white] (u_1A){$u_{i1}$}; 
        \draw (-2.9,-1.6) node [circ1, fill = white] (u_2A){$u_{i2}$}; 
        \draw (-1.2,-1.6) node [circ1, fill = white] (u_4A){$u_{i4}$}; 
        \draw (1.2,-1.6) node [circ1, fill = white] (z_5A){$z_{i5}$}; 

        \draw[->] (z_1A) -- (z_2A); \draw[->] (z_2A) -- (z_3A); \draw[->] (z_3A) -- (z_4A); 
        \draw[->] (z_4A) -- (u_3A); \draw[->] (u_3A) -- (u_1A); \draw[->] (u_3A) -- (u_2A);
        \draw[->] (u_3A) -- (u_4A); \draw[->] (u_3A) -- (z_5A);
        
        \node[step_label] at (-1.8,-1.5) {Step 1: Draw $z_{i4}^d$};
    \end{scope}

    \begin{scope}[xshift=8cm, yshift=0cm]
        \draw (0,-0.2) node [circ1, fill = white] (u_3B){$u_{i3}$}; 
        \draw (-1.2,0.4) node [circ1, fill = green!10!white] (z_4B){$z_{i4}^d$};
        \draw (-2.4,1) node [circ1, fill = blue!20!white] (z_3B){$z_{i3}^d$};
        \draw (-3.6,1.6) node [circ1, fill = blue!20!white] (z_2B){$z_{i2}^d$};
        \draw (-4.8,2.2) node [circ1, fill = blue!20!white] (z_1B){$z_{i1}^d$};
        \draw (-4.8,-1.6) node [circ1, fill = white] (u_1B){$u_{i1}$}; 
        \draw (-2.9,-1.6) node [circ1, fill = white] (u_2B){$u_{i2}$}; 
        \draw (-1.2,-1.6) node [circ1, fill = white] (u_4B){$u_{i4}$}; 
        \draw (1.2,-1.6) node [circ1, fill = white] (z_5B){$z_{i5}$}; 

        \draw[->] (z_1B) -- (z_2B); \draw[->] (z_2B) -- (z_3B); \draw[->] (z_3B) -- (z_4B); 
        \draw[->] (z_4B) -- (u_3B); \draw[->] (u_3B) -- (u_1B); \draw[->] (u_3B) -- (u_2B);
        \draw[->] (u_3B) -- (u_4B); \draw[->] (u_3B) -- (z_5B);
        
        \node[step_label] at (-1.8,-1.5) {Step 2: Draw $z_{i3}^d$, $z_{i2}^d$ and $z_{i1}^d$ sequentially};
    \end{scope}

    \begin{scope}[xshift=0cm, yshift=-6.5cm]
        \draw (0,-0.2) node [circ1, fill = blue!20!white] (u_3C){$u_{i3}^d$}; 
        \draw (-1.2,0.4) node [circ1, fill = green!10!white] (z_4C){$z_{i4}^d$};
        \draw (-2.4,1) node [circ1, fill = green!10!white] (z_3C){$z_{i3}^d$};
        \draw (-3.6,1.6) node [circ1, fill = green!10!white] (z_2C){$z_{i2}^d$};
        \draw (-4.8,2.2) node [circ1, fill = green!10!white] (z_1C){$z_{i1}^d$};
        \draw (-4.8,-1.6) node [circ1, fill = white] (u_1C){$u_{i1}$}; 
        \draw (-2.9,-1.6) node [circ1, fill = white] (u_2C){$u_{i2}$}; 
        \draw (-1.2,-1.6) node [circ1, fill = white] (u_4C){$u_{i4}$}; 
        \draw (1.2,-1.6) node [circ1, fill = white] (z_5C){$z_{i5}$}; 

        \draw[->] (z_1C) -- (z_2C); \draw[->] (z_2C) -- (z_3C); \draw[->] (z_3C) -- (z_4C); 
        \draw[->] (z_4C) -- (u_3C); \draw[->] (u_3C) -- (u_1C); \draw[->] (u_3C) -- (u_2C);
        \draw[->] (u_3C) -- (u_4C); \draw[->] (u_3C) -- (z_5C);

        \node[step_label] at (-1.8,-1.5) {Step 3: Draw $u_{i3}^d$};
    \end{scope}

    \begin{scope}[xshift=8cm, yshift=-6.5cm]
        \draw (0,-0.2) node [circ1, fill =green!10!white] (u_3D){$u_{i3}^d$}; 
        \draw (-1.2,0.4) node [circ1, fill = green!10!white] (z_4D){$z_{i4}^d$};
        \draw (-2.4,1) node [circ1, fill = green!10!white] (z_3D){$z_{i3}^d$};
        \draw (-3.6,1.6) node [circ1, fill = green!10!white] (z_2D){$z_{i2}^d$};
        \draw (-4.8,2.2) node [circ1, fill = green!10!white] (z_1D){$z_{i1}^d$};
        \draw (-4.8,-1.6) node [circ1, fill = gray!10!white] (u_1D){$u_{i1}$}; 
        \draw (-2.9,-1.6) node [circ1, fill = gray!10!white] (u_2D){$u_{i2}$}; 
        \draw (-1.2,-1.6) node [circ1, fill = gray!10!white] (u_4D){$u_{i4}$}; 
        \draw (1.2,-1.6) node [circ1, fill = gray!10!white] (z_5D){$z_{i5}$}; 

        \draw[->] (z_1D) -- (z_2D); \draw[->] (z_2D) -- (z_3D); \draw[->] (z_3D) -- (z_4D); 
        \draw[->] (z_4D) -- (u_3D); \draw[->] (u_3D) -- (u_1D); \draw[->] (u_3D) -- (u_2D);
        \draw[->] (u_3D) -- (u_4D); \draw[->] (u_3D) -- (z_5D);

        \node[draw, dashed, ellipse, red, thick, inner sep=2pt, rotate=-30, fit=(z_4D) (u_3D)] {};
        \draw (0.7, 1) node [red](){$y_i^d$}; 
        
        \node[step_label] at (-1.8,-1.5) {Step 4: Compute probabilities for $u_{i1}$, $u_{i2}$, $u_{i3}$ and $z_{i5}$};
    \end{scope}

\end{tikzpicture}

%% file: Tables/tab_ghk.tex
\resizebox{\textwidth}{!}{
\begin{threeparttable}
\centering
\smallskip

\begin{tabular}{lccccccc}
\toprule
                                                                                    &                                                      &  & (1)                                                            & (2)                                                            & (3)                                                            & (4)                                                            & (5)                                                                        \\
Market Size                               &                                                      &  & 8 prods                                                        & 8 prods                                                        & 32 prods                                                       & 32 prods                                                       & 32 prods                                                                   \\ \addlinespace
Estimation Method                         &                                                      &  & Rank-based                                                         & Policy-based                                                        & Rank-based                                                         & Policy-based                                                        & \begin{tabular}[c]{@{}c@{}}Policy-based \\ (with same total draws)\end{tabular} \\ \midrule
                                                                                    & \begin{tabular}[c]{@{}c@{}}True\\ value\end{tabular} &  & \begin{tabular}[c]{@{}c@{}}Estimates\\ (Std. Dev)\end{tabular} & \begin{tabular}[c]{@{}c@{}}Estimates\\ (Std. Dev)\end{tabular} & \begin{tabular}[c]{@{}c@{}}Estimates\\ (Std. Dev)\end{tabular} & \begin{tabular}[c]{@{}c@{}}Estimates\\ (Std. Dev)\end{tabular} & \begin{tabular}[c]{@{}c@{}}Estimates\\ (Std. Dev)\end{tabular}             \\ \addlinespace
Outside Option: $\gamma^{outside}$                                                  & -0.3                                                 &  & \begin{tabular}[c]{@{}c@{}}-0.290\\ (0.046)\end{tabular}       & \begin{tabular}[c]{@{}c@{}}-0.304\\ (0.044)\end{tabular}       & \begin{tabular}[c]{@{}c@{}}-0.132\\ (0.066)\end{tabular}       & \begin{tabular}[c]{@{}c@{}}-0.175\\ (0.075)\end{tabular}       & \begin{tabular}[c]{@{}c@{}}0.004\\ (0.076)\end{tabular}                    \\ \addlinespace
Attributes Coefficients                                                             &                                                      &  &                                                                &                                                                &                                                                &                                                                &                                                                            \\
\quad Attribute 1: $\gamma_1$                                                       & 1                                                    &  & \begin{tabular}[c]{@{}c@{}}0.994\\ (0.091)\end{tabular}        & \begin{tabular}[c]{@{}c@{}}1.015\\ (0.097)\end{tabular}        & \begin{tabular}[c]{@{}c@{}}0.944\\ (0.034)\end{tabular}        & \begin{tabular}[c]{@{}c@{}}0.976\\ (0.037)\end{tabular}        & \begin{tabular}[c]{@{}c@{}}0.939\\ (0.046)\end{tabular}                    \\
\quad Attribute 2: $\gamma_2$                                                       & 0.5                                                  &  & \begin{tabular}[c]{@{}c@{}}0.495\\ (0.066)\end{tabular}        & \begin{tabular}[c]{@{}c@{}}0.516\\ (0.065)\end{tabular}        & \begin{tabular}[c]{@{}c@{}}0.472\\ (0.023)\end{tabular}        & \begin{tabular}[c]{@{}c@{}}0.485\\ (0.025)\end{tabular}        & \begin{tabular}[c]{@{}c@{}}0.471\\ (0.034)\end{tabular}                    \\
\quad Attribute 3: $\gamma_3$                                                       & 0.2                                                  &  & \begin{tabular}[c]{@{}c@{}}0.189\\ (0.043)\end{tabular}        & \begin{tabular}[c]{@{}c@{}}0.196\\ (0.043)\end{tabular}        & \begin{tabular}[c]{@{}c@{}}0.194\\ (0.019)\end{tabular}        & \begin{tabular}[c]{@{}c@{}}0.201\\ (0.021)\end{tabular}        & \begin{tabular}[c]{@{}c@{}}0.190\\ (0.019)\end{tabular}                    \\
\quad Attribute 4: $\gamma_4$                                                       & 0.7                                                  &  &                                                                &                                                                & \begin{tabular}[c]{@{}c@{}}0.661\\ (0.031)\end{tabular}        & \begin{tabular}[c]{@{}c@{}}0.682\\ (0.029)\end{tabular}        & \begin{tabular}[c]{@{}c@{}}0.661\\ (0.028)\end{tabular}                    \\
\quad Attribute 5: $\gamma_5$                                                       & 0.4                                                  &  &                                                                &                                                                & \begin{tabular}[c]{@{}c@{}}0.375\\ (0.022)\end{tabular}        & \begin{tabular}[c]{@{}c@{}}0.387\\ (0.049)\end{tabular}        & \begin{tabular}[c]{@{}c@{}}0.375\\ (0.051)\end{tabular}                    \\
                                                                                    &                                                      &  &                                                                &                                                                &                                                                &                                                                &                                                                            \\
Price Coefficient: $\beta$                                                          & -0.8                                                 &  & \begin{tabular}[c]{@{}c@{}}-0.788\\ (0.028)\end{tabular}       & \begin{tabular}[c]{@{}c@{}}-0.802\\ (0.028)\end{tabular}       & \begin{tabular}[c]{@{}c@{}}-0.753\\ (0.016)\end{tabular}       & \begin{tabular}[c]{@{}c@{}}-0.776\\ (0.016)\end{tabular}       & \begin{tabular}[c]{@{}c@{}}-0.748\\ (0.020)\end{tabular}                   \\
Log Search Cost: $\Bar{c}$                                                          & -3                                                   &  & \begin{tabular}[c]{@{}c@{}}-2.977\\ (0.051)\end{tabular}       & \begin{tabular}[c]{@{}c@{}}-3.008\\ (0.052)\end{tabular}       & \begin{tabular}[c]{@{}c@{}}-3.011\\ (0.052)\end{tabular}       & \begin{tabular}[c]{@{}c@{}}-3.175\\ (0.048)\end{tabular}       & \begin{tabular}[c]{@{}c@{}}-3.279\\ (0.054)\end{tabular}                   \\
                                                                                    &                                                      &  &                                                                &                                                                &                                                                &                                                                &                                                                            \\
Log-Likelihood (True value)                                                         &                                                      &  & -6751                                                          & -6733                                                          & -24921                                                         & -25029                                                         & -26355                                                                     \\
Log-Likelihood (Estimates)                                                          &                                                      &  & -6747                                                          & -6730                                                          & -24900                                                         & -25007                                                         & -26296                                                                     \\
RMSE                                                                                &                                                      &  & 0.053                                                          & 0.054                                                          & 0.074                                                          & 0.084                                                          & 0.154                                                                      \\
                                                                                    &                                                      &  &                                                                &                                                                &                                                                &                                                                &                                                                            \\
Average Iter Running Time (s)                                                       &                                                      &  & 0.27                                                           & 0.29                                                           & 1.05                                                           & 1.56                                                           & 1.11                                                                       \\
Mean Convergence Time (s)                                                           &                                                      &  & 88                                                             & 88                                                             & 534                                                            & 928                                                           & 705                                                                        \\ 
Total Draws Made                                                       &                                                      &  & 1176                                                            & 3461                                                            & 2095                                                            & 16395                                                          & 2097                                                                        \\\addlinespace
\begin{tabular}[c]{@{}l@{}}Lines of MATLAB code\\ for likelihood computation\end{tabular} &                                                      &  & 35                                                             & 89                                                             & 35                                                             & 89                                                             & 89                                                                         \\ \bottomrule
\end{tabular} 
\smallskip
\begin{tablenotes}
    \item \footnotesize \emph{Notes:} We simulate 50 datasets for 2,000 consumers under different random seeds and report the average results across these 50 estimations. In columns (1) to (4), each estimation is based on 500 groups of simulation draws. In column (5), the number of draws per estimation is adjusted to match the total number of simulated values in column (3), which implies approximately 65 groups per estimation on average, since each draw under Policy-based involves roughly 8 times the simulation dimensionality of Rank-based. Standard deviations across simulations are reported in parentheses. 
\end{tablenotes}
\end{threeparttable}}

%% file: TikzPics/figure_recover.tex
\begin{tikzpicture}[scale=0.55]
    \tikzstyle{every node}=[font=\normalsize, scale = 0.85]
    \draw (-5,0) -- (-3.2,3.2) node[circ2, fill = blue!20!white, text width = 0.5cm] (I_a_1){$I_1$}; 
    \draw (-5,0) -- (-3.2,1.6) node[circ2, fill = green!10!white, text width = 0.5cm] (I_b_1){$I_2$}; 
    \draw (-5,0) -- (-3.2,0) node[circ2, fill = green!10!white, text width = 0.5cm] (I_c_1){$I_3$}; 
    \draw (-5,0) -- (-3.2,-1.6) node[circ2, fill = green!10!white, text width = 0.5cm] (I_d_1){$I_4$}; 
    \draw (-5,0) -- (-3.2,-3.2) node[circ2, fill = green!10!white, text width = 0.5cm] (I_e_1){$I_5$}; 

    \draw (I_a_1) -- (-1.6,3.2) node[circ2, fill = white, text width = 0.5cm] (B_P_a_1){$P_1$};  
    \draw (I_b_1) -- (-1.6,1.6) node[circ2, fill = white, text width = 0.5cm] (B_P_b_1){$P_2$};  
    \draw (I_c_1) -- (-1.6,0) node[circ2, fill = white, text width = 0.5cm] (B_P_c_1){$P_3$};  
    \draw (I_d_1) -- (-1.6,-1.6) node[circ2, fill = white, text width = 0.5cm] (B_P_d_1){$P_4$};  
    \draw (I_e_1) -- (-1.6,-3.2) node[circ2, fill = white, text width = 0.5cm] (B_P_e_1){$P_5$};  

    \draw (-3, -4.5) node[font=\normalsize] (Step_1){Step 1};
    
    \draw (0,0) -- (1.8,3.2) node[circ3, fill = white, text width = 0.5cm] (I_a_2){}; 
    \draw (0,0) -- (1.8,1.6) node[circ2, fill = gray!20!white, text width = 0.5cm] (I_b_2){$I_2$}; 
    \draw (0,0) -- (1.8,0) node[circ2, fill = gray!20!white, text width = 0.5cm] (I_c_3){$I_3$}; 
    \draw (0,0) -- (1.8,-1.6) node[circ2, fill = blue!20!white, text width = 0.5cm] (I_d_2){$I_4$}; 
    \draw (0,0) -- (1.8,-3.2) node[circ2, fill = green!10!white, text width = 0.5cm] (I_e_2){$I_5$}; 

    \draw (I_a_2) -- (3.4,3.2) node[circ2, fill = green!10!white, text width = 0.5cm] (B_P_a_1){$P_1$};  
    \draw (I_b_2) -- (3.4,1.6) node[circ2, fill = green!10!white, text width = 0.5cm] (B_P_b_1){$P_2$};  
    \draw (I_c_2) -- (3.4,0) node[circ2, fill = green!10!white, text width = 0.5cm] (B_P_c_1){$P_3$};  
    \draw (I_d_2) -- (3.4,-1.6) node[circ2, fill = white, text width = 0.5cm] (B_P_d_1){$P_4$};  
    \draw (I_e_2) -- (3.4,-3.2) node[circ2, fill = white, text width = 0.5cm] (B_P_e_1){$P_5$};  

    \draw (2, -4.5) node[font=\normalsize] (Step_2){Step 2};

    \draw (5,0) -- (6.8,3.2) node[circ3, fill = white, text width = 0.5cm] (I_a_3){}; 
    \draw (5,0) -- (6.8,1.6) node[circ3, fill = white, text width = 0.5cm] (I_b_3){}; 
    \draw (5,0) -- (6.8,0) node[circ3, fill = white, text width = 0.5cm] (I_c_3){}; 
    \draw (5,0) -- (6.8,-1.6) node[circ3, fill = white, text width = 0.5cm] (I_d_3){}; 
    \draw (5,0) -- (6.8,-3.2) node[circ2, fill = green!10!white, text width = 0.5cm] (I_e_3){$I_5$}; 

    \draw (I_a_3) -- (8.4,3.2) node[circ2, fill = green!10!white, text width = 0.5cm] (B_P_a_3){$P_1$};
    \draw (I_b_3) -- (8.4,1.6) node[circ2, fill = green!10!white, text width = 0.5cm] (B_P_b_3){$P_2$};
    \draw (I_c_3) -- (8.4,0) node[circ2, fill = blue!20!white, text width = 0.5cm] (B_P_c_3){$P_3$}; 
    \draw (I_d_3) -- (8.4,-1.6) node[circ2, fill = green!10!white, text width = 0.5cm] (B_P_d_3){$P_4$};
    \draw (I_e_3) -- (8.4,-3.2) node[circ2, fill = white, text width = 0.5cm] (B_P_e_3){$P_5$};

    \draw (7, -4.5) node[font=\normalsize] (Step_3){Step 3};

    \draw (2, -6) node[font=\normalsize] {The Branching Project Representation};

    \draw (14.5,2) node[circ1, fill = blue!20!white, text width = 0.4cm] (I_B_2){$I_2$}; 
    \draw (19.5,2) node[circ1, fill = blue!20!white, text width = 0.4cm] (I_C_2){$I_3$}; 
    \draw (17,0.8) node[circ1, fill = blue!20!white, text width = 0.4cm] (I_D_2){$I_4$}; 
        
    \draw (14,-3.2) node[circ1, fill = green!10!white, text width = 0.4cm] (P_A_2){$P_1$}; 
    \draw (16,-3.2) node[circ1, fill = green!10!white, text width = 0.4cm] (P_B_2){$P_2$}; 
    \draw (18,-3.2) node[circ1, fill = green!10!white, text width = 0.4cm] (P_D_2){$P_4$}; 
    \draw (20,-3.2) node[circ1, fill = green!10!white, text width = 0.4cm] (I_E_2){$I_5$}; 
    \draw (17,3.2) node[circ1, fill = blue!20!white, text width = 0.4cm] (I_A_2){$I_1$}; 
    \draw (17,-1.2) node[circ1, fill = blue!20!white, text width = 0.4cm] (P_C_2){$P_3$}; 
    \draw [->, red, thick] (I_A_2) -- (I_B_2);
    \draw [->, red, thick] (I_A_2) -- (I_C_2);
    \draw [->, red, thick] (I_B_2) -- (I_D_2);
    \draw [->, red, thick] (I_C_2) -- (I_D_2);
    \draw [->, red, thick] (I_D_2) -- (P_C_2);
    \draw [->, red, thick] (P_C_2) -- (P_A_2);
    \draw [->, red, thick] (P_C_2) -- (P_B_2);
    \draw [->, red, thick] (P_C_2) -- (P_D_2);
    \draw [->, red, thick] (P_C_2) -- (I_E_2);

    \draw (17, -6) node[font=\normalsize] {The Ranking Representation};
\end{tikzpicture}

%% file: TikzPics/figure_ept.tex
\begin{tikzpicture}[scale=0.55]
    \tikzstyle{every node}=[font=\normalsize, scale = 0.85]

    \draw (0,0) -- (1.8,3.2) node[circ2, fill = green!10!white, text width = 0.5cm] (E_a_2){$E_1$}; 
    \draw (0,0) -- (1.8,1.6) node[circ2, fill = green!10!white, text width = 0.5cm] (E_b_2){$E_2$}; 
    \draw (0,0) -- (1.8,0) node[circ2, fill = blue!20!white, text width = 0.5cm] (I_c_2){$I_3$}; 
    \draw (0,0) -- (1.8,-1.6) node[circ2, fill = green!10!white, text width = 0.5cm] (E_d_2){$E_4$}; 
    \draw (0,0) -- (1.8,-3.2) node[circ2, fill = green!10!white, text width = 0.5cm] (E_e_2){$E_5$}; 

    \draw (I_c_2) -- (3.6,0) node[circ2, fill = white, text width = 0.5cm] (B_P_c_2){$P_3$}; 

    \draw (2.5, -4.5) node[font=\normalsize] (Step_1){Step 1};

    \draw (6,0) -- (7.8,3.2) node[circ2, fill = green!10!white, text width = 0.5cm] (E_a_3){$E_1$}; 
    \draw (6,0) -- (7.8,1.6) node[circ2, fill = green!10!white, text width = 0.5cm] (E_b_3){$E_2$}; 
    \draw (6,0) -- (7.8,0) node[circ3, fill = white, text width = 0.5cm] (I_c_3){}; 
    \draw (6,0) -- (7.8,-1.6) node[circ2, fill = green!10!white, text width = 0.5cm] (E_d_3){$E_4$}; 
    \draw (6,0) -- (7.8,-3.2) node[circ2, fill = green!10!white, text width = 0.5cm] (E_e_3){$E_5$}; 

    \draw (I_c_3) -- (9.6,0) node[circ2, fill = blue!20!white, text width = 0.5cm] (B_P_c_3){$P_3$};  

    \draw (8.5, -4.5) node[font=\normalsize] (Step_2){Step 2};

    \draw (17.5,2) node[circ1, fill = blue!20!white, text width = 0.4cm] (I_C_2){$I_3$}; 
    \draw (20.5,2) node[circ1, fill = blue!20!white, text width = 0.4cm] (P_C_2){$P_3$}; 
        
    \draw (16,-2.5) node[circ1, fill = green!10!white, text width = 0.4cm] (E_A_2){$E_1$}; 
    \draw (18,-2.5) node[circ1, fill = green!10!white, text width = 0.4cm] (E_B_2){$E_2$}; 
    \draw (20,-2.5) node[circ1, fill = green!10!white, text width = 0.4cm] (E_D_2){$E_4$}; 
    \draw (22,-2.5) node[circ1, fill = green!10!white, text width = 0.4cm] (E_E_2){$E_5$}; 
    \draw [->, red, thick] (I_C_2) -- (E_A_2);
    \draw [->, red, thick] (I_C_2) -- (E_B_2);
    \draw [->, red, thick] (I_C_2) -- (E_D_2);
    \draw [->, red, thick] (I_C_2) -- (E_E_2);
    \draw [->, red, thick] (P_C_2) -- (E_A_2);
    \draw [->, red, thick] (P_C_2) -- (E_B_2);
    \draw [->, red, thick] (P_C_2) -- (E_D_2);
    \draw [->, red, thick] (P_C_2) -- (E_E_2);
    
\end{tikzpicture}

%% file: Tables/tab_firstview_main.tex
\resizebox{\textwidth}{!}{
\begin{threeparttable}
\centering
\smallskip
\begin{tabular}{cccccccc}
\toprule
                            &          & (1)                                                      & (2)                                                      & (3)                                                      & (4)                                                      & (5)                                                      & Full Info                                                \\ \midrule
    & \begin{tabular}[c]{@{}c@{}}True\\ value\end{tabular} & \begin{tabular}[c]{@{}c@{}}Estimates\\ (Std. Dev)\end{tabular} & \begin{tabular}[c]{@{}c@{}}Estimates\\ (Std. Dev)\end{tabular} & \begin{tabular}[c]{@{}c@{}}Estimates\\ (Std. Dev)\end{tabular} & \begin{tabular}[c]{@{}c@{}}Estimates\\ (Std. Dev)\end{tabular} & \begin{tabular}[c]{@{}c@{}}Estimates\\ (Std. Dev)\end{tabular} & \begin{tabular}[c]{@{}c@{}}Estimates\\ (Std. Dev)\end{tabular}             \\ \addlinespace
$\gamma_1$                  & 1        & \begin{tabular}[c]{@{}c@{}}0.987\\ (0.090)\end{tabular}  & \begin{tabular}[c]{@{}c@{}}0.978\\ (0.041)\end{tabular}  & \begin{tabular}[c]{@{}c@{}}1.022\\ (0.039)\end{tabular}  & \begin{tabular}[c]{@{}c@{}}0.977\\ (0.049)\end{tabular}  & \begin{tabular}[c]{@{}c@{}}0.972\\ (0.049)\end{tabular}  & \begin{tabular}[c]{@{}c@{}}0.973\\ (0.037)\end{tabular}  \\ \addlinespace
$\gamma_2$                  & 0.5      & \begin{tabular}[c]{@{}c@{}}0.494\\ (0.052)\end{tabular}  & \begin{tabular}[c]{@{}c@{}}0.490\\ (0.028)\end{tabular}  & \begin{tabular}[c]{@{}c@{}}0.512\\ (0.029)\end{tabular}  & \begin{tabular}[c]{@{}c@{}}0.487\\ (0.034)\end{tabular}  & \begin{tabular}[c]{@{}c@{}}0.485\\ (0.040)\end{tabular}  & \begin{tabular}[c]{@{}c@{}}0.484\\ (0.027)\end{tabular}  \\ \addlinespace
$\gamma_3$                  & 0.2      & \begin{tabular}[c]{@{}c@{}}0.217\\ (0.039)\end{tabular}  & \begin{tabular}[c]{@{}c@{}}0.202\\ (0.024)\end{tabular}  & \begin{tabular}[c]{@{}c@{}}0.211\\ (0.025)\end{tabular}  & \begin{tabular}[c]{@{}c@{}}0.211\\ (0.031)\end{tabular}  & \begin{tabular}[c]{@{}c@{}}0.207\\ (0.028)\end{tabular}  & \begin{tabular}[c]{@{}c@{}}0.199\\ (0.023)\end{tabular}  \\ \addlinespace
$\beta$                     & -0.6     & \begin{tabular}[c]{@{}c@{}}-0.565\\ (0.078)\end{tabular} & \begin{tabular}[c]{@{}c@{}}-0.587\\ (0.037)\end{tabular} & \begin{tabular}[c]{@{}c@{}}-0.614\\ (0.035)\end{tabular} & \begin{tabular}[c]{@{}c@{}}-0.582\\ (0.044)\end{tabular} & \begin{tabular}[c]{@{}c@{}}-0.579\\ (0.040)\end{tabular} & \begin{tabular}[c]{@{}c@{}}-0.583\\ (0.034)\end{tabular} \\ \addlinespace
$\log \bar{c}$                   & -1.5     & \begin{tabular}[c]{@{}c@{}}-2.093\\ (0.868)\end{tabular} & \begin{tabular}[c]{@{}c@{}}-1.477\\ (0.037)\end{tabular} & \begin{tabular}[c]{@{}c@{}}-1.433\\ (0.032)\end{tabular} & \begin{tabular}[c]{@{}c@{}}-1.461\\ (0.085)\end{tabular} & \begin{tabular}[c]{@{}c@{}}-1.496\\ (0.040)\end{tabular} & \begin{tabular}[c]{@{}c@{}}-1.482\\ (0.034)\end{tabular} \\
                            &          &                                                          &                                                          &                                      &                                                          &                                                          &                                                          \\
Log-Likelihood (True value) &          & -3846                                                    & -7937                                                   & -7299                                                   & -6224                                                   & -6598                                                   & -9026  \\ 
Log-Likelihood (Estimates)  &          & -3839                                                    & -7934                                                   & -7294                                                   & -6221                                                   & -6595                                                   & -9022                                                   \\  \addlinespace
RMSE                        &          & 0.293                                                    & 0.034                                                    & 0.044                                                    & 0.054                                                 &  0.041                                                  & 0.034                                                   \\                                                   \bottomrule

\end{tabular}
\smallskip
\begin{tablenotes}
    \item \footnotesize \emph{Notes:} We simulate data for 2,000 consumers and report results averaged over 50 estimations using different random seeds, each based on 100 simulation draws. Standard deviations of the estimates across simulations are reported in parentheses. 
\end{tablenotes}
\end{threeparttable}}

%% file: TikzPics/figure_spd.tex
\begin{tikzpicture}[scale=0.45]
    \tikzstyle{every node}=[font=\normalsize, scale = 0.75]
    \draw (-7,0) -- (-5.8,3.2) node[circ2, fill = blue!20!white, text width = 0.5cm] (I_a_1){$I_1$}; 
    \draw (-7,0) -- (-5.8,1.2) node[circ2, fill = green!10!white, text width = 0.5cm] (I_b_1){$I_2$}; 
    \draw (-7,0) -- (-5.8,-1.6) node[circ2, fill = green!10!white, text width = 0.5cm] (D_1_1){$D_\mathrm{I}$}; 

    \draw (I_a_1) -- (-3.2,3.2) node[circ2, fill = white, text width = 0.5cm] (P_a_1){$P_1$}; 
    \draw (D_1_1) -- (-4.8,-0.2) node[circ2, fill = white, text width = 0.5cm] (I_c_1){$I_3$}; 
    \draw (D_1_1) -- (-4.8,-3) node[circ2, fill = white, text width = 0.5cm] (D_2_1){$D_\mathrm{II}$};  
    \draw (I_c_1) -- (-3.2,-0.2) node[circ2, fill = white, text width = 0.5cm] (P_c_1){$P_3$}; 
    
    \draw (-5, -4.5) node[font=\normalsize] (Step_1){Step 1};
    
    \draw (-1.5,0) -- (-0.3,3.2) node[circ3, fill = white, text width = 0.5cm] (I_a_2){}; 
    \draw (-1.5,0) -- (-0.3,1.2) node[circ2, fill = green!10!white, text width = 0.5cm] (I_b_2){$I_2$}; 
    \draw (-1.5,0) -- (-0.3,-1.6) node[circ2, fill = blue!20!white, text width = 0.5cm] (D_1_2){$D_\mathrm{I}$}; 

    \draw (I_a_2) -- (2.3,3.2) node[circ2, fill = green!10!white, text width = 0.5cm] (P_a_2){$P_1$};  
    \draw (D_1_2) -- (0.7,-0.2) node[circ2, fill = white, text width = 0.5cm] (I_c_2){$I_3$};  
    \draw (D_1_2) -- (0.7,-3) node[circ2, fill = white, text width = 0.5cm] (D_2_2){$D_\mathrm{II}$};  
    \draw (I_c_2) -- (2.3,-0.2) node[circ2, fill = white, text width = 0.5cm] (P_c_2){$P_3$}; 

    \draw (0.5, -4.5) node[font=\normalsize] (Step_2){Step 2};

    \draw (4,0) -- (5.2,3.2) node[circ3, fill = white, text width = 0.5cm] (I_a_3){}; 
    \draw (4,0) -- (5.2,1.2) node[circ2, fill = green!10!white, text width = 0.5cm] (I_b_3){$I_2$}; 
    \draw (4,0) -- (5.2,-1.6) node[circ3, fill = white, text width = 0.5cm] (D_1_3){}; 

    \draw (I_a_3) -- (7.8,3.2) node[circ2, fill = green!10!white, text width = 0.5cm] (P_a_3){$P_1$};  
    \draw (D_1_3) -- (6.2,-0.2) node[circ2, fill = blue!20!white, text width = 0.5cm] (I_c_3){$I_3$};  
    \draw (D_1_3) -- (6.2,-3) node[circ2, fill = green!10!white, text width = 0.5cm] (D_2_3){$D_\mathrm{II}$};  
    \draw (I_c_3) -- (7.8,-0.2) node[circ2, fill = white, text width = 0.5cm] (P_c_3){$P_3$}; 

    \draw (6, -4.5) node[font=\normalsize] (Step_3){Step 3};

    \draw (9.5,0) -- (10.7,3.2) node[circ3, fill = white, text width = 0.5cm] (I_a_4){}; 
    \draw (9.5,0) -- (10.7,1.2) node[circ2, fill = green!10!white, text width = 0.5cm] (I_b_4){$I_2$}; 
    \draw (9.5,0) -- (10.7,-1.6) node[circ3, fill = white, text width = 0.5cm] (D_1_4){}; 

    \draw (I_a_4) -- (13.3,3.2) node[circ2, fill = green!10!white, text width = 0.5cm] (P_a_4){$P_1$};  
    \draw (D_1_4) -- (11.7,-0.2) node[circ3, fill = white, text width = 0.5cm] (I_c_4){};  
    \draw (D_1_4) -- (11.7,-3) node[circ2, fill = green!10!white, text width = 0.5cm] (D_2_4){$D_\mathrm{II}$};  
    \draw (I_c_4) -- (13.3,-0.2) node[circ2, fill = blue!20!white, text width = 0.5cm] (P_c_4){$P_3$}; 

    \draw (11.5, -4.5) node[font=\normalsize] (Step_4){Step 4};

    \draw (3.5, -5.7) node[font=\normalsize] {The Branching Project Representation};

    \draw (19,3.2) node[circ1, fill = blue!20!white, text width = 0.4cm] (I_A){$I_1$}; 
    \draw (19,0) node[circ1, fill = blue!20!white, text width = 0.4cm] (D_1){$D_\mathrm{I}$}; 
    \draw (22,0) node[circ1, fill = blue!20!white, text width = 0.4cm] (I_C){$I_3$}; 
    \draw (25,0) node[circ1, fill = blue!20!white, text width = 0.4cm] (P_C){$P_3$}; 
        
    \draw (19,-3.2) node[circ1, fill = green!10!white, text width = 0.4cm] (P_A){$P_1$}; 
    \draw (22,-3.2) node[circ1, fill = green!10!white, text width = 0.4cm] (I_B){$I_2$}; 
    \draw (25,-3.2) node[circ1, fill = green!10!white, text width = 0.4cm] (D_2){$D_\mathrm{II}$}; 
    \draw [->, red, thick] (I_A) -- (D_1);
    \draw [->, red, thick] (D_1) -- (P_A);
    \draw [->, red, thick] (D_1) -- (I_B);
    \draw [->, red, thick] (I_C) -- (P_A);
    \draw [->, red, thick] (I_C) -- (I_B);
    \draw [->, red, thick] (I_C) -- (D_2);
    \draw [->, red, thick] (P_C) -- (P_A);
    \draw [->, red, thick] (P_C) -- (I_B);
    \draw [->, red, thick] (P_C) -- (D_2);

    \draw (22, -5.7) node[font=\normalsize] {The Ranking Representation};
\end{tikzpicture}

%% file: Tables/tab_spd.tex
\begin{threeparttable}
\centering
\smallskip
\begin{tabularx}{0.9\linewidth}{@{}Y@{}}
\begin{tabular}{cccrlc}
\toprule
&                        & True value    & \multicolumn{2}{c}{Estimates} &       \\ \midrule
&$\gamma_1$              & 0.3           & 0.292 & (0.034)  &   \\ \addlinespace
&$\gamma_2$              & 0.2           & 0.180 & (0.058)  &   \\ \addlinespace
&$\gamma_3$              & 0.1           & 0.096 & (0.037)  &   \\ \addlinespace
&$\beta$                 & -0.6          & -0.572 & (0.017) &   \\ \addlinespace
&$c^0$                   & -2            & -1.953 & (0.047) &   \\ \addlinespace
&$c^1$                   & -2.5          & -2.474 & (0.052) &   \\ 
                         &               &        &         &  \\ 
&Log-Likelihood (True value)      &      & \multicolumn{2}{c}{-20892} &   \\ 
&Log-Likelihood (Estimates)       &      & \multicolumn{2}{c}{-20885} &   \\ \addlinespace
&RMSE                    &               & \multicolumn{2}{c}{0.048}  &   \\ 
\bottomrule 
\end{tabular}
\end{tabularx}
\smallskip
\begin{tablenotes}
    \item \footnotesize \emph{Notes:} We simulate data for 2,000 consumers and report results averaged over 100 estimations using different random seeds, each based on 1,000 simulation draws. Standard deviations of the estimates across simulations are reported in parentheses. 
\end{tablenotes}
\end{threeparttable}

%% file: Tables/tab_literature.tex
\begin{threeparttable}[H]
\caption{Overview of Recent Empirical Studies applying Sequential Search Models}
\centering 
\footnotesize{
\begin{tabular}{llllll}
\toprule
\textbf{Study} & \textbf{Model} & \textbf{Data} & \textbf{Estimation Methods} & \textbf{In This Paper} \\
\midrule
\cite{chen2017sequential}                           & Weitzman                                       & Search Order and Purchase & Crude Frequency Simulator                               & \multirow{8}{*}{Sections \ref{sec:OSR_representation} - \ref{sec:empirics}}              \\
\cite{ghose2019modeling}                            & Weitzman                                       & Search Order and Purchase & Crude Frequency Simulator                               &                                           \\
\cite{yavorsky2021consumer}                         & Weitzman                                       & Search Order and Purchase & Kernel-smoothed Frequency Simulator                     &                                           \\
\cite{jiang2021consumer}                            & Weitzman                                       & Search Order and Purchase & Rank-based GHK                                               &                                           \\
\cite{morozov2021estimation}                        & Weitzman                                       & Search Order and Purchase & Importance Sampling                                      &                                           \\
\cite{chung2025simulated}                           & Weitzman                                       & Search Order and Purchase & Rank-based GHK                                               &                                           \\
\cite{morozov2023measuring}                         & Weitzman                                       & Search Order and Purchase & Bayesian MCMC                                            &                                           \\
\cite{onzo2025bayesian}                             & Weitzman                                       & Search Order and Purchase & Bayesian MCMC                                            &                                           \\
\cite{ursu2025online}                               & Weitzman                                       & Search Order and Purchase & Kernel-smoothed Frequency Simulator               &                                           \\
\midrule
\cite{honka2017simultaneous}                   & Weitzman                                       & Searched Set and Purchase & Kernel-smoothed Frequency Simulator                     & \multirow{6}{*}{Section \ref{subsec:extension_incomplete}} \\
\cite{ursu2018power}                           & Weitzman                                       & Searched Set and Purchase & Kernel-smoothed Frequency Simulator                     &                                           \\
\cite{jolivet2019consumer}                     & Weitzman                                       & Purchase                  & Maxmin Frequency Simulator                              &                                           \\
\cite{moraga2023consumer}                      & Weitzman                                       & \begin{tabular}[c]{@{}l@{}}Purchase, with moment \\ conditions from search order \end{tabular} & Multinomial Logit  &                                           \\
\cite{kaye2024personalization}                 & Weitzman                                       & Searched Set and Purchase & Kernel-smoothed Frequency Simulator                     &                                           \\
\cite{compiani2024online}                      & Weitzman                                       & Searched Set and Purchase & Exploded Logit                                          &                                           \\
\midrule
\cite{zhang2024product}                        & Sequential Search with Discovery                   & Search Order and Purchase & Kernel-smoothed Frequency Simulator                     &  Section \ref{subsec:extension_discovery}                                         \\
\midrule
\cite{gibbard2023search}                      & Two-stage Sequential Search                     & Searched Set and Purchase & Rank-based GHK                                              & Appendix \ref{app:ts_implementation}                                \\
\cite{greminger2024heterogeneous}             & Sequential Search with Discovery                & Searched Set and Purchase & \begin{tabular}[c]{@{}l@{}}Crude Frequency Simulator, with most \\ conditional probabilities computed analytically\end{tabular}                           & Appendix \ref{app:spd_incomplete_implementation}                                \\
\midrule
\cite{ursu2023search}                         & Other Additional Action                         & Search Order and Purchase & Kernel-smoothed Frequency Simulator                     &       \multirow{4}{*}{Section \ref{sec:discussion}}     \\
\cite{klein2024Do}                            & Multi-round Weitzman                            & Search Order and Purchase & Modified Rank-based GHK        &            \\
\cite{elberg2019dynamic}                      & Belief Transition                               & Purchase                  & Kernel-smoothed Frequency Simulator                           &                   \\
\cite{gardete2024multiattribute}                      & Belief Transition & Search Order and Purchase & Exploded Logit                                          &                                           \\
\bottomrule
\end{tabular}}
\vspace{-0.1cm}
\begin{tablenotes}
\footnotesize
\item \emph{Notes:} The table may be incomplete. We include only studies that develop or apply empirical methods for models satisfying the Independence assumption. Accordingly, we exclude research involving belief updating or interdependent actions, such as forward-looking consumer learning.
\end{tablenotes}
\end{threeparttable}

%% file: TikzPics/figure_branching.tex
\begin{tikzpicture}[scale=0.6]
    \tikzstyle{every node}=[font=\small, scale = 0.76]
    \draw (-2.5,9) node[rect1, fill = white, text width = 1.8cm] (A_I_1){Which to inspect?}; 
    \draw (A_I_1) -- (-5.5,7) node[circ2, fill = green!10!white, text width = 0.5cm] (A_I_a_1){$I_1$}; 
    \draw (A_I_1) -- (-4,7) node[circ2, fill = green!10!white, text width = 0.5cm] (A_I_b_1){$I_2$}; 
    \draw (A_I_1) -- (-2.5,7) node[circ2, fill = blue!20!white, text width = 0.5cm] (A_I_c_1){$I_3$}; 
    \draw (A_I_1) -- (-1,7) node[circ2, fill = green!10!white, text width = 0.5cm] (A_I_d_1){$I_4$}; 
    \draw (A_I_1) -- (0.5,7) node[circ2, fill = green!10!white, text width = 0.5cm] (A_I_e_1){$I_5$}; 
    
    \draw (A_I_c_1) -- (-2.5,5.5) node[rect1, fill = white, text width = 1.8cm] (A_S_1){Stop?}; 
    \draw (A_S_1) -- (-5,4) node[rect1, fill = white, text width = 1.8cm] (A_I_2){Which to Inspect?}; 
    \draw (A_I_2) -- (-7.3,2) node[circ2, fill = green!10!white, text width = 0.5cm] (A_I_a_2){$I_1$}; 
    \draw (A_I_2) -- (-5.8,2) node[circ2, fill = green!10!white, text width = 0.5cm] (A_I_b_2){$I_2$}; 
    \draw (A_I_2) -- (-4.2,2) node[circ2, fill = green!10!white, text width = 0.5cm] (A_I_d_2){$I_4$}; 
    \draw (A_I_2) -- (-2.7,2) node[circ2, fill = blue!20!white, text width = 0.5cm] (A_I_e_2){$I_5$}; 
    \draw (A_S_1) -- (0,4) node[rect1, fill = white, text width = 1.8cm] (A_P_2){Which to Purchase?}; 
    \draw (A_P_2) -- (0,2) node[circ2, fill = green!10!white, text width = 0.5cm] (A_P_c_2){$P_3$}; 
    
    \draw (A_I_e_2) -- (-2.7,0.5) node[rect1, fill = white, text width = 1.8cm] (A_S_2){Stop?}; 
    \draw (A_S_2) -- (-5.7,-1) node[rect1, fill = white, text width = 1.8cm] (A_I_3){Which to Inspect?}; 
    \draw (A_I_3) -- (-7.2,-3) node[circ2, fill = green!10!white, text width = 0.5cm] (A_I_a_3){$I_1$}; 
    \draw (A_I_3) -- (-5.7,-3) node[circ2, fill = green!10!white, text width = 0.5cm] (A_I_b_3){$I_2$}; 
    \draw (A_I_3) -- (-4.2,-3) node[circ2, fill = blue!20!white, text width = 0.5cm] (A_I_d_3){$I_4$}; 
    \draw (A_S_2) -- (0.3,-1) node[rect1, fill = white, text width = 1.8cm] (A_P_3){Which to Purchase?}; 
    \draw (A_P_3) -- (-0.5,-3) node[circ2, fill = green!10!white, text width = 0.5cm] (A_P_c_3){$P_3$}; 
    \draw (A_P_3) -- (1.1,-3) node[circ2, fill = green!10!white, text width = 0.5cm] (A_P_e_3){$P_5$}; 

    \draw (A_I_d_3) -- (-4.2,-4.5) node[rect1, fill = white, text width = 1.8cm] (A_S_3){Stop?}; 
    \draw (A_S_3) -- (-7.2,-6) node[rect1, fill = white, text width = 1.8cm] (A_I_4){Which to Inspect?}; 
    \draw (A_I_4) -- (-8,-8) node[circ2, fill = green!10!white, text width = 0.5cm] (A_I_a_4){$I_1$}; 
    \draw (A_I_4) -- (-6.4,-8) node[circ2, fill = green!10!white, text width = 0.5cm] (A_I_b_4){$I_2$}; 
    \draw (A_S_3) -- (-1.2,-6) node[rect1, fill = white, text width = 1.8cm] (A_P_4){Which to Purchase?}; 
    \draw (A_P_4) -- (-2.7,-8) node[circ2, fill = green!10!white, text width = 0.5cm] (A_P_c_4){$P_3$}; 
    \draw (A_P_4) -- (-1.2,-8) node[circ2, fill = blue!20!white, text width = 0.5cm] (A_P_d_4){$P_4$}; 
    \draw (A_P_4) -- (0.3,-8) node[circ2, fill = green!10!white, text width = 0.5cm] (A_P_e_4){$P_5$}; 

    \draw (A_P_d_4) -- (-1.2,-10) node[rect1, fill = white, text width = 1.8cm] (A_P){End}; 

    \draw (-2.5,-11.5) node[font = \large](){The Sequential Search Process};
    
    \draw (7.5,10) -- (4.5,8.5) node[circ2, fill = green!10!white, text width = 0.5cm] (B_I_a_1){$I_1$}; 
    \draw (7.5,10) -- (6,8.5) node[circ2, fill = green!10!white, text width = 0.5cm] (B_I_b_1){$I_2$}; 
    \draw (7.5,10) -- (7.5,8.5) node[circ2, fill = blue!20!white, text width = 0.5cm] (B_I_c_1){$I_3$}; 
    \draw (7.5,10) -- (9,8.5) node[circ2, fill = green!10!white, text width = 0.5cm] (B_I_d_1){$I_4$}; 
    \draw (7.5,10) -- (10.5,8.5) node[circ2, fill = green!10!white, text width = 0.5cm] (B_I_e_1){$I_5$}; 
    \draw (B_I_a_1) -- (4.5,7) node[circ2, fill = white, text width = 0.5cm] (B_P_a_1){$P_1$}; 
    \draw (B_I_b_1) -- (6,7) node[circ2, fill = white, text width = 0.5cm] (B_P_b_1){$P_2$}; 
    \draw (B_I_c_1) -- (7.5,7) node[circ2, fill = white, text width = 0.5cm] (B_P_c_1){$P_3$}; 
    \draw (B_I_d_1) -- (9,7) node[circ2, fill = white, text width = 0.5cm] (B_P_d_1){$P_4$}; 
    \draw (B_I_e_1) -- (10.5,7) node[circ2, fill = white, text width = 0.5cm] (B_P_e_1){$P_5$}; 

    \draw (7.5,5) -- (4.5,3.5) node[circ2, fill = green!10!white, text width = 0.5cm] (B_I_a_2){$I_1$}; 
    \draw (7.5,5) -- (6,3.5) node[circ2, fill = green!10!white, text width = 0.5cm] (B_I_b_2){$I_2$}; 
    \draw (7.5,5) -- (7.5,3.5) node[circ3, fill = white, text width = 0.5cm] (B_I_c_2){}; 
    \draw (7.5,5) -- (9,3.5) node[circ2, fill = green!10!white, text width = 0.5cm] (B_I_d_2){$I_4$}; 
    \draw (7.5,5) -- (10.5,3.5) node[circ2, fill = blue!20!white, text width = 0.5cm] (B_I_e_2){$I_5$}; 
    \draw (B_I_a_2) -- (4.5,2) node[circ2, fill = white, text width = 0.5cm] (B_P_a_2){$P_1$}; 
    \draw (B_I_b_2) -- (6,2) node[circ2, fill = white, text width = 0.5cm] (B_P_b_2){$P_2$}; 
    \draw (B_I_c_2) -- (7.5,2) node[circ2, fill = green!10!white, text width = 0.5cm] (B_P_c_2){$P_3$}; 
    \draw (B_I_d_2) -- (9,2) node[circ2, fill = white, text width = 0.5cm] (B_P_d_2){$P_4$}; 
    \draw (B_I_e_2) -- (10.5,2) node[circ2, fill = white, text width = 0.5cm] (B_P_e_2){$P_5$}; 

    \draw (7.5,0) -- (4.5,-1.5) node[circ2, fill = green!10!white, text width = 0.5cm] (B_I_a_3){$I_1$}; 
    \draw (7.5,0) -- (6,-1.5) node[circ2, fill = green!10!white, text width = 0.5cm] (B_I_b_3){$I_2$}; 
    \draw (7.5,0) -- (7.5,-1.5) node[circ3, fill = white, text width = 0.5cm] (B_I_c_3){}; 
    \draw (7.5,0) -- (9,-1.5) node[circ2, fill = blue!20!white, text width = 0.5cm] (B_I_d_3){$I_4$}; 
    \draw (7.5,0) -- (10.5,-1.5) node[circ3, fill = white, text width = 0.5cm] (B_I_e_3){}; 
    \draw (B_I_a_3) -- (4.5,-3) node[circ2, fill = white, text width = 0.5cm] (B_P_a_3){$P_1$}; 
    \draw (B_I_b_3) -- (6,-3) node[circ2, fill = white, text width = 0.5cm] (B_P_b_3){$P_2$}; 
    \draw (B_I_c_3) -- (7.5,-3) node[circ2, fill = green!10!white, text width = 0.5cm] (B_P_c_3){$P_3$}; 
    \draw (B_I_d_3) -- (9,-3) node[circ2, fill = white, text width = 0.5cm] (B_P_d_3){$P_4$}; 
    \draw (B_I_e_3) -- (10.5,-3) node[circ2, fill = green!10!white, text width = 0.5cm] (B_P_e_3){$P_5$}; 

    \draw (7.5,-5) -- (4.5,-6.5) node[circ2, fill = green!10!white, text width = 0.5cm] (B_I_a_4){$I_1$}; 
    \draw (7.5,-5) -- (6,-6.5) node[circ2, fill = green!10!white, text width = 0.5cm] (B_I_b_4){$I_2$}; 
    \draw (7.5,-5) -- (7.5,-6.5) node[circ3, fill = white, text width = 0.5cm] (B_I_c_4){}; 
    \draw (7.5,-5) -- (9,-6.5) node[circ3, fill = white, text width = 0.5cm] (B_I_d_4){}; 
    \draw (7.5,-5) -- (10.5,-6.5) node[circ3, fill = white, text width = 0.5cm] (B_I_e_4){}; 
    \draw (B_I_a_4) -- (4.5,-8) node[circ2, fill = white, text width = 0.5cm] (B_P_a_4){$P_1$}; 
    \draw (B_I_b_4) -- (6,-8) node[circ2, fill = white, text width = 0.5cm] (B_P_b_4){$P_2$}; 
    \draw (B_I_c_4) -- (7.5,-8) node[circ2, fill = green!10!white, text width = 0.5cm] (B_P_c_4){$P_3$}; 
    \draw (B_I_d_4) -- (9,-8) node[circ2, fill = blue!20!white, text width = 0.5cm] (B_P_d_4){$P_4$}; 
    \draw (B_I_e_4) -- (10.5,-8) node[circ2, fill = green!10!white, text width = 0.5cm] (B_P_e_4){$P_5$}; 

    \draw (B_P_d_4) -- (9,-10) node[rect1, fill = white, text width = 1.8cm] (B_P){End}; 

    \draw (7.5,-11.5) node[font = \large](){The Branching Project Representation};

    \draw (19.5,-5) node[circ2, fill = blue!20!white, text width = 0.5cm] (P_D){$P_4$}; 
    \draw (16,-1.5) node[circ2, fill = blue!20!white, text width = 0.5cm] (I_D){$I_4$}; 
    \draw (15.4,-8) node[circ2, fill = green!10!white, text width = 0.5cm] (I_A){$I_1$}; 
    \draw [red, ->, thick] (I_D) -- (I_A); 
    \draw [red, ->, thick] (P_D) -- (I_A); 
    \draw (16.8,-8) node[circ2, fill = green!10!white, text width = 0.5cm] (I_B){$I_2$}; 
    \draw [red, ->, thick] (I_D) -- (I_B); 
    \draw [red, ->, thick] (P_D) -- (I_B); 
    \draw (18.2,-8) node[circ2, fill = green!10!white, text width = 0.5cm] (P_C){$P_3$}; 
    \draw [red, ->, thick] (I_D) -- (P_C); 
    \draw [red, ->, thick] (P_D) -- (P_C); 
    \draw (19.6,-8) node[circ2, fill = green!10!white, text width = 0.5cm] (P_E){$P_5$}; 
    \draw [red, ->, thick] (I_D) -- (P_E); 
    \draw [red, ->, thick] (P_D) -- (P_E); 

    \draw (16,3.5) node[circ2, fill = blue!20!white, text width = 0.5cm] (I_E){$I_5$}; 
    \draw [red, ->, thick] (I_E) -- (I_D); 
    \draw (16,8.5) node[circ2, fill = blue!20!white, text width = 0.5cm] (I_C){$I_3$}; 
    \draw [red, ->, thick] (I_C) -- (I_E); 

    \draw (17.5,-11.5) node [font = \large](){The Ranking Representation};
\end{tikzpicture}

%% file: Tables/tab_bruteforce.tex
\begin{threeparttable}
\centering
\smallskip
\begin{tabularx}{0.9\linewidth}{@{}Y@{}}
\begin{tabular}{cccc}
\toprule
& Cosine Similarity     & K-S Stat      &  R-Squared      \\ \midrule 
& 99.9999\%             & 0.000469      & 0.999997 \\ 
\bottomrule 
\end{tabular}
\end{tabularx}
\smallskip
\end{threeparttable}

%% file: Tables/tab_iden_Chung.tex
\begin{threeparttable}
\centering
\smallskip
\begin{tabular}{ccccccc}
\toprule
                     & True value  ($\sigma_\varepsilon = 1$) & \multicolumn{5}{c}{Estimates with exponential search cost: $c_{ij} \sim Exp (\lambda)$}                                                                                       \\ \cline{3-7} \addlinespace 
                     &            & $\hat{\sigma}_\varepsilon=0.5$  & $\hat{\sigma}_\varepsilon=0.75$    & $\hat{\sigma}_\varepsilon=1$    & $\hat{\sigma}_\varepsilon=1.25$  
                     & $\hat{\sigma}_\varepsilon=1.5$ \\ \midrule
$\gamma_1$           & 1                                     & 0.496 (0.50)                      & 0.744 (0.74)                    & 0.992 (0.99)                 & 1.240 (1.24)                     & 1.488 (1.49)                   \\\addlinespace
$\gamma_2$           & 0.5                                   & 0.243 (0.49)                      & 0.364 (0.73)                    & 0.486 (0.97)                 & 0.607 (1.21)                    & 0.728 (1.46)                   \\\addlinespace
$\gamma_3$           & -0.2                                  & -0.102 (0.51)                     & -0.154 (0.77)                   & -0.205 (1.03)                & -0.256 (1.28)                   & -0.307 (1.54)                  \\\addlinespace
$\bar{\beta}$        & -0.6                                  & -0.297 (0.50)                     & -0.445 (0.74)                   & -0.594 (0.99)                & -0.742 (1.24)                   & -0.891 (1.49)                  \\\addlinespace
$\sigma_\beta$       & 0.2                                   & 0.106 (0.53)                      & 0.159 (0.80)                    & 0.212 (1.06)                 & 0.265 (1.33)                    & 0.318 (1.59)                   \\\addlinespace
$1/\lambda$          & 0.8                                   & 0.391 (0.49)                      & 0.587 (0.73)                    & 0.783 (0.98)                 & 0.979 (1.22)                    & 1.174 (1.47)                   \\\addlinespace
                     &            &                           &                              &                           &                             &                             \\
Log-L                &            & 27571.73                  & 27571.73                     & 27571.73                  & 27571.73                     & 27571.73                    \\ \addlinespace \toprule
                     & True value  ($\sigma_\varepsilon = 1$) & \multicolumn{5}{c}{Estimates with log-normal search cost: $\log c_{ij} \sim N(c_0,1)$}                                                                                       \\ \cline{3-7} \addlinespace
                     &            & $\hat{\sigma}_\varepsilon=0.5$  & $\hat{\sigma}_\varepsilon=0.75$    & $\hat{\sigma}_\varepsilon=1$    & $\hat{\sigma}_\varepsilon=1.25$  
                     & $\hat{\sigma}_\varepsilon=1.5$ \\ \midrule
$\gamma_1$           & 1                                     & 0.491 (0.49)                   & 0.736 (0.74)                    & 0.983 (0.98)                 & 1.227 (1.23)                    & 1.473 (1.47)                   \\\addlinespace
$\gamma_2$           & 0.5                                   & 0.249 (0.50)                   & 0.374 (0.75)                    & 0.498 (1.00)                 & 0.622 (1.24)                    & 0.747 (1.49)                   \\\addlinespace
$\gamma_3$           & -0.2                                  & -0.099 (0.50)                  & -0.149 (0.75)                   & -0.199 (1.00)                & -0.249 (1.25)                   & -0.298 (1.49)                  \\\addlinespace
$\bar{\beta}$        & -0.6                                  & -0.294 (0.49)                  & -0.441 (0.74)                   & -0.589 (0.98)                & -0.736 (1.23)                   & -0.883 (1.47)                  \\\addlinespace
$\sigma_\beta$       & 0.2                                   & 0.102 (0.51)                   & 0.153 (0.77)                    & 0.205 (1.03)                 & 0.256 (1.28)                    & 0.306 (1.53)                   \\\addlinespace
$c_0$                & -1.5                                  & -2.219 (1.48)                  & -1.817 (1.21)                   & -1.530 (1.02)                 & -1.306 (0.87)                   & -1.124 (0.75)                  \\\addlinespace
\multicolumn{1}{l}{} &            &                           &                              &                           &                              &                             \\
Log-L                &            & 24535.65                  & 24534.74                     & 24534.75                  & 24534.75                     & 24534.75          \\ \bottomrule         
\end{tabular}
\smallskip
\begin{tablenotes}
    \item \footnotesize \emph{Notes:} We simulate data for 5,000 consumers and report results averaged over 20 estimations using different random seeds, each with 800 error draws. Standard deviations of the mean estimates across simulations are omitted in the table. The ratios of the mean estimates to the true values are reported in parentheses. 
\end{tablenotes}
\end{threeparttable}

%% file: Tables/tab_iden_hete.tex
\begin{threeparttable}
\centering
\smallskip
\begin{tabularx}{0.9\linewidth}{@{}Y@{}}
\begin{tabular}{cccrlc}
\toprule
&                        & True value    & \multicolumn{2}{c}{Estimates} &       \\ \midrule
&$\gamma_1$              & 1             & 0.987 & (0.025)  &   \\ \addlinespace
&$\gamma_2$              & 0.5           & 0.496 & (0.021)  &   \\ \addlinespace
&$\gamma_3$              & 0.2           & 0.203 & (0.021)  &   \\ \addlinespace
&$\bar{\beta}$           & -0.6          & -0.589 & (0.029) &   \\ \addlinespace
&$\sigma_\beta$          & 0.2           & 0.162 & (0.089) &   \\ \addlinespace
&$\log c$                & -1.5          & -1.502 & (0.021) &   \\ 
                         &               &        &         &  \\ 
&Log-Likelihood (True value)      &      & \multicolumn{2}{c}{-22365} &   \\ \addlinespace
&Log-Likelihood (Estimates)       &      & \multicolumn{2}{c}{-22361} &   \\ 
\bottomrule 
\end{tabular}
\end{tabularx}
\smallskip
\begin{tablenotes}
    \item \footnotesize \emph{Notes:} We simulate data for 2,000 consumers and report results averaged over 50 estimations using different random seeds, each based on 800 simulation draws. Standard deviations of the estimates across simulations are reported in parentheses. 
\end{tablenotes}
\end{threeparttable}

%% file: TikzPics/figure_nopath.tex
\begin{tikzpicture}[scale=0.55]
    \tikzstyle{every node}=[font=\normalsize, scale = 0.85]

    \draw (-4.5,0) -- (-2.7,3.2) node[circ2, fill = gray!20!white, text width = 0.5cm] (I_a_1){$I_1$}; 
    \draw (-4.5,0) -- (-2.7,1.6) node[circ2, fill = gray!20!white, text width = 0.5cm] (I_b_1){$I_2$}; 
    \draw (-4.5,0) -- (-2.7,0) node[circ2, fill = blue!20!white, text width = 0.5cm] (E_c_1){$E_3$}; 
    \draw (-4.5,0) -- (-2.7,-1.6) node[circ2, fill = gray!20!white, text width = 0.5cm] (I_d_1){$I_4$}; 
    \draw (-4.5,0) -- (-2.7,-3.2) node[circ2, fill = green!10!white, text width = 0.5cm] (I_e_1){$I_5$}; 

    \draw (I_a_1) -- (-0.9,3.2) node[circ2, fill = green!10!white, text width = 0.5cm] (P_a_1){$P_1$}; 
    \draw (I_b_1) -- (-0.9,1.6) node[circ2, fill = green!10!white, text width = 0.5cm] (P_b_1){$P_2$}; 
    \draw (E_c_1) -- (-0.9,0) node[circ2, fill = white, text width = 0.5cm] (P_c_1){$P_3$}; 
    \draw (I_d_1) -- (-0.9,-1.6) node[circ2, fill = green!10!white, text width = 0.5cm] (P_d_1){$P_4$};
    \draw (I_e_1) -- (-0.9,-3.2) node[circ2, fill = white, text width = 0.5cm] (P_e_1){$P_5$}; 

    \draw (0.5,0) -- (2.3,3.2) node[circ3, fill = white, text width = 0.5cm] (I_a_2){}; 
    \draw (0.5,0) -- (2.3,1.6) node[circ3, fill = white, text width = 0.5cm] (I_b_2){}; 
    \draw (0.5,0) -- (2.3,0) node[circ3, fill = white, text width = 0.5cm] (E_c_2){}; 
    \draw (0.5,0) -- (2.3,-1.6) node[circ3, fill = white, text width = 0.5cm] (I_d_2){}; 
    \draw (0.5,0) -- (2.3,-3.2) node[circ2, fill = green!10!white, text width = 0.5cm] (I_e_2){$I_5$}; 

    \draw (I_a_2) -- (4.1,3.2) node[circ2, fill = green!10!white, text width = 0.5cm] (P_a_2){$P_1$}; 
    \draw (I_b_2) -- (4.1,1.6) node[circ2, fill = green!10!white, text width = 0.5cm] (P_b_2){$P_2$}; 
    \draw (E_c_2) -- (4.1,0) node[circ2, fill = blue!20!white, text width = 0.5cm] (P_c_2){$P_3$}; 
    \draw (I_d_2) -- (4.1,-1.6) node[circ2, fill = green!10!white, text width = 0.5cm] (P_d_2){$P_4$};
    \draw (I_e_2) -- (4.1,-3.2) node[circ2, fill = white, text width = 0.5cm] (P_e_2){$P_5$}; 
    
    \draw (-2, -4.5) node[font=\normalsize] (Step_1){Step 1};

    \draw (0, -6) node[font=\normalsize] {The Branching Project Representation};

    \draw (13,0) node[circ1, fill = blue!20!white, text width = 0.4cm] (E_C_2){$E_3$}; 
    \draw (10,2.5) node[circ1, fill = blue!20!white, text width = 0.4cm] (I_A_2){$I_1$}; 
    \draw (12,2.5) node[circ1, fill = blue!20!white, text width = 0.4cm] (I_B_2){$I_2$}; 
    \draw (14,2.5) node[circ1, fill = blue!20!white, text width = 0.4cm] (I_D_2){$I_4$}; 

    \draw (16,0) node[circ1, fill = blue!20!white, text width = 0.4cm] (P_C_2){$P_3$}; 
    
    \draw (10,-3.5) node[circ1, fill = green!10!white, text width = 0.4cm] (P_A_2){$P_1$}; 
    \draw (12,-3.5) node[circ1, fill = green!10!white, text width = 0.4cm] (P_B_2){$P_2$}; 
    \draw (14,-3.5) node[circ1, fill = green!10!white, text width = 0.4cm] (P_D_2){$P_4$}; 
    \draw (16,-3.5) node[circ1, fill = green!10!white, text width = 0.4cm] (I_E_2){$I_5$}; 
    \draw [->, red, thick] (P_C_2) -- (P_A_2);
    \draw [->, red, thick] (P_C_2) -- (P_B_2);
    \draw [->, red, thick] (P_C_2) -- (P_D_2);
    \draw [->, red, thick] (P_C_2) -- (I_E_2);
    \draw [->, red, thick] (E_C_2) -- (P_A_2);
    \draw [->, red, thick] (E_C_2) -- (P_B_2);
    \draw [->, red, thick] (E_C_2) -- (P_D_2);
    \draw [->, red, thick] (E_C_2) -- (I_E_2);
    \draw [->, red, thick] (I_A_2) -- (E_C_2);
    \draw [->, red, thick] (I_B_2) -- (E_C_2);
    \draw [->, red, thick] (I_D_2) -- (E_C_2);

    \draw (3, -4.5) node[font=\normalsize] (Step_2){Step 2};
    
    \draw (13, -6) node[font=\normalsize] {The Ranking Representation};
\end{tikzpicture}

%% file: TikzPics/figure_nopurchase.tex
\begin{tikzpicture}[scale=0.55]
    \tikzstyle{every node}=[font=\normalsize, scale = 0.85]
    \draw (-10,0) -- (-8.2,3.2) node[circ2, fill = blue!20!white, text width = 0.5cm] (I_a_1){$I_1$}; 
    \draw (-10,0) -- (-8.2,1.6) node[circ2, fill = green!10!white, text width = 0.5cm] (I_b_1){$I_2$};
    \draw (-10,0) -- (-8.2,0) node[circ2, fill = green!10!white, text width = 0.5cm] (I_c_1){$I_3$}; 
    \draw (-10,0) -- (-8.2,-1.6) node[circ2, fill = green!10!white, text width = 0.5cm] (I_d_1){$I_4$}; 
    \draw (-10,0) -- (-8.2,-3.2) node[circ2, fill = green!10!white, text width = 0.5cm] (I_e_1){$I_5$};

    \draw (I_a_1) -- (-6.4,3.2) node[circ2, fill = white, text width = 0.5cm] (P_a_1){$P_1$};  
    \draw (I_b_1) -- (-6.4,1.6) node[circ2, fill = white, text width = 0.5cm] (P_b_1){$P_2$}; 
    \draw (I_c_1) -- (-6.4,0) node[circ2, fill = white, text width = 0.5cm] (P_a_1){$P_3$};  
    \draw (I_d_1) -- (-6.4,-1.6) node[circ2, fill = white, text width = 0.5cm] (P_b_1){$P_4$}; 
    \draw (I_e_1) -- (-6.4,-3.2) node[circ2, fill = white, text width = 0.5cm] (P_e_1){$P_5$}; 

    \draw (-7.5, -4.5) node[font=\normalsize] (Step_1){Step 1};
    
    \draw (-5,0) -- (-3.2,3.2) node[circ3, fill = white, text width = 0.5cm] (I_a_2){}; 
    \draw (-5,0) -- (-3.2,1.6) node[circ2, fill = blue!20!white, text width = 0.5cm] (I_b_2){$I_2$};
    \draw (-5,0) -- (-3.2,0) node[circ2, fill = green!10!white, text width = 0.5cm] (I_c_2){$I_3$}; 
    \draw (-5,0) -- (-3.2,-1.6) node[circ2, fill = green!10!white, text width = 0.5cm] (I_d_2){$I_4$}; 
    \draw (-5,0) -- (-3.2,-3.2) node[circ2, fill = green!10!white, text width = 0.5cm] (I_e_2){$I_5$}; 

    \draw (I_a_2) -- (-1.4,3.2) node[circ2, fill = green!10!white, text width = 0.5cm] (P_a_2){$P_1$};  
    \draw (I_b_2) -- (-1.4,1.6) node[circ2, fill = white, text width = 0.5cm] (P_b_2){$P_2$}; 
    \draw (I_c_2) -- (-1.4,0) node[circ2, fill = white, text width = 0.5cm] (P_c_2){$P_3$};  
    \draw (I_d_2) -- (-1.4,-1.6) node[circ2, fill = white, text width = 0.5cm] (P_d_2){$P_4$}; 
    \draw (I_e_2) -- (-1.4,-3.2) node[circ2, fill = white, text width = 0.5cm] (P_e_2){$P_5$}; 

    \draw (-2.5, -4.5) node[font=\normalsize] (Step_2){Step 2};

    \draw (0,0) -- (1.8,3.2) node[circ3, fill = white, text width = 0.5cm] (I_a_3){}; 
    \draw (0,0) -- (1.8,1.6) node[circ3, fill = white, text width = 0.5cm] (I_b_3){};
    \draw (0,0) -- (1.8,0) node[circ2, fill = blue!20!white, text width = 0.5cm] (I_c_3){$I_3$}; 
    \draw (0,0) -- (1.8,-1.6) node[circ2, fill = green!10!white, text width = 0.5cm] (I_d_3){$I_4$}; 
    \draw (0,0) -- (1.8,-3.2) node[circ2, fill = green!10!white, text width = 0.5cm] (I_e_3){$I_5$}; 

    \draw (I_a_3) -- (3.6,3.2) node[circ2, fill = green!10!white, text width = 0.5cm] (P_a_3){$P_1$};  
    \draw (I_b_3) -- (3.6,1.6) node[circ2, fill = green!10!white, text width = 0.5cm] (P_b_3){$P_2$}; 
    \draw (I_c_3) -- (3.6,0) node[circ2, fill = white, text width = 0.5cm] (P_c_3){$P_3$};  
    \draw (I_d_3) -- (3.6,-1.6) node[circ2, fill = white, text width = 0.5cm] (P_d_3){$P_4$}; 
    \draw (I_e_3) -- (3.6,-3.2) node[circ2, fill = white, text width = 0.5cm] (P_e_3){$P_5$}; 

    \draw (2.5, -4.5) node[font=\normalsize] (Step_3){Step 3};

    \draw (5,0) -- (6.8,3.2) node[circ3, fill = white, text width = 0.5cm] (I_a_4){}; 
    \draw (5,0) -- (6.8,1.6) node[circ3, fill = white, text width = 0.5cm] (I_b_4){};
    \draw (5,0) -- (6.8,0) node[circ3, fill = white, text width = 0.5cm] (I_c_4){}; 
    \draw (5,0) -- (6.8,-1.6) node[circ2, fill = blue!20!white, text width = 0.5cm] (I_d_4){$I_4$}; 
    \draw (5,0) -- (6.8,-3.2) node[circ2, fill = green!10!white, text width = 0.5cm] (I_e_4){$I_5$}; 

    \draw (I_a_4) -- (8.6,3.2) node[circ2, fill = green!10!white, text width = 0.5cm] (P_a_4){$P_1$};  
    \draw (I_b_4) -- (8.6,1.6) node[circ2, fill = green!10!white, text width = 0.5cm] (P_b_4){$P_2$}; 
    \draw (I_c_4) -- (8.6,0) node[circ2, fill = green!10!white, text width = 0.5cm] (P_c_4){$P_3$};  
    \draw (I_d_4) -- (8.6,-1.6) node[circ2, fill = white, text width = 0.5cm] (P_d_4){$P_4$}; 
    \draw (I_e_4) -- (8.6,-3.2) node[circ2, fill = white, text width = 0.5cm] (P_e_4){$P_5$}; 

    \draw (7.5, -4.5) node[font=\normalsize] (Step_4){Step 4};

    \draw (10,0) -- (11.8,3.2) node[circ3, fill = white, text width = 0.5cm] (I_a_5){}; 
    \draw (10,0) -- (11.8,1.6) node[circ3, fill = white, text width = 0.5cm] (I_b_5){};
    \draw (10,0) -- (11.8,0) node[circ3, fill = white, text width = 0.5cm] (I_c_5){}; 
    \draw (10,0) -- (11.8,-1.6) node[circ3, fill = white, text width = 0.5cm] (I_d_5){}; 
    \draw (10,0) -- (11.8,-3.2) node[circ2, fill = gray!20!white, text width = 0.5cm] (I_e_5){$I_5$}; 

    \draw (I_a_5) -- (13.6,3.2) node[circ2, fill = green!10!white, text width = 0.5cm] (P_a_5){$P_1$};  
    \draw (I_b_5) -- (13.6,1.6) node[circ2, fill = green!10!white, text width = 0.5cm] (P_b_5){$P_2$}; 
    \draw (I_c_5) -- (13.6,0) node[circ2, fill = green!10!white, text width = 0.5cm] (P_c_5){$P_3$};  
    \draw (I_d_5) -- (13.6,-1.6) node[circ2, fill = green!10!white, text width = 0.5cm] (P_d_5){$P_4$}; 
    \draw (I_e_5) -- (13.6,-3.2) node[circ2, fill = white, text width = 0.5cm] (P_e_5){$P_5$}; 

    \draw (12.5, -4.5) node[font=\normalsize] (Step_5){Step 5};
    
    \draw (2.5, -6) node[font=\normalsize] {The Branching Project Representation};

    \draw (-1,-8) node[circ1, fill = blue!20!white, text width = 0.4cm] (I_A_2){$I_1$}; 
    \draw (1,-8.9) node[circ1, fill = blue!20!white, text width = 0.4cm] (I_B_2){$I_2$}; 
    \draw (3,-9.8) node[circ1, fill = blue!20!white, text width = 0.4cm] (I_C_2){$I_3$}; 
    \draw (5,-10.7) node[circ1, fill = blue!20!white, text width = 0.4cm] (I_D_2){$I_4$}; 
    \draw (0,-13) node[circ1, fill = green!10!white, text width = 0.4cm] (P_D_2){$P_4$};
    \draw (2,-13) node[circ1, fill = green!10!white, text width = 0.4cm] (P_A_2){$P_1$};
    \draw (4,-13) node[circ1, fill = green!10!white, text width = 0.4cm] (P_B_2){$P_2$}; 
    \draw (6,-13) node[circ1, fill = green!10!white, text width = 0.4cm] (P_C_2){$P_3$}; 
    \draw (3,-15.3) node[circ1, fill = gray!20!white, text width = 0.4cm] (I_E_2){$I_5$}; 
    
    \draw [->, red, thick] (I_A_2) -- (I_B_2);
    \draw [->, red, thick] (I_B_2) -- (I_C_2);
    \draw [->, red, thick] (I_C_2) -- (I_D_2);
    \draw [->, red, thick] (I_D_2) -- (P_A_2);
    \draw [->, red, thick] (I_D_2) -- (P_B_2);
    \draw [->, red, thick] (I_D_2) -- (P_C_2);

    \draw [blue, thick, dashed] (-1, -13.9) -- (-1, -12.1) -- (7, -12.1) -- (7, -13.9) -- (-1, -13.9); 
    \draw [blue, thick, ->] (3,-13.9) -- (I_E_2); 
    
    \draw (3, -17) node[font=\normalsize] {The Ranking Representation};
\end{tikzpicture}

%% file: TikzPics/figure_firstview.tex
\begin{tikzpicture}[scale=0.55]
    \tikzstyle{every node}=[font=\normalsize, scale = 0.85]
    \draw (-5,0) -- (-3.2,3.2) node[circ2, fill = blue!20!white, text width = 0.5cm] (I_a_1){$I_1$}; 
    \draw (-5,0) -- (-3.2,1.6) node[circ2, fill = green!10!white, text width = 0.5cm] (E_b_1){$E_2$}; 
    \draw (-5,0) -- (-3.2,0) node[circ2, fill = green!10!white, text width = 0.5cm] (I_c_1){$I_3$}; 
    \draw (-5,0) -- (-3.2,-1.6) node[circ2, fill = green!10!white, text width = 0.5cm] (E_d_1){$E_4$}; 
    \draw (-5,0) -- (-3.2,-3.2) node[circ2, fill = green!10!white, text width = 0.5cm] (E_e_1){$E_5$}; 

    \draw (I_a_1) -- (-1.6,3.2) node[circ2, fill = white, text width = 0.5cm] (B_P_a_1){$P_1$};  
    \draw (I_c_1) -- (-1.6,0) node[circ2, fill = white, text width = 0.5cm] (B_P_c_1){$P_3$};  

    \draw (-3, -4.5) node[font=\normalsize] (Step_1){Step 1};
    
    \draw (0,0) -- (1.8,3.2) node[circ3, fill = white, text width = 0.5cm] (I_a_2){}; 
    \draw (0,0) -- (1.8,1.6) node[circ2, fill = green!10!white, text width = 0.5cm] (E_b_2){$E_2$}; 
    \draw (0,0) -- (1.8,0) node[circ2, fill = blue!20!white, text width = 0.5cm] (I_c_2){$I_3$}; 
    \draw (0,0) -- (1.8,-1.6) node[circ2, fill = green!10!white, text width = 0.5cm] (E_d_2){$E_4$}; 
    \draw (0,0) -- (1.8,-3.2) node[circ2, fill = green!10!white, text width = 0.5cm] (E_e_2){$E_5$}; 

    \draw (I_a_2) -- (3.4,3.2) node[circ2, fill = green!10!white, text width = 0.5cm] (B_P_a_2){$P_1$};  
    \draw (I_c_2) -- (3.4,0) node[circ2, fill = white, text width = 0.5cm] (B_P_c_2){$P_3$}; 

    \draw (2, -4.5) node[font=\normalsize] (Step_2){Step 2};

    \draw (5,0) -- (6.8,3.2) node[circ3, fill = white, text width = 0.5cm] (I_a_3){}; 
    \draw (5,0) -- (6.8,1.6) node[circ2, fill = green!10!white, text width = 0.5cm] (E_b_3){$E_2$}; 
    \draw (5,0) -- (6.8,0) node[circ3, fill = white, text width = 0.5cm] (I_c_3){}; 
    \draw (5,0) -- (6.8,-1.6) node[circ2, fill = green!10!white, text width = 0.5cm] (E_d_3){$E_4$}; 
    \draw (5,0) -- (6.8,-3.2) node[circ2, fill = green!10!white, text width = 0.5cm] (E_e_3){$E_5$}; 

    \draw (I_a_3) -- (8.4,3.2) node[circ2, fill = green!10!white, text width = 0.5cm] (B_P_a_3){$P_1$};  
    \draw (I_c_3) -- (8.4,0) node[circ2, fill = blue!20!white, text width = 0.5cm] (B_P_c_3){$P_3$};  

    \draw (7, -4.5) node[font=\normalsize] (Step_3){Step 3};

    \draw (2, -6) node[font=\normalsize] {The Branching Project Representation};

    \draw (15.5,0) node[circ1, fill = blue!20!white, text width = 0.4cm] (I_C_2){$I_3$}; 
    \draw (18.5,0) node[circ1, fill = blue!20!white, text width = 0.4cm] (P_C_2){$P_3$}; 
        
    \draw (14,-3.2) node[circ1, fill = green!10!white, text width = 0.4cm] (P_A_2){$P_1$}; 
    \draw (16,-3.2) node[circ1, fill = green!10!white, text width = 0.4cm] (E_B_2){$E_2$}; 
    \draw (18,-3.2) node[circ1, fill = green!10!white, text width = 0.4cm] (E_D_2){$E_4$}; 
    \draw (20,-3.2) node[circ1, fill = green!10!white, text width = 0.4cm] (E_E_2){$E_5$}; 
    \draw (14,3.2) node[circ1, fill = blue!20!white, text width = 0.4cm] (I_A_2){$I_1$}; 
    \draw [->, red, thick] (I_A_2) -- (I_C_2);
    \draw [->, red, thick] (I_C_2) -- (P_A_2);
    \draw [->, red, thick] (I_C_2) -- (E_B_2);
    \draw [->, red, thick] (I_C_2) -- (E_D_2);
    \draw [->, red, thick] (I_C_2) -- (E_E_2);
    \draw [->, red, thick] (P_C_2) -- (P_A_2);
    \draw [->, red, thick] (P_C_2) -- (E_B_2);
    \draw [->, red, thick] (P_C_2) -- (E_D_2);
    \draw [->, red, thick] (P_C_2) -- (E_E_2);

    \draw (17, -6) node[font=\normalsize] {The Ranking Representation};
\end{tikzpicture}

%% file: TikzPics/figure_partial.tex
\begin{tikzpicture}[scale=0.55]
    \tikzstyle{every node}=[font=\normalsize, scale = 0.85]
    \draw (-5,0) -- (-3.2,3.2) node[circ2, fill = blue!20!white, text width = 0.5cm] (I_a_1){$I_1$}; 
    \draw (-5,0) -- (-3.2,1.6) node[circ2, fill = green!10!white, text width = 0.5cm] (I_b_1){$I_2$};
    \draw (-5,0) -- (-3.2,0) node[circ2, fill = green!10!white, text width = 0.5cm] (E_c_1){$E_3$}; 
    \draw (-5,0) -- (-3.2,-1.6) node[circ2, fill = green!10!white, text width = 0.5cm] (E_d_1){$E_4$}; 
    \draw (-5,0) -- (-3.2,-3.2) node[circ2, fill = green!10!white, text width = 0.5cm] (I_e_1){$I_5$};

    \draw (I_a_1) -- (-1.4,3.2) node[circ2, fill = white, text width = 0.5cm] (P_a_1){$P_1$};  
    \draw (I_b_1) -- (-1.4,1.6) node[circ2, fill = white, text width = 0.5cm] (P_b_1){$P_2$}; 
    \draw (I_e_1) -- (-1.4,-3.2) node[circ2, fill = white, text width = 0.5cm] (P_e_1){$P_5$}; 

    \draw (-3, -4.5) node[font=\normalsize] (Step_1){Step 1};
    
    \draw (0,0) -- (1.8,3.2) node[circ3, fill = white, text width = 0.5cm] (I_a_2){}; 
    \draw (0,0) -- (1.8,1.6) node[circ2, fill = blue!20!white, text width = 0.5cm] (I_b_2){$I_2$};
    \draw (0,0) -- (1.8,0) node[circ2, fill = green!10!white, text width = 0.5cm] (E_c_2){$E_3$}; 
    \draw (0,0) -- (1.8,-1.6) node[circ2, fill = green!10!white, text width = 0.5cm] (E_d_2){$E_4$}; 
    \draw (0,0) -- (1.8,-3.2) node[circ2, fill = green!10!white, text width = 0.5cm] (I_e_2){$I_5$}; 

    \draw (I_a_2) -- (3.6,3.2) node[circ2, fill = green!10!white, text width = 0.5cm] (P_a_2){$P_1$};  
    \draw (I_b_2) -- (3.6,1.6) node[circ2, fill = white, text width = 0.5cm] (P_b_2){$P_2$}; 
    \draw (I_e_2) -- (3.6,-3.2) node[circ2, fill = white, text width = 0.5cm] (P_e_2){$P_5$}; 

    \draw (2, -4.5) node[font=\normalsize] (Step_2){Step 2};

    \draw (5,0) -- (6.8,3.2) node[circ3, fill = white, text width = 0.5cm] (I_a_3){}; 
    \draw (5,0) -- (6.8,1.6) node[circ3, fill = white, text width = 0.5cm] (I_b_3){};
    \draw (5,0) -- (6.8,0) node[circ2, fill = green!10!white, text width = 0.5cm] (E_c_3){$E_3$}; 
    \draw (5,0) -- (6.8,-1.6) node[circ2, fill = green!10!white, text width = 0.5cm] (E_d_3){$E_4$}; 
    \draw (5,0) -- (6.8,-3.2) node[circ2, fill = gray!20!white, text width = 0.5cm] (I_e_3){$I_5$}; 

    \draw (I_a_3) -- (8.6,3.2) node[circ2, fill = green!10!white, text width = 0.5cm] (P_a_3){$P_1$};  
    \draw (I_b_3) -- (8.6,1.6) node[circ2, fill = green!10!white, text width = 0.5cm] (P_b_3){$P_2$};  
    \draw (I_e_3) -- (8.6,-3.2) node[circ2, fill = white, text width = 0.5cm] (P_e_3){$P_5$}; 
    
    \draw (7.4, -4.5) node[font=\normalsize] (Step_3){Step 3};

    \draw (2, -6) node[font=\normalsize] {The Branching Project Representation};

    \draw (14,3.2) node[circ1, fill = blue!20!white, text width = 0.4cm] (I_A_2){$I_1$}; 
    \draw (17,1.2) node[circ1, fill = blue!20!white, text width = 0.4cm] (I_B_2){$I_2$}; 
    \draw (14,-1) node[circ1, fill = green!10!white, text width = 0.4cm] (P_A_2){$P_1$}; 
    \draw (16,-1) node[circ1, fill = green!10!white, text width = 0.4cm] (E_C_2){$E_3$}; 
    \draw (18,-1) node[circ1, fill = green!10!white, text width = 0.4cm] (E_D_2){$E_4$};
    \draw (20,-1) node[circ1, fill = green!10!white, text width = 0.4cm] (P_B_2){$P_2$}; 
    \draw (17,-3.2) node[circ1, fill = gray!20!white, text width = 0.4cm] (I_E_2){$I_5$}; 
    
    \draw [->, red, thick] (I_A_2) -- (I_B_2);
    \draw [->, red, thick] (I_B_2) -- (P_A_2);
    \draw [->, red, thick] (I_B_2) -- (E_C_2);
    \draw [->, red, thick] (I_B_2) -- (E_D_2);

    \draw [blue, thick, dashed] (13, -1.9) -- (13, -0.1) -- (21, -0.1) -- (21, -1.9) -- (13, -1.9); 
    \draw [blue, thick, ->] (17,-1.9) -- (I_E_2); 

    \draw (17, -6) node[font=\normalsize] {The Ranking Representation};
\end{tikzpicture}

%% file: Tables/tab_firstview_app.tex
\begin{threeparttable}
\centering
\smallskip
\begin{tabular}{ccccccc}
\toprule
                            & True val & \multicolumn{5}{c}{Estimates}                                                                                                                                                                                                                                                                        \\ \cline{3-7} 
                            &          & (1)                                                      & (2)                                                      & (3)                                                      & (4)                                                      & Full Info                                                \\ \midrule
$\gamma_1$                  & 1        & \begin{tabular}[c]{@{}c@{}}1.219\\ (0.090)\end{tabular}  & \begin{tabular}[c]{@{}c@{}}0.992\\ (0.017)\end{tabular}  & \begin{tabular}[c]{@{}c@{}}1.010\\ (0.021)\end{tabular}  & \begin{tabular}[c]{@{}c@{}}0.986\\ (0.022)\end{tabular}  & \begin{tabular}[c]{@{}c@{}}0.992\\ (0.017)\end{tabular}  \\ \addlinespace
$\gamma_2$                  & 0.5      & \begin{tabular}[c]{@{}c@{}}0.580\\ (0.045)\end{tabular}  & \begin{tabular}[c]{@{}c@{}}0.491\\ (0.009)\end{tabular}  & \begin{tabular}[c]{@{}c@{}}0.495\\ (0.009)\end{tabular}  & \begin{tabular}[c]{@{}c@{}}0.493\\ (0.011)\end{tabular}  & \begin{tabular}[c]{@{}c@{}}0.492\\ (0.008)\end{tabular}  \\ \addlinespace
$\gamma_3$                  & 0.2      & \begin{tabular}[c]{@{}c@{}}0.205\\ (0.025)\end{tabular}  & \begin{tabular}[c]{@{}c@{}}0.180\\ (0.011)\end{tabular}  & \begin{tabular}[c]{@{}c@{}}0.186 \\ (0.015)\end{tabular} & \begin{tabular}[c]{@{}c@{}}0.172\\ (0.013)\end{tabular}  & \begin{tabular}[c]{@{}c@{}}0.181\\ (0.011)\end{tabular}  \\ \addlinespace
$\bar{\beta}$               & -0.6     & \begin{tabular}[c]{@{}c@{}}-0.741\\ (0.060)\end{tabular} & \begin{tabular}[c]{@{}c@{}}-0.601\\ (0.021)\end{tabular} & \begin{tabular}[c]{@{}c@{}}-0.611\\ (0.026)\end{tabular} & \begin{tabular}[c]{@{}c@{}}-0.599\\ (0.026)\end{tabular} & \begin{tabular}[c]{@{}c@{}}-0.601\\ (0.021)\end{tabular} \\ \addlinespace
$\sigma_\beta$              & 0.2      & \begin{tabular}[c]{@{}c@{}}0.186\\ (0.112)\end{tabular}  & \begin{tabular}[c]{@{}c@{}}0.192\\ (0.038)\end{tabular}  & \begin{tabular}[c]{@{}c@{}}0.202\\ (0.067)\end{tabular}  & \begin{tabular}[c]{@{}c@{}}0.229\\ (0.059)\end{tabular}  & \begin{tabular}[c]{@{}c@{}}0.192\\ (0.037)\end{tabular}  \\\addlinespace
$1/\lambda$                 & 0.8      & \begin{tabular}[c]{@{}c@{}}1.617\\ (0.339)\end{tabular}  & \begin{tabular}[c]{@{}c@{}}0.781\\ (0.009)\end{tabular}  & \begin{tabular}[c]{@{}c@{}}0.825\\ (0.020)\end{tabular}  & \begin{tabular}[c]{@{}c@{}}0.768\\ (0.011)\end{tabular}  & \begin{tabular}[c]{@{}c@{}}0.781\\ (0.009)\end{tabular}  \\ \addlinespace
                            &          &                                                          &                                                          &                                                          &                                                          &                                                          \\
Log-Likelihood (True value) &          & -8894                                                    & -22792                                                   & -16724                                                   & -18915                                                   & -26639                                                   \\ \addlinespace
Log-Likelihood (Estimates)  &          & -8891                                                    & -22789                                                   & -16722                                                   & -18911                                                   & -26635                                                   \\ \addlinespace
RMSE                        &          & 0.357                                                    & 0.022                                                    & 0.032                                                    & 0.034                                                    & 0.022                                                    \\ \addlinespace
Median Convergence Time (s) &          & 1218.9                                                   & 805.3                                                    & 1612.5                                                   & 1497.5                                                   & 879.2                                                    \\ \bottomrule
\end{tabular}
\smallskip
\begin{tablenotes}
    \item \footnotesize \emph{Notes:} We simulate data for 2,000 consumers and report results averaged over 50 estimations using different random seeds, each based on 800 simulation draws. Standard deviations of the estimates across simulations are reported in parentheses. 
\end{tablenotes}
\end{threeparttable}

%% file: Tables/tab_additional.tex
\begin{threeparttable}
\centering
\smallskip
\begin{tabular}{cccccc}
\hline
                   & True value & \multicolumn{2}{c}{Estimates (Start from true value)}                                                               & \multicolumn{2}{c}{Estimates (Start from all zero)}                                                                 \\ \cline{3-6} 
                   &            & Full Info                                                & Additional Info                                          & Full Info                                                & Additional Info                                          \\ \midrule
$\gamma_1$         & 1          & \begin{tabular}[c]{@{}c@{}}1.040\\ (0.006)\end{tabular}  & \begin{tabular}[c]{@{}c@{}}1.040\\ (0.006)\end{tabular}  & \begin{tabular}[c]{@{}c@{}}0.858\\ (0.252)\end{tabular}  & \begin{tabular}[c]{@{}c@{}}0.911\\ (0.264)\end{tabular}  \\ \addlinespace
$\gamma_2$         & 0.5        & \begin{tabular}[c]{@{}c@{}}0.491\\ (0.003)\end{tabular}  & \begin{tabular}[c]{@{}c@{}}0.491\\ (0.003)\end{tabular}  & \begin{tabular}[c]{@{}c@{}}0.287\\ (0.208)\end{tabular}  & \begin{tabular}[c]{@{}c@{}}0.298\\ (0.218)\end{tabular}  \\ \addlinespace
$\gamma_3$         & -0.2       & \begin{tabular}[c]{@{}c@{}}-0.255\\ (0.004)\end{tabular} & \begin{tabular}[c]{@{}c@{}}-0.254\\ (0.003)\end{tabular} & \begin{tabular}[c]{@{}c@{}}-0.088\\ (0.110)\end{tabular} & \begin{tabular}[c]{@{}c@{}}-0.103\\ (0.156)\end{tabular} \\ \addlinespace
$\bar{\beta}$      & -0.6       & \begin{tabular}[c]{@{}c@{}}-0.668\\ (0.007)\end{tabular} & \begin{tabular}[c]{@{}c@{}}-0.668\\ (0.006)\end{tabular} & \begin{tabular}[c]{@{}c@{}}-0.438\\ (0.188)\end{tabular} & \begin{tabular}[c]{@{}c@{}}-0.467\\ (0.233)\end{tabular} \\ \addlinespace
$\sigma_\beta$     & 0.2        & \begin{tabular}[c]{@{}c@{}}0.202\\ (0.048)\end{tabular}  & \begin{tabular}[c]{@{}c@{}}0.206\\ (0.040)\end{tabular}  & \begin{tabular}[c]{@{}c@{}}0.373\\ (0.287)\end{tabular}  & \begin{tabular}[c]{@{}c@{}}0.330\\ (0.243)\end{tabular}  \\ \addlinespace
$\bar{c}_0$        & -1.5       & \begin{tabular}[c]{@{}c@{}}-1.505\\ (0.004)\end{tabular} & \begin{tabular}[c]{@{}c@{}}-1.504\\ (0.005)\end{tabular} & \begin{tabular}[c]{@{}c@{}}-1.468\\ (0.113)\end{tabular} & \begin{tabular}[c]{@{}c@{}}-1.434\\ (0.118)\end{tabular} \\ 
                   &            &                                                          &                                                          &                                                          &                                                          \\
Log-L (True value) &            & \begin{tabular}[c]{@{}c@{}}-9082\\ (5.39)\end{tabular}   & \begin{tabular}[c]{@{}c@{}}-9221\\ (5.36)\end{tabular}   & \begin{tabular}[c]{@{}c@{}}-9083\\ (4.77)\end{tabular}   & \begin{tabular}[c]{@{}c@{}}-9222\\ (4.71)\end{tabular}   \\ \addlinespace
Log-L (Estimates)  &            & \begin{tabular}[c]{@{}c@{}}-9079\\ (5.37)\end{tabular}   & \begin{tabular}[c]{@{}c@{}}-9218\\ (5.43)\end{tabular}   & \begin{tabular}[c]{@{}c@{}}-9207\\ (87.53)\end{tabular}  & \begin{tabular}[c]{@{}c@{}}-9341\\ (110.02)\end{tabular} \\ \bottomrule
\end{tabular}
\smallskip
\begin{tablenotes}
    \item \footnotesize \emph{Notes:} Data are simulated for 2,000 consumers, and the reported results are obtained after averaging 50 estimations with different seeds and with 1,000 error draws each. The standard deviation of the mean estimate across these simulations is reported in parentheses.
\end{tablenotes}
\end{threeparttable}

%% file: Tables/tab_incomplete_outside.tex
\begin{threeparttable}
\centering
\smallskip
\begin{tabular}{ccccc}
\toprule
                            &          & (1)                                                      & (2)                                                      & (3)                                                      \\  \midrule
                                & \begin{tabular}[c]{@{}c@{}}True\\ value\end{tabular} & \begin{tabular}[c]{@{}c@{}}Estimates\\ (Std. Dev)\end{tabular} & \begin{tabular}[c]{@{}c@{}}Estimates\\ (Std. Dev)\end{tabular} & \begin{tabular}[c]{@{}c@{}}Estimates\\ (Std. Dev)\end{tabular}             \\ \addlinespace
$\gamma^{outside}$          & 0.5      & \begin{tabular}[c]{@{}c@{}}0.534\\ (0.163)\end{tabular}  & \begin{tabular}[c]{@{}c@{}}0.689\\ (0.319)\end{tabular}  & \begin{tabular}[c]{@{}c@{}}0.498\\ (0.045)\end{tabular}  \\ \addlinespace 
$\gamma_1$                  & 1        & \begin{tabular}[c]{@{}c@{}}1.070\\ (0.901)\end{tabular}  & \begin{tabular}[c]{@{}c@{}}0.991\\ (0.036)\end{tabular}  & \begin{tabular}[c]{@{}c@{}}0.994\\ (0.033)\end{tabular}  \\ \addlinespace 
$\gamma_2$                  & 0.5      & \begin{tabular}[c]{@{}c@{}}0.529\\ (0.044)\end{tabular}  & \begin{tabular}[c]{@{}c@{}}0.493\\ (0.021)\end{tabular}  & \begin{tabular}[c]{@{}c@{}}0.495\\ (0.018)\end{tabular}  \\ \addlinespace 
$\gamma_3$                  & 0.2      & \begin{tabular}[c]{@{}c@{}}0.221\\ (0.035)\end{tabular}  & \begin{tabular}[c]{@{}c@{}}0.205\\ (0.021)\end{tabular}  & \begin{tabular}[c]{@{}c@{}}0.203\\ (0.021)\end{tabular} \\ \addlinespace 
$\beta$               & -0.6     & \begin{tabular}[c]{@{}c@{}}-0.634\\ (0.066)\end{tabular} & \begin{tabular}[c]{@{}c@{}}-0.587\\ (0.035)\end{tabular} & \begin{tabular}[c]{@{}c@{}}-0.592\\ (0.033)\end{tabular} \\ \addlinespace 
$\eta_0$                   & -1.8     & \begin{tabular}[c]{@{}c@{}}-3.720\\ (2.612)\end{tabular} & \begin{tabular}[c]{@{}c@{}}-1.901\\ (0.135)\end{tabular} & \begin{tabular}[c]{@{}c@{}}-1.796\\ (0.042)\end{tabular} \\ \addlinespace 
$\eta_1$                   & 0.1      & \begin{tabular}[c]{@{}c@{}}-0.052\\ (0.527)\end{tabular} & \begin{tabular}[c]{@{}c@{}}0.103\\ (0.009)\end{tabular} & \begin{tabular}[c]{@{}c@{}}0.100\\ (0.007)\end{tabular} \\
                            &          &                                                          &                                                          &                                      \\
Log-Likelihood (True value) &          & -8081                                                    & -12569                                                   & -15221                                                   \\ \addlinespace 
Log-Likelihood (Estimates)  &          & -8074                                                    & -12566                                                   & -15218                                                   \\ \bottomrule
\end{tabular}
\smallskip
\begin{tablenotes}
    \item \footnotesize \emph{Notes:} We simulate data for 4000 consumers and report results averaged over 50 estimations using different random seeds, each based on 50 simulation draws. Standard deviations of the estimates across simulations are reported in parentheses. 
\end{tablenotes}
\end{threeparttable}

%% file: Tables/tab_est_compare_expedia.tex
\begin{threeparttable}
  \begin{small}
  \setlength{\tabcolsep}{5pt}
  \begin{tabular}{l ccc ccc}
    \toprule
    & \multicolumn{3}{c}{\textbf{Panel A: Destination 1}} & \multicolumn{3}{c}{\textbf{Panel B: Destination 2}} \\
    \cmidrule(lr){2-4} \cmidrule(lr){5-7}
     & \makecell{GHK with \\ Searched Set} & \makecell{GHK with \\ Searched Set \\ and Purchase} & \makecell{KSFS from \\ Ursu (2018)} & \makecell{GHK with \\ Searched Set} & \makecell{GHK with \\ Searched Set \\ and Purchase} & \makecell{KSFS from \\ Ursu (2018)} \\
    \midrule
    Hotel Stars     & 0.508*** & 0.583*** & 0.375*** & 0.166*** & 0.225*** & 0.088*** \\
                    & (0.029)  & (0.036)  & (0.014)  & (0.029) & (0.035) & (0.028) \\[0.5ex]
    Review Scores   & $-$0.161*** & $-$0.538*** & $-$0.247*** & 0.006 & $-$0.001 & $-$0.045** \\
                    & (0.039)  & (0.055)  & (0.034)  & (0.021) & (0.023) & (0.022) \\[0.5ex]
    Location Scores & 0.056* & $-$0.195*** & $-$0.148*** & 0.197*** & 0.199*** & 0.102*** \\
                    & (0.034)  & (0.053)  & (0.030)  & (0.011) & (0.014) & (0.016) \\[0.5ex]
    Chain            & $-$0.026 & $-$0.050 & $-$0.009 & 0.004 & $-$0.062 & $-$0.010 \\
                    & (0.027)  & (0.036)  & (0.035)  & (0.030) & (0.042) & (0.047) \\[0.5ex]
    Promotion       & 0.232*** & 0.143*** & 0.158*** & 0.105*** & 0.041 & 0.065 \\
                    & (0.026)  & (0.033)  & (0.031)  & (0.039) & (0.042) & (0.045) \\[0.5ex]
    Price (\$100)   & $-$0.537*** & $-$0.386*** & $-$0.287*** & $-$0.334*** & $-$0.439*** & $-$0.231*** \\
                    & (0.026)  & (0.033)  & (0.029)  & (0.023) & (0.021) & (0.029) \\
    \midrule
    \midrule
    & \multicolumn{3}{c}{\textbf{Panel C: Destination 3}} & \multicolumn{3}{c}{\textbf{Panel D: Destination 4}} \\
    \cmidrule(lr){2-4} \cmidrule(lr){5-7}
    Variable & \makecell{GHK with \\ Searched Set} & \makecell{GHK with \\ Searched Set \\ and Purchase} & \makecell{KSFS from \\ Ursu (2018)} & \makecell{GHK with \\ Searched Set} & \makecell{GHK with \\ Searched Set \\ and Purchase} & \makecell{KSFS from \\ Ursu (2018)} \\
    \midrule
    Hotel Stars     & 0.146*** & 0.265*** & 0.098*** & 0.167*** & 0.245*** & 0.194*** \\
                    & (0.038)  & (0.045)  & (0.028)  & (0.037)  & (0.040)  & (0.024) \\[0.5ex]
    Review Scores   & 0.015    & 0.047    & $-$0.049* & 0.104** & $-$0.134*** & $-$0.081** \\
                    & (0.033)  & (0.036)  & (0.026)  & (0.041)  & (0.041)  & (0.029) \\[0.5ex]
    Location Scores & 0.120*** & 0.123*** & $-$0.004 & 0.126*** & 0.073*** & 0.044*** \\
                    & (0.025)  & (0.028)  & (0.025)  & (0.014)  & (0.016)  & (0.016) \\[0.5ex]
    Chain            & $-$0.080 & $-$0.112* & $-$0.181*** & 0.005 & $-$0.066 & $-$0.028 \\
                    & (0.052)  & (0.059)  & (0.056)  & (0.048)  & (0.055)  & (0.057) \\[0.5ex]
    Promotion       & 0.059    & 0.187*** & 0.028    & 0.041    & 0.039    & 0.023 \\
                    & (0.045)  & (0.055)  & (0.062)  & (0.047)  & (0.055)  & (0.055) \\[0.5ex]
    Price (\$100)   & $-$0.122*** & $-$0.144*** & $-$0.121*** & $-$0.247*** & $-$0.268*** & $-$0.186*** \\
                    & (0.024)  & (0.028)  & (0.042)  & (0.028)  & (0.027)  & (0.033) \\
    \bottomrule
  \end{tabular}
  \end{small}
  \begin{tablenotes}[flushleft]
    \small
    \item \textit{Notes:} Standard errors are in parentheses. ***, **, and * denote significance at the 1\%, 5\%, and 10\% levels, respectively. For GHK estimates, standard errors are computed using the BHHH method. KSFS estimation results from \citet{ursu2018power} are reported by rounding the values in the original Table 8 to three decimal places.
  \end{tablenotes}
\end{threeparttable}

%% file: TikzPics/figure_twostage.tex
\begin{tikzpicture}[scale=0.55]
    \tikzstyle{every node}=[font=\normalsize, scale = 0.8]
    \draw (-12,0) -- (-10.5,2.4) node[circ2, fill = blue!20!white, text width = 0.5cm] (B_a_1){$\mathcal{B}_1$};
    \draw (-12,0) -- (-10.5,0) node[circ2, fill = green!10!white, text width = 0.5cm] (B_b_1){$\mathcal{B}_2$}; 
    \draw (-12,0) -- (-10.5,-2.4) node[circ2, fill = green!10!white, text width = 0.5cm] (B_c_1){$\mathcal{B}_3$}; 

    \draw (B_a_1) -- (-9,2.4) node[circ2, fill = white, text width = 0.5cm] (C_a_1){$\mathcal{C}_1$};  
    \draw (B_b_1) -- (-9,0) node[circ2, fill = white, text width = 0.5cm] (C_b_1){$\mathcal{C}_2$}; 
    \draw (B_c_1) -- (-9,-2.4) node[circ2, fill = white, text width = 0.5cm] (C_c_1){$\mathcal{C}_3$};  

    \draw (C_a_1) -- (-7.5,2.4) node[circ2, fill = white, text width = 0.5cm] (A_a_1){$\mathcal{A}_1$};  
    \draw (C_b_1) -- (-7.5,0) node[circ2, fill = white, text width = 0.5cm] (A_b_1){$\mathcal{A}_2$}; 
    \draw (C_c_1) -- (-7.5,-2.4) node[circ2, fill = white, text width = 0.5cm] (A_c_1){$\mathcal{A}_3$};  

    \draw (-9, -4) node[font=\normalsize] (Step_1){Step 1};
    
    \draw (-6,0) -- (-4.5,2.4) node[circ3, fill = white, text width = 0.5cm] (B_a_2){};
    \draw (-6,0) -- (-4.5,0) node[circ2, fill = green!10!white, text width = 0.5cm] (B_b_2){$\mathcal{B}_2$}; 
    \draw (-6,0) -- (-4.5,-2.4) node[circ2, fill = green!10!white, text width = 0.5cm] (B_c_2){$\mathcal{B}_3$}; 

    \draw (B_a_2) -- (-3,2.4) node[circ2, fill = blue!20!white, text width = 0.5cm] (C_a_2){$\mathcal{C}_1$};  
    \draw (B_b_2) -- (-3,0) node[circ2, fill = white, text width = 0.5cm] (C_b_2){$\mathcal{C}_2$}; 
    \draw (B_c_2) -- (-3,-2.4) node[circ2, fill = white, text width = 0.5cm] (C_c_2){$\mathcal{C}_3$};  

    \draw (C_a_2) -- (-1.5,2.4) node[circ2, fill = white, text width = 0.5cm] (A_a_2){$\mathcal{A}_1$};  
    \draw (C_b_2) -- (-1.5,0) node[circ2, fill = white, text width = 0.5cm] (A_b_2){$\mathcal{A}_2$}; 
    \draw (C_c_2) -- (-1.5,-2.4) node[circ2, fill = white, text width = 0.5cm] (A_c_2){$\mathcal{A}_3$};  

    \draw (-3, -4) node[font=\normalsize] (Step_2){Step 2};

    \draw (0,0) -- (1.5,2.4) node[circ3, fill = white, text width = 0.5cm] (B_a_3){};
    \draw (0,0) -- (1.5,0) node[circ2, fill = blue!20!white, text width = 0.5cm] (B_b_3){$\mathcal{B}_2$}; 
    \draw (0,0) -- (1.5,-2.4) node[circ2, fill = green!10!white, text width = 0.5cm] (B_c_3){$\mathcal{B}_3$}; 

    \draw (B_a_3) -- (3,2.4) node[circ3, fill = white, text width = 0.5cm] (C_a_3){};  
    \draw (B_b_3) -- (3,0) node[circ2, fill = white, text width = 0.5cm] (C_b_3){$\mathcal{C}_2$}; 
    \draw (B_c_3) -- (3,-2.4) node[circ2, fill = white, text width = 0.5cm] (C_c_3){$\mathcal{C}_3$};  

    \draw (C_a_3) -- (4.5,2.4) node[circ2, fill = green!10!white, text width = 0.5cm] (A_a_3){$\mathcal{A}_1$};  
    \draw (C_b_3) -- (4.5,0) node[circ2, fill = white, text width = 0.5cm] (A_b_3){$\mathcal{A}_2$}; 
    \draw (C_c_3) -- (4.5,-2.4) node[circ2, fill = white, text width = 0.5cm] (A_c_3){$\mathcal{A}_3$};  

    \draw (3, -4) node[font=\normalsize] (Step_3){Step 3};

    \draw (6,0) -- (7.5,2.4) node[circ3, fill = white, text width = 0.5cm] (B_a_4){};
    \draw (6,0) -- (7.5,0) node[circ3, fill = white, text width = 0.5cm] (B_b_4){}; 
    \draw (6,0) -- (7.5,-2.4) node[circ2, fill = green!10!white, text width = 0.5cm] (B_c_4){$\mathcal{B}_3$}; 

    \draw (B_a_4) -- (9,2.4) node[circ3, fill = white, text width = 0.5cm] (C_a_4){};  
    \draw (B_b_4) -- (9,0) node[circ2, fill = blue!20!white, text width = 0.5cm] (C_b_4){$\mathcal{C}_2$}; 
    \draw (B_c_4) -- (9,-2.4) node[circ2, fill = white, text width = 0.5cm] (C_c_4){$\mathcal{C}_3$};  

    \draw (C_a_4) -- (10.5,2.4) node[circ2, fill = green!10!white, text width = 0.5cm] (A_a_4){$\mathcal{A}_1$};  
    \draw (C_b_4) -- (10.5,0) node[circ2, fill = white, text width = 0.5cm] (A_b_4){$\mathcal{A}_2$}; 
    \draw (C_c_4) -- (10.5,-2.4) node[circ2, fill = white, text width = 0.5cm] (A_c_4){$\mathcal{A}_3$};  

    \draw (9, -4) node[font=\normalsize] (Step_4){Step 4};

    \draw (12,0) -- (13.5,2.4) node[circ3, fill = white, text width = 0.5cm] (B_a_5){};
    \draw (12,0) -- (13.5,0) node[circ3, fill = white, text width = 0.5cm] (B_b_5){}; 
    \draw (12,0) -- (13.5,-2.4) node[circ2, fill = green!10!white, text width = 0.5cm] (B_c_5){$\mathcal{B}_3$}; 

    \draw (B_a_5) -- (15,2.4) node[circ3, fill = white, text width = 0.5cm] (C_a_5){};  
    \draw (B_b_5) -- (15,0) node[circ3, fill = white, text width = 0.5cm] (C_b_5){}; 
    \draw (B_c_5) -- (15,-2.4) node[circ2, fill = white, text width = 0.5cm] (C_c_5){$\mathcal{C}_3$};  

    \draw (C_a_5) -- (16.5,2.4) node[circ2, fill = green!10!white, text width = 0.5cm] (A_a_5){$\mathcal{A}_1$};  
    \draw (C_b_5) -- (16.5,0) node[circ2, fill = blue!20!white, text width = 0.5cm] (A_b_5){$\mathcal{A}_2$}; 
    \draw (C_c_5) -- (16.5,-2.4) node[circ2, fill = white, text width = 0.5cm] (A_c_5){$\mathcal{A}_3$};  

    \draw (15, -4) node[font=\normalsize] (Step_5){Step 5};
    
    \draw (3, -5.5) node[font=\normalsize] {The Branching Project Representation};

    \draw (-2,-7.5) node[circ1, fill = blue!20!white, text width = 0.4cm] (B_A_2){$\mathcal{B}_1$}; 
    \draw (2,-7.5) node[circ1, fill = blue!20!white, text width = 0.4cm] (C_A_2){$\mathcal{C}_1$}; 
    \draw (0,-10) node[circ1, fill = blue!20!white, text width = 0.4cm] (B_B_2){$\mathcal{B}_2$}; 
    \draw (4,-10) node[circ1, fill = blue!20!white, text width = 0.4cm] (C_B_2){$\mathcal{C}_2$}; 
    \draw (8,-10) node[circ1, fill = blue!20!white, text width = 0.4cm] (A_B_2){$\mathcal{A}_2$};
    \draw (6,-12.5) node[circ1, fill = green!10!white, text width = 0.4cm] (A_A_2){$\mathcal{A}_1$};
    \draw (2,-12.5) node[circ1, fill = green!10!white, text width = 0.4cm] (B_C_2){$\mathcal{B}_3$}; 
    
    \draw [->, red, thick] (B_A_2) -- (B_B_2);
    \draw [->, red, thick] (C_A_2) -- (B_B_2);
    \draw [->, red, thick] (B_B_2) -- (A_A_2);
    \draw [->, red, thick] (B_B_2) -- (B_C_2);
    \draw [->, red, thick] (C_B_2) -- (A_A_2);
    \draw [->, red, thick] (C_B_2) -- (B_C_2);
    \draw [->, red, thick] (A_B_2) -- (A_A_2);
    \draw [->, red, thick] (A_B_2) -- (B_C_2);
    
    \draw (3, -14.3) node[font=\normalsize] {The Ranking Representation};
\end{tikzpicture}

%% file: TikzPics/figure_route.tex
\begin{tikzpicture}[scale = 1]
\tikzstyle{every node}=[font=\small, scale = 0.85]; 

\draw (0,0) node[circ1, fill = green!10!white, text width = 0.4cm] (I_A_2){$I_1$}; 
\draw (1.8,0) node[circ1, fill = blue!10!white, text width = 0.4cm] (I_B_2){$I_2$}; 
\draw (0.9,-1.2) node[circ1, fill = blue!10!white, text width = 0.4cm] (I_C_2){$I_3$}; 
\draw (2.7,-1.2) node[circ1, fill = blue!10!white, text width = 0.4cm] (I_D_2){$I_4$}; 
\draw (3.6,-1.8) node[circ1, fill = blue!10!white, text width = 0.4cm] (P_C_2){$P_3$};

\draw[->] (I_A_2) -- (I_B_2); \draw[->] (I_A_2) -- (I_C_2); 
\draw[->] (I_B_2) -- (I_D_2); \draw[->] (I_C_2) -- (I_D_2); \draw[->] (I_D_2) -- (P_C_2);

\draw (0,-3.2) node[circ1, fill = green!10!white, text width = 0.4cm] (P_A_2){$P_1$}; 
\draw (1.2,-3.2) node[circ1, fill = green!10!white, text width = 0.4cm] (P_B_2){$P_2$}; 
\draw (2.4,-3.2) node[circ1, fill = green!10!white, text width = 0.4cm] (P_D_2){$P_4$}; 
\draw[->] (P_C_2) -- (P_D_2); \draw[->] (P_C_2) -- (P_B_2); \draw[->] (P_C_2) -- (P_A_2); 
\draw (4.5,-3.2) node[circ1, fill = green!10!white, text width = 0.4cm] (I_E_2){$I_5$}; 
\draw[->] (P_C_2) -- (I_E_2);

\draw[blue, thick, ->] (I_D_2.north) .. controls +(-0.1,0.4) .. (I_B_2.east);
\draw[blue, thick, ->] (I_D_2.south west) .. controls +(-0.7,-0.15) .. (I_C_2.south east);
\draw[blue, thick, ->] (P_C_2.north) .. controls +(-0.1,0.2) .. (I_D_2.east);
\draw (2,-4.2) node[font = \normalsize, align = center]{Proposed GHK sampling order 1\\(from bottom to top)}; 

\draw (8,0) node[circ1, fill = green!10!white, text width = 0.4cm] (I_A_3){$I_1$}; 
\draw (9.8,0) node[circ1, fill = blue!10!white, text width = 0.4cm] (I_B_3){$I_2$}; 
\draw (8.9,-1.2) node[circ1, fill = blue!10!white, text width = 0.4cm] (I_C_3){$I_3$}; 
\draw (10.7,-1.2) node[circ1, fill = blue!10!white, text width = 0.4cm] (I_D_3){$I_4$}; 
\draw (11.6,-1.8) node[circ1, fill = blue!10!white, text width = 0.4cm] (P_C_3){$P_3$};

\draw[->] (I_A_3) -- (I_B_3); \draw[->] (I_A_3) -- (I_C_3); 
\draw[->] (I_B_3) -- (I_D_3); \draw[->] (I_C_3) -- (I_D_3); \draw[->] (I_D_3) -- (P_C_3);

\draw (8,-3.2) node[circ1, fill = green!10!white, text width = 0.4cm] (P_A_3){$P_1$}; 
\draw (9.2,-3.2) node[circ1, fill = green!10!white, text width = 0.4cm] (P_B_3){$P_2$}; 
\draw (10.4,-3.2) node[circ1, fill = green!10!white, text width = 0.4cm] (P_D_3){$P_4$}; 
\draw[->] (P_C_3) -- (P_D_3); \draw[->] (P_C_3) -- (P_B_3); \draw[->] (P_C_3) -- (P_A_3); 
\draw (12.5,-3.2) node[circ1, fill = green!10!white, text width = 0.4cm] (I_E_3){$I_5$}; 
\draw[->] (P_C_3) -- (I_E_3);

\draw[blue, thick, ->] (I_D_3.north) .. controls +(-0.1,0.4) .. (I_B_3.east);
\draw[blue, thick, ->] (I_D_3.south west) .. controls +(-0.7,-0.15) .. (I_C_3.south east);
\draw[blue, thick, ->] (I_D_3.east) .. controls +(0.1,0) and +(-0.1,0.3) .. (P_C_3.north);
\draw (10,-4.2) node[font = \normalsize, align = center]{Proposed GHK sampling order 2\\(from middle to peripheral)}; 

\draw (0,-6) node[circ1, fill = green!10!white, text width = 0.4cm] (I_A_4){$I_1$}; 
\draw (1.8,-6) node[circ1, fill = blue!10!white, text width = 0.4cm] (I_B_4){$I_2$}; 
\draw (0.9,-7.2) node[circ1, fill = blue!10!white, text width = 0.4cm] (I_C_4){$I_3$}; 
\draw (2.7,-7.2) node[circ1, fill = blue!10!white, text width = 0.4cm] (I_D_4){$I_4$}; 
\draw (3.6,-7.8) node[circ1, fill = blue!10!white, text width = 0.4cm] (P_C_4){$P_3$};

\draw[->] (I_A_4) -- (I_B_4); \draw[->] (I_A_4) -- (I_C_4); 
\draw[->] (I_B_4) -- (I_D_4); \draw[->] (I_C_4) -- (I_D_4); \draw[->] (I_D_4) -- (P_C_4);

\draw (0,-9.2) node[circ1, fill = blue!10!white, text width = 0.4cm] (P_A_4){$P_1$}; 
\draw (1.2,-9.2) node[circ1, fill = blue!10!white, text width = 0.4cm] (P_B_4){$P_2$}; 
\draw (2.4,-9.2) node[circ1, fill = blue!10!white, text width = 0.4cm] (P_D_4){$P_4$}; 
\draw[->] (P_C_4) -- (P_D_4); \draw[->] (P_C_4) -- (P_B_4); \draw[->] (P_C_4) -- (P_A_4); 
\draw (4.5,-9.2) node[circ1, fill = green!10!white, text width = 0.4cm] (I_E_4){$I_5$}; 
\draw[->] (P_C_4) -- (I_E_4);

\draw[blue, thick, ->] (I_D_4.north) .. controls +(-0.1,0.4) .. (I_B_4.east);
\draw[blue, thick, ->] (I_D_4.south west) .. controls +(-0.7,-0.15) .. (I_C_4.south east);
\draw[blue, thick, ->] (P_C_4.north) .. controls +(-0.1,0.2) .. (I_D_4.east);
\draw[blue, thick, ->] (P_A_4.north) .. controls +(0,0.6) and +(-0.6,-0.2) .. (P_C_4.west);
\draw[blue, thick, ->] (P_B_4.north) .. controls +(0,0.6) and +(-0.4,-0.1) .. (P_C_4.south west);
\draw[blue, thick, ->] (P_D_4.east) .. controls +(0.6,0.2) .. (P_C_4.south);
\draw (2,-10.2) node[font = \normalsize, align = center]{Policy-based GHK sampling order}; 

\draw (8,-6) node[circ1, fill = green!10!white, text width = 0.4cm] (I_A_1){$I_1$}; 
\draw (9.8,-6) node[circ1, fill = blue!10!white, text width = 0.4cm] (I_B_1){$I_2$}; 
\draw (8.9,-7.2) node[circ1, fill = blue!10!white, text width = 0.4cm] (I_C_1){$I_3$}; 
\draw (10.7,-7.2) node[circ1, fill = blue!10!white, text width = 0.4cm] (I_D_1){$I_4$}; 
\draw (11.6,-7.8) node[circ1, fill = blue!10!white, text width = 0.4cm] (P_C_1){$P_3$}; 

\draw[->] (I_A_1) -- (I_B_1); \draw[->] (I_A_1) -- (I_C_1); 
\draw[->] (I_B_1) -- (I_D_1); \draw[->] (I_C_1) -- (I_D_1); \draw[->] (I_D_1) -- (P_C_1); 

\draw (8,-9.2) node[circ1, fill = green!10!white, text width = 0.4cm] (P_A_1){$P_1$}; 
\draw (9.2,-9.2) node[circ1, fill = green!10!white, text width = 0.4cm] (P_B_1){$P_2$}; 
\draw (10.4,-9.2) node[circ1, fill = green!10!white, text width = 0.4cm] (P_D_1){$P_4$}; 
\draw[->] (P_C_1) -- (P_D_1); \draw[->] (P_C_1) -- (P_B_1); \draw[->] (P_C_1) -- (P_A_1); 
\draw (12.5,-9.2) node[circ1, fill = green!10!white, text width = 0.4cm] (I_E_1){$I_5$}; 
\draw[->] (P_C_1) -- (I_E_1); 

\draw[blue, thick, ->] (I_B_1.east) .. controls +(0.3,-0.15) .. (I_D_1.north);
\draw[blue, thick, ->] (I_C_1.south east) .. controls +(0.7,-0.15) .. (I_D_1.south west);
\draw[blue, thick, ->] (P_C_1.north) .. controls +(-0.1,0.2) .. (I_D_1.east);
\draw (10,-10.2) node[font = \normalsize, align = center]{Proposed GHK sampling order 3 \\ (from peripheral to middle)}; 

\end{tikzpicture}

%% file: TikzPics/figure_route_imple.tex
\begin{tikzpicture}[scale = 1]
\tikzstyle{every node}=[font=\small, scale = 0.85]; 
\tikzset{redcurve/.style={->, red, thick, >=stealth}};

\begin{scope}[shift={(0,0)}]
    \draw (0,0) node[circ1, fill = green!10!white, text width = 0.4cm] (I_A_2){$I_1$}; 
    \draw (1.8,0) node[circ1, fill = blue!20!white, text width = 0.4cm] (I_B_2){$I_2$}; 
    \draw (0.9,-1.2) node[circ1, fill = blue!20!white, text width = 0.4cm] (I_C_2){$I_3$}; 
    \draw (2.7,-1.2) node[circ1, fill = blue!20!white, text width = 0.4cm] (I_D_2){$I_4$}; 
    \draw (3.6,-1.8) node[circ1, fill = blue!20!white, text width = 0.4cm] (P_C_2){$P_3$};

    \draw (0,-3.2) node[circ1, fill = green!10!white, text width = 0.4cm] (P_A_2){$P_1$}; 
    \draw (1.2,-3.2) node[circ1, fill = green!10!white, text width = 0.4cm] (P_B_2){$P_2$}; 
    \draw (2.4,-3.2) node[circ1, fill = green!10!white, text width = 0.4cm] (P_D_2){$P_4$}; 
    \draw (4.5,-3.2) node[circ1, fill = green!10!white, text width = 0.4cm] (I_E_2){$I_5$}; 

    \draw[blue, thick, ->] (P_C_2.north) .. controls +(-0.1,0.2) .. (I_D_2.east);
    \draw[blue, thick, ->] (I_D_2.north) .. controls +(-0.1,0.4) .. (I_B_2.east);
    \draw[blue, thick, ->] (I_D_2.south west) .. controls +(-0.7,-0.15) .. (I_C_2.south east);
    \draw[blue, thick, ->] (I_B_2.west) .. controls +(-0.4,0.2) .. (I_A_2.east);
    \draw[blue, thick, ->] (I_C_2.north west) .. controls +(-0.3,0.3) .. (I_A_2.south);
    
    \draw[blue, thick, ->] (P_C_2.east) .. controls +(0.5,-0.4) .. (I_E_2.north);

    \draw[redcurve] (I_A_2) .. controls +(-0.3,-1) .. (P_A_2);
    \draw[redcurve] (I_B_2) .. controls +(0.3,-1) .. (P_B_2);
    \draw[redcurve] (I_D_2) .. controls +(0.3,-0.8) .. (P_D_2);
    \draw[redcurve] (I_C_2.south) .. controls +(0.5,-1.2) and +(-0.2,0) .. (P_C_2.west);
    
    \draw (2,-4.5) node[font = \normalsize, align = center]{Proposed GHK sampling order 1}; 
\end{scope}

\begin{scope}[shift={(8,0)}]

    \draw (0,0) node[circ1, fill = green!10!white, text width = 0.4cm] (I_A_3){$I_1$}; 
    \draw (1.8,0) node[circ1, fill = blue!20!white, text width = 0.4cm] (I_B_3){$I_2$}; 
    \draw (0.9,-1.2) node[circ1, fill = blue!20!white, text width = 0.4cm] (I_C_3){$I_3$}; 
    \draw (2.7,-1.2) node[circ1, fill = blue!20!white, text width = 0.4cm] (I_D_3){$I_4$}; 
    \draw (3.6,-1.8) node[circ1, fill = blue!20!white, text width = 0.4cm] (P_C_3){$P_3$};

    \draw (0,-3.2) node[circ1, fill = green!10!white, text width = 0.4cm] (P_A_3){$P_1$}; 
    \draw (1.2,-3.2) node[circ1, fill = green!10!white, text width = 0.4cm] (P_B_3){$P_2$}; 
    \draw (2.4,-3.2) node[circ1, fill = green!10!white, text width = 0.4cm] (P_D_3){$P_4$}; 
    \draw (4.5,-3.2) node[circ1, fill = green!10!white, text width = 0.4cm] (I_E_3){$I_5$}; 

    \draw[blue, thick, ->] (I_D_3.east) .. controls +(0.1,0) and +(-0.1,0.3) .. (P_C_3.north);
    \draw[blue, thick, ->] (I_D_3.north) .. controls +(-0.1,0.4) .. (I_B_3.east);
    \draw[blue, thick, ->] (I_D_3.south west) .. controls +(-0.7,-0.15) .. (I_C_3.south east);
    \draw[blue, thick, ->] (I_B_3.west) .. controls +(-0.4,0.2) .. (I_A_3.east);
    \draw[blue, thick, ->] (I_C_3.north west) .. controls +(-0.3,0.3) .. (I_A_3.south);

    \draw[blue, thick, ->] (P_C_3.east) .. controls +(0.5,-0.4) .. (I_E_3.north);

    \draw[redcurve] (I_A_3) .. controls +(-0.3,-1) .. (P_A_3);
    \draw[redcurve] (I_B_3) .. controls +(0.3,-1) .. (P_B_3);
    \draw[redcurve] (I_D_3) .. controls +(0.3,-0.8) .. (P_D_3);
    \draw[redcurve] (I_C_3.south) .. controls +(0.5,-1.2) and +(-0.2,0) .. (P_C_3.west);
    
    \draw (2,-4.5) node[font = \normalsize, align = center]{Proposed GHK sampling order 2}; 
\end{scope}

\end{tikzpicture}